\documentclass[oldversion]{aa}  
\pdfoutput=1
\usepackage{graphicx}
\usepackage{txfonts}
\usepackage{natbib}
\usepackage{wasysym}
\bibpunct{(}{)}{;}{a}{}{,} 
\begin{document}
\title{Spectroastrometry of rotating gas disks for the detection of supermassive black holes in galactic nuclei.}
\subtitle{I. Method and simulations}
\author{Alessio Gnerucci\inst{1}, 
            Alessandro Marconi\inst{1},
           Alessandro Capetti\inst{2},
           David J. Axon\inst{3,4},
           Andrew Robinson\inst{3}
}
\offprints{A. Gnerucci}

\institute{Dipartimento di Fisica e Astronomia, Universit\`a degli Studi di Firenze, Firenze, Italy\\
             \email{gnerucci@arcetri.astro.it, marconi@arcetri.astro.it}
             \and INAF - Osservatorio Astronomico di Torino, Strada Osservatorio 20, 10025 Pino Torinese, Italy\\
             \email{capetti@oato.inaf.it}
             \and Physics Department, Rochester Institute of Technology, 85 Lomb Memorial Drive, Rochester, NY 14623, USA\\
             \email{djasps@rit.edu, axrsps@rit.edu}
             \and School of Mathematical \& Physical Sciences, University of Sussex, Falmer, Brighton, BN2 9BH, UK
            }

\date{Received ; accepted}

\abstract{
This is the first in a series of papers in which we study the application of spectroastrometry in the context of gas kinematical studies aimed at measuring the mass of supermassive black holes. The spectroastrometrical method consists in measuring the photocenter of light emission in different wavelength or velocity channels. In particular we explore the potential of spectroastrometry of gas emission lines in galaxy nuclei to constrain the kinematics of rotating gas disks and to measure the mass of putative supermassive black holes.
By means of detailed simulations and test cases, we show that the fundamental advantage of spectroastrometry is that it can provide information on the gravitational potential of a galaxy on scales significantly smaller ($\sim 1/10$) than the limit imposed by the spatial resolution of the observations.
We then describe a simple method to infer detailed kinematical informations from spectroastrometry in longslit spectra and to measure the mass of nuclear mass concentrations. Such method can be applied straightforwardly to integral field spectra, which do not have the complexities due to a partial spatial covering of the source in the case of longslit spectra.
}
  
\authorrunning{Gnerucci et al.}
\titlerunning{Spectroastrometry of rotating gas disks and supermassive black holes.}

  \maketitle


\section{Introduction}\label{s1}

One of the fundamental open questions of modern astrophysics is understanding the physical processes that transformed the nearly homogeneous primordial medium into the present-day universe, characterized by a wealth of complex structures such as galaxies and clusters of galaxies.
Understanding how galaxies formed and how they become the complex systems we observe today is therefore a major theoretical and observational effort. In the last few years, strong evidence has emerged for the existence of tight links between supermassive black holes (BH), nuclear activity and galaxy evolution.
These links reveal what is now called as co-evolution of black holes and their host galaxies.
Strong evidence is provided by the discovery of 'relic' BHs in the center of most nearby galaxies, and that BH masses ($M_\mathrm{BH}\simeq 10^6-10^{10}\, \mathrm{M}_\odot$) are tightly proportional to  structural parameters of the host spheroid like mass, luminosity and stellar velocity dispersion (e.g.~Kormendy \& Richstone 1995, Gebhardt et al. 2000, Ferrarese \& Merritt 2000, Marconi \& Hunt 2003, Ferrarese \& Ford 2005, Graham \& Driver 2008 and references therein).
Moreover, while it is widely accepted that Active Galactic Nuclei (AGN) are powered by accretion of matter on a supermassive BH, it has been possible to show that BH growth is mostly due to accretion of matter during AGN activity, and therefore that most galaxies went through a phase of strong nuclear activity (Soltan 1982, Yu \& Tremaine, Marconi et al. 2004). It is believed that the physical mechanism responsible for this co-evolution of BHs an  their galaxies is probably the feedback by the AGN, i.e.~the accreting BH, on the host galaxy  (Silk \& Rees 1998, Fabian 1999, Granato et al. 2004, Di Matteo et al. 2005, Menci et al. 2006, Bower et al. 2006).

In order to proceed further it is important to secure the most evident sign of co-evolution, the correlations between BH mass and galaxy properties which can be achieved by increasing the number, accuracy and mass range of existing measurements. Supermassive BHs are detected and their masses measured by studying the kinematics of gas or stars in galaxy nuclei and, currently, about $\sim 50$ BH mass measurements most of which in the $10^7-10^9\,\mathrm{M}_\odot$ range and only very few measurements below and above those limits (see, e.g., the most recent compilation by Graham 2008).

This paper, the first in a series, deals with BH mass measurements from gas kinematics and presents a new method, based on spectroastrometry, which can provide a simple but accurate way to estimate BH mass and which partly overcomes the limitations due to spatial resolution which plague the 'classical' gas (or stellar) kinematical methods.

In Sect. \ref{s2} we introduce the standard method for gas
kinematical studies, that is based on the gas rotation curves, and briefly
discuss its characteristics and limitations. In Sect. \ref{s3} we introduce the new gas kinematical method based on
spectroastrometry. We explain the basis of this approach that consist on
measuring ``spectroastrometric curves'' (Sect. \ref{s31}) and show
simulations based on a model of the gas dynamics from which we want to learn
how the spectroastrometric curve changes in function of some parameters of the
model (Sect. \ref{s32}). In Sect. \ref{s4} we explain the practical application of the method. We start by presenting a method for using simultaneously several spectroastrometric curves of the same source (Sect. \ref{s42}) and we consider its application on noisy data (Sect. \ref{s43}). In Sect. \ref{s44} we present a trivial fitting method for recovering the values of the various model parameters using long slit spectroscopy with noisy data. Finally, in Sect. \ref{ifu} we describe the practical application of the method with Integral Field Units (IFU's) and in Sect. \ref{s5} we draw our conclusions.
In appendix \ref{a1} we discuss in details of the spectroastrometric measurements with special regard to the determination of the light centroids and in appendix \ref{a3} we describe in detail the method to recover 2D spectroastrometric maps from multiple slit spectra.

\section{The standard gas kinematical method}\label{s2}
  \begin{figure*}[!ht]
  \centering
  \includegraphics[width=0.9\linewidth]{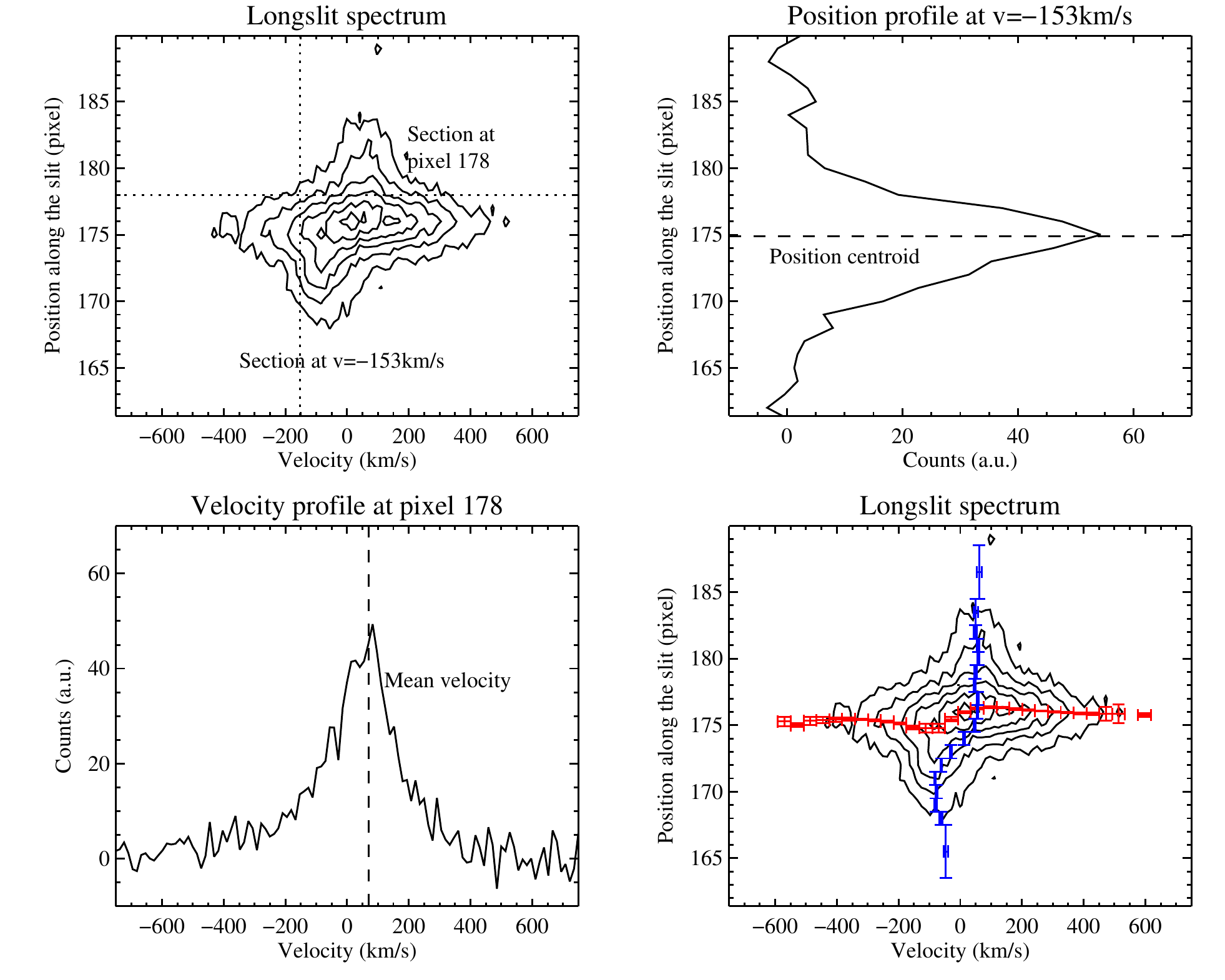}
  \caption{Example of the measurements on which the construction of rotation and spectroastrometric curves is based. Upper left panel: the longslit spectrum of a given continuum-subtracted emission line (isophotes) with superimposed two particular sections at fixed slit position and velocity (dotted lines). Bottom left panel: the profile relative to the Sect. at fixed slit position with superimposed the position of its centroid (dashed line). Upper right panel: the profile relative to the Sect. at fixed velocity with superimposed the position of its centroid (dashed line). Bottom right panel: the position-velocity diagram (PVD) of the upper left panel with superimposed rotation (blue dots with error bars) and spectroastrometric curves (red points with error bars). 
  In all panels velocities are measured relative to the systemic velocity of the galaxy $V_{SYS}$ which is set to the zero point of the velocity scale.}
  \label{fig01}
  \end{figure*}

The standard method of gas kinematics is based on spatially resolved spectra
of the nuclear region of a galaxy. This method consists in recovering the
rotation curve of a given gas emission line from a longslit spectrum. 
The modeling assumes that
the gas is rotating in a thin disc configuration (neglecting hydrodynamical
effects) under the effect of the gravitational potential of the stellar mass
and of a pointlike dark mass $M_{BH}$ that is the BH. The value of $M_{BH}$
and other unknown parameters of the model are obtained by a fitting the
rotation curves 
(see \cite{Marconi:2006} and references therein for details). Because of the many unknown parameters of the model, to better constrain the fit, usually many spectra of the galactic nucleus are obtained each with a different orientation of the spectroscopic slit and a simultaneous fit of all of them is performed.

Clearly, the ability to detect the presence of a BH and measuring its mass strongly depends on the signal-to-noise ratio of the data and only poor constraints on $M_{BH}$ can be obtained from low $S/N$ data. However, even with high $S/N$ data, the
fundamental limit of the standard gas kinematical method (``rotation curves method'' hereafter) resides in the
ability to spatially resolve the region where the gravitational potential of the BH dominates  with respect to the contribution of the stars. Similar
considerations apply equally to stellar dynamical measurements.

As a rule of thumb, this region corresponds to the so-called
 sphere of influence of the BH.
The radius of the sphere of influence
($r_{BH}$) can be estimated as (\citealt{binney-tremaine} ):

\begin{equation}
r_{BH}=\frac{GM_{BH}}{\sigma^2_{\star}} = 0.7
\arcsec\,\left(\frac{M_{BH}}{10^8\,M_\odot}\right)\,\left(\frac{\sigma_\star}{200\,km/s}\right)^{-2}
\,\left(\frac{D}{3\,{\rm Mpc}}\right)^{-1}
\label{1}
\end{equation}
where $G$ is the gravitational constant, $M_{BH}$ the BH mass and
$\sigma_{\star}$ is the velocity dispersion of the stars in the galaxy. For a
galaxy with BH mass $M_{BH}\sim10^8M_{\astrosun}$ and stellar velocity dispersion
$\sigma_{star}\sim200km/s$ we obtain $r_{BH}\sim11.2$pc which, for a very
nearby galaxy at distance $D\sim3$ Mpc, provides an apparent size of $\sim0.7\arcsec$. If a typical very good spatial resolution available from ground based observations is of the order of $\sim0.5\arcsec$ (this Full Width Half Maximum (hereafter FWHM) value should then be compared with twice $r_{BH}$), we can notice that the sphere of influence is only marginally resolved even for very nearby galaxies with moderately large BHs. For a  galaxy distance of $30$\,Mpc the sphere of influence becomes marginally resolved even with the Hubble Space Telescope (HST) which provides the best spatial resolution currently available from space.
Therefore, the rotation curves method can detect only BHs with moderately high masses and located in nearby galaxies; this is a strong limitation from the point of view of a ``demographic'' study of BHs in galactic nuclei, because the adopted ``investigation tool'' cannot reveal the entire BH population. 

\section{The spectroastrometric gas kinematical method}\label{s3}
\begin{figure*}[!ht]
\centering
\includegraphics[width=0.9\linewidth]{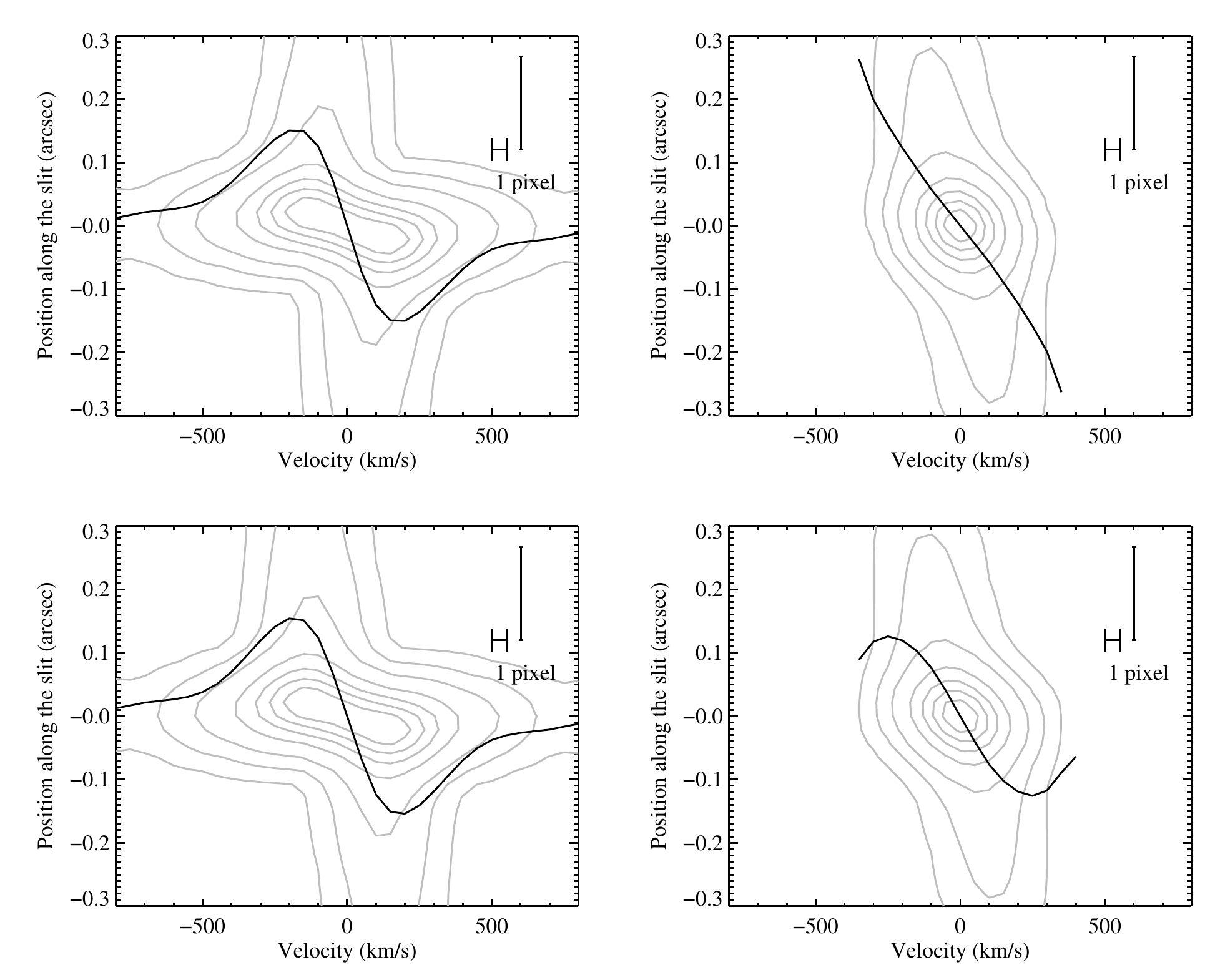}
\caption{ Spectroastrometric curve (black solid line) obtained from a
 simulated spectrum for a rotating gas disk with various values of the
 central BH and stellar mass. In gray we plot the isophotes of the simulated
 line spectrum. Upper left panel: case of $M_{BH}=10^8M_{\astrosun}$ BH
   without mass contribution from stars.  Upper right panel: stars only, with
   $M_{BH}=0$.  Lower panels: $M_{BH}=10^{8}M_{\astrosun}$ (left) and
   $M_{BH}=10^{6.5}M_{\astrosun}$ (right) with included the same stellar mass
   distribution as in model in the upper right panel. Note that the
   differences in the kinematics observed in the left panels (both BH
   dominated, but with and without stellar contributions) are marginal, and
   only visible i.e.\ beyond $\simeq 0\farcs2$ from the nucleus.}
     \label{fig05a}
\end{figure*}

The use of spectroastrometry was originally introduced by \cite{beckers:1982},
\cite{christy:1983} and \cite{Aime:1988} to detect unresolved binaries.
However these earlier studies required specialist instrumentation, and it was
not until the work of \cite{Bailey:1998} that this method was exploited using
standard common user instrumentation: a longslit CCD spectrograph.

Subsequently spectroastrometry has been used by several authors to study pre
main sequence binaries \citep{Baines:2004, Porter:2004, Porter:2005} and the
presence of inflow or outflow or the disk structure on the gas surrounding pre
main sequence stars \citep{takami:2003, Whelan:2005}. More recently,
\cite{Brannigan:2006} discussed the presence and the detection of artifact in
the output of the method and \cite{Pontoppidan:2008} used spectroastrometry on
$4.7\mu m$ CO lines to study the kinematical properties of proto-planetary disks.

The fundamental advantage of the spectroastrometric
method is that, in principle, it can provide position measurements
on scales smaller than the spatial resolution of the observations.

We will now explain the general principle of the spectroastrometric method with a simple example:  consider two point-like sources
located at a distance smaller than the spatial resolution of the telescope;
these sources will be seen as spatially unresolved with their relative
distance not measurable from a conventional image.  However, if in the two
sources are present spectral features, such as absorption or emission lines at
different wavelengths, the light profiles extracted from a longslit spectrum
at these wavelengths will show the two sources separately.  From the
difference in the centroid of the light profiles at these two wavelengths one
can estimate the separation between the two sources even if this is much
smaller that the spatial resolution. This ``overcoming" of the spatial
resolution limit is made possible by the ``spectral" separation of the two
sources.

While this method has been applied to unresolved pointlike sources as binary
stars and protostellar systems, it has never been applied to the problem of
measuring BH masses from the nuclear gas emission in galaxies. Motivated by
the need to overcome the shortcomings of the standard gas kinematic method
described above in Sect. \ref{s2}, in this paper we study the application of
the spectroastrometric method to gas kinematical studies of the mass of BH's
in galactic nuclei. We will start presenting 
the spectroastrometric method based on the same
longslit spectra used for the rotation curves method but instead of the
rotation curves we recover the ``spectroastrometric curve".
However, as shown in Sect. \ref{ifu}, the application of
spectroastrometry to integral field data is even 
more simple and straightforward than to longslit spectra.

Here we focus on the theory of the method and the development of a practical
framework for its application, exploring its capabilities and limitations
using simulated data in order to understand how the spectroastrometric curve
is affected by the object itself or by the instrumental setup. In subsequent
papers we will apply the method to the kinematical data of real galaxies to yield
new improved BH mass measurements.

In this paper we concentrate on the application of the spectroastrometric method to continuum subtracted spectra in order to focus exclusively on the gas kinematics. Indeed, among other things, the underlying continuum only dilutes or modifies the spectroastrometric signal expected for the spatial distribution and kinematics of emission line gas.

\subsection{The spectroastrometric curve}\label{s31}

Here we explain how the spectroastrometric curve is obtained and its main
differences with ``classical" rotation curves.

A longslit spectrum of a continuum subtracted emission line provides a pixel
array whose axes map the dispersion and slit directions, the so-called
position-velocity diagram (hereafter PVD). The slit axis maps the observed
position along the slit. The dispersion axis maps the wavelength of emission
from which the line of sight velocity can be derived.  The upper left panel in
Fig. \ref{fig01} shows the isophotes of an emission line from a PVD.  Ideally, for infinite S/N and perfectly circularly rotating gas, the
rotation curve denotes the mean gas velocity as a function of
the position along the slit. In practice, that curve is obtained by fitting the observed line profiles along the slit with gaussian functions which provide an estimate of the average velocity after discarding components which are clearly not circularly rotating. The lower left panel of Fig. \ref{fig01}
displays the line profile extracted at the slit position marked by the
horizontal dotted line in the PVD.  The spectroastrometric curve provides the
mean position of the emitting gas as a function of velocity. In ideal data that curve is derived by taking the light profile of the line at given velocities along the
dispersion direction and measuring the corresponding mean emission centroids. In practice, the finite S/N of real data requires more complex measurements, which are described below.
The upper right panel of Fig. \ref{fig01} displays the light profile extracted
at the velocity marked by the vertical dotted line in the PVD.  The two curves
can be compared in lower right panel superimposed on the isophotes in the PVD
diagram. Clearly, the two methods analyze the same spectrum from complementary
points of view.
\begin{figure*}[!ht]
\centering
\includegraphics[width=0.9\linewidth]{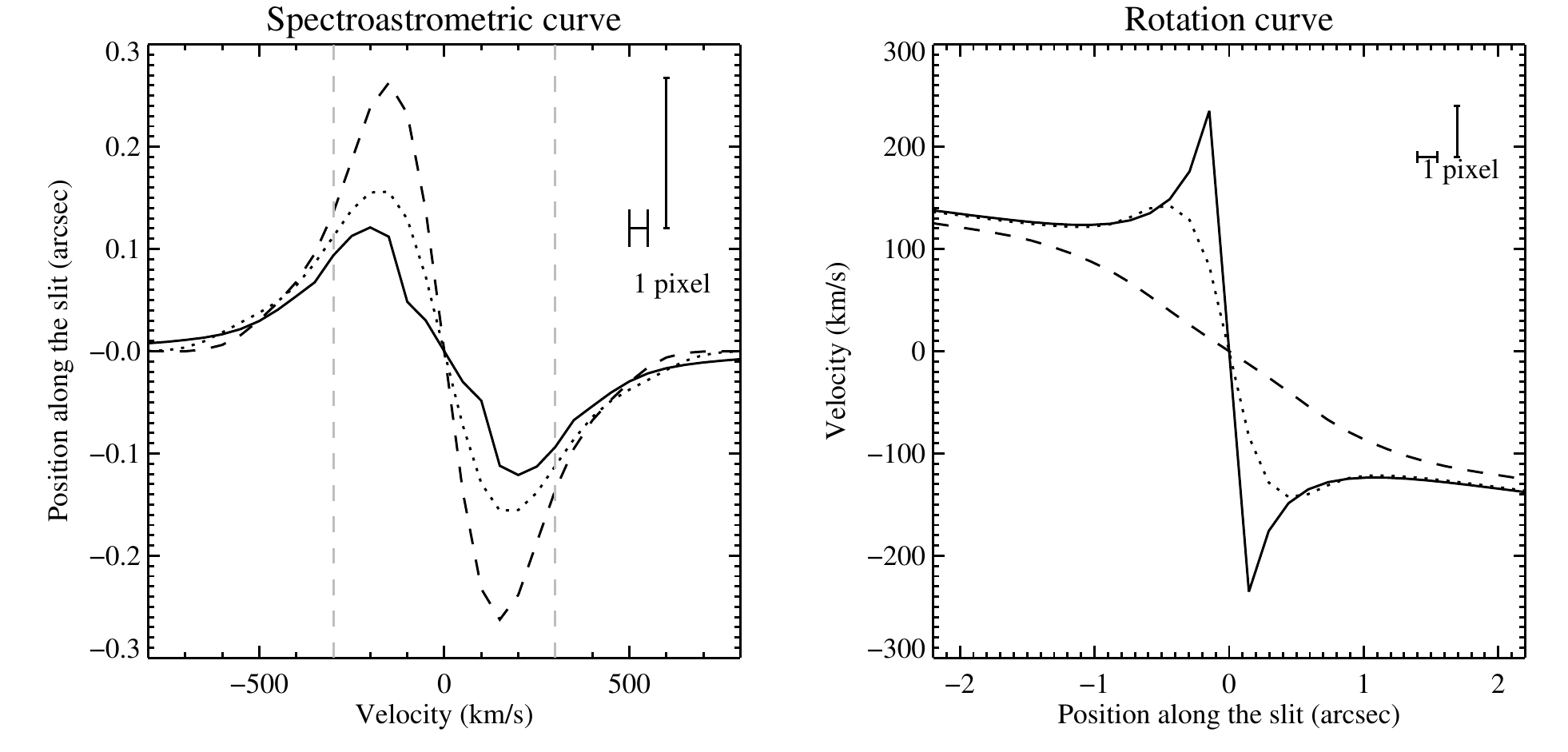}
\caption{Left panel: spectroastrometric curves from spectra differing only in
 spatial resolution. Solid line: spatial resolution of $0.1\arcsec$.  Dotted
 line: spatial resolution of $0.5\arcsec$. Dashed line: spatial resolution of
 $1.0\arcsec$. The vertical long-dashed lines denote the "high velocities". Right panel: corresponding rotation curves. }
     \label{fig11}
\end{figure*}

\begin{figure*}[!ht]
\centering
\includegraphics[width=0.9\linewidth]{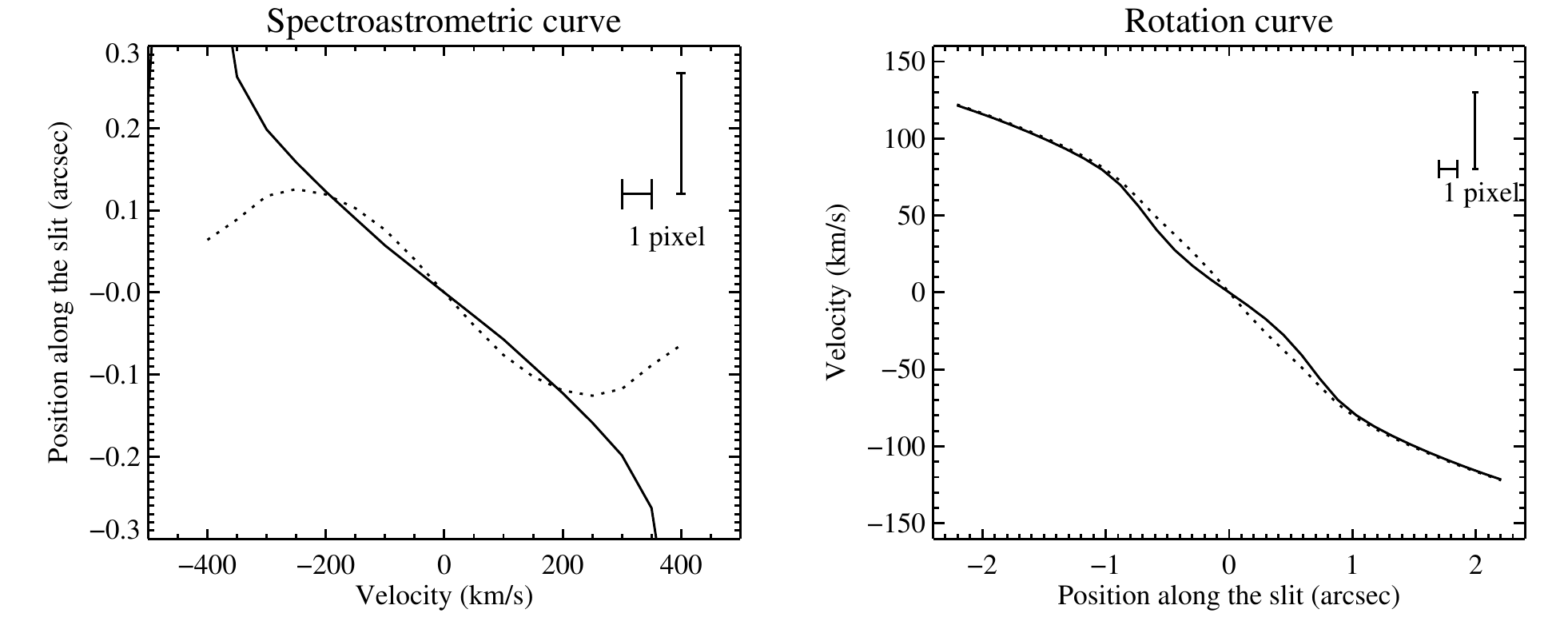}
\caption{Left panel: spectroastrometric curves for spectra differing in the
 value BH mass. Solid line: $M_{BH}=0$. Dotted line:
 $M_{BH}=10^{6.5}M_{\astrosun}$. Right panel: rotation curves for the same
 models.}
     \label{fig12}
\end{figure*}

\subsection{Simulations}\label{s32}

In the following we show the results of the tests based on simulations of
rotating gas disks, outlining the effect of varying the free parameters of the
simulation and of the spectroscopic observations.

The model we use in our simulations is based on a thin gas disk rotating in a
plane under the effect of the gravitational potential produced by a mass
distribution of stars in a galaxy nucleus and by a pointlike dark mass, the
central BH $M_{BH}$. Therefore, we assume that the gas is circularly rotating in the disk plane with the rotational velocity uniquely determined by the combination of $M_{BH}$ and the stellar mass distribution. This model depends on several parameters: the dynamical
parameters (the BH mass, the shape of the mass density function of the stars
and the systemic velocity of the galaxy) and the geometrical parameters that
establish the position and orientation of the gas disk (the distance of the
galaxy or the angular distance scale, the inclination of the disk plane with
respect to the line of sight and the orientation of the line of modes of the
disk). The model takes into account the effect of the shape of the intrinsic
light distribution of the emission line on the sky plane, that is modeled
analytically with a combination of gaussian or exponential functions. The model takes also into
account the effect of the instrumental Point Spread Function that is modeled
with a gaussian function with a given FWHM, and the effect of the others
instrumental setup parameters: the spectral resolution, the slit width and
position angle, the detector's pixel size. With this model we can simulate a
longslit spectrum of a gas emission line in a particular galaxy nucleus
\citep[see][and references therein for a detailed description]{Marconi:2006}.

For simplicity,  in order to illustrate the technique, we make use of a basic
reference model which is meant to closely match the physical parameters of
a BH of mass $10^8M_{\astrosun}$ in an elliptical galaxy in the local universe
observed with a typical longslit spectrograph, like ISAAC at the VLT \citep{moorwood:1999}. The key model parameters are:
\begin{itemize}
\item The distance of the galaxy is set to $3.5\,$Mpc that corresponds to an
 angular distance scale of 17 pc/ \arcsec (e.g. the same of the galaxy
 Centaurus-A). 
\item The disk inclination is set to $35^{\circ}$.
\item The disk line of nodes position angle (with respect to the North direction) is set to $0^{\circ}$ 
\item The amplitude of rotational velocity in the disk due to the stellar mass
 component at $\pm1\arcsec$ is set to $\pm$ 200 km/s.
\item The FWHM of the spatial PSF is set to $0.5\arcsec$ as for typical high
 quality ground based observations.
\item The spectral resolution is set to 10 km/s.
\item The detector's angular pixel size is set to $0.15\arcsec\times0.15\arcsec$, resulting in a spatial oversampling of $\sim 3.5$.
\item The spectrograph slit is usually set parallel to the disk line of nodes unless otherwise specified.
\end{itemize}

We have chosen a basic set of parameters which results in a resolved BH sphere of influence with the aim of clearly showing the spectroastrometric features of disk rotation. In the rest of the paper we will of course consider more extreme sets of parameter values which will result in non-resolved BH spheres of influence in order to show the full power of spectroastrometry.

\begin{figure*}[!ht]
\centering
\includegraphics[width=0.9\linewidth]{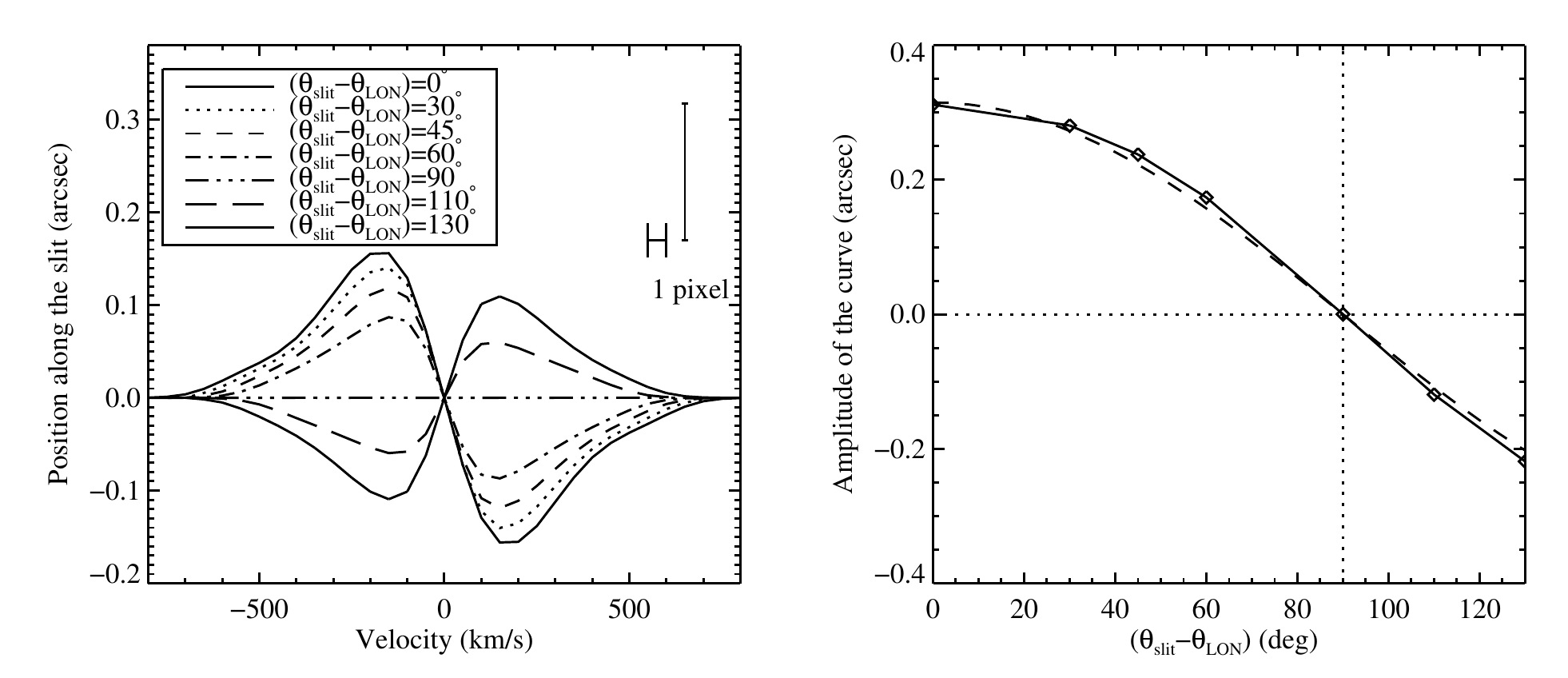}
\caption{Left panel: simulated spectroastrometric curves from spectra
 differing only in slit position angle (referred to the disk line of nodes
 $\theta_{slit}-\theta_{LON}$). Right panel: measured amplitude of the
 spectroastrometric curve (distance between the maximum and the minimum) as a
 function of the slit position angle referred to the disk line of nodes; the
 dashed line denotes a cosine function opportunely rescaled to match the
 value of the amplitude of $\theta_{slit}-\theta_{LON}=0$.}
     \label{fig06}
\end{figure*}

\subsubsection{BH mass}\label{s321}

We first consider the effect of changing the value of the central BH mass on
the spectroastrometric curve.

We simulated the spectrum of a particular rotating gas disk model with
different values of the mass of the central BH, and then we derived the
spectroastrometric curves from these simulated spectra (see Fig.
\ref{fig05a}).

The spectroastrometric curve in Fig. \ref{fig05a} is centered at the systemic
velocity of the galaxy.  In these noiseless simulations, used for illustration
purposes, the spectroastrometric curves are drawn only when the mean flux of the light profile
along the slit is larger than $10^{-3}$ of the maximum flux of the spectrum.
This is meant to give a representation of the low signal-to-noise regions that
should be used in real data for the measurement of the centroid of the light
profile, without considering unrealistically low flux levels.  We defer a discussion of the practical cut off fluxes imposed by the
signal-to-noise in real data to Sect. \ref{s43}.

In the $M_{BH}=0$ case (Fig. \ref{fig05a} upper right panel) the gas kinematics
is due only to the gravitational potential of the stellar mass distribution
and the spectroastrometric curve is monotonically decreasing at increasing
velocity. In all other cases (Fig. \ref{fig05a} other panels), the points at high  velocities ($v\lesssim-300$km/s and $v \gtrsim$300km/s)
tend to approach the $0\arcsec$ value for increasing velocities (we consider
``increasing'' referring to $|v-V_{sys}|$; $V_{sys}=0$ for this simulation),
giving the curve the characteristic ``S"-shape. The reason for this behavior
is that when the gas kinematics is dominated by the BH's gravitational
potential, the gas at high velocity is located close to the BH and the light
centroid of the high velocity gas has to be near to the BH position.

Operatively the ``high velocity'' points are those where the line emission along the slit is spatially unresolved. In Appendix \ref{a1} we describe how to find these points as a by product of centroid determination. We can also recall that the presence of a turnover in the spectroastrometric curve is the signature of the presence of a BH. Therefore, the "high velocity" range is also characterized by a $\sim 75\%$ drop of the spectroastrometric signal with respect to the maximum shift reached at the turnover point.

The velocity in a circular orbit of radius $r$ around a pointlike mass (the
BH) is well approximated by a keplerian law with $v= (GM_{BH}/r)^{1/2}$ in the
innermost region of the gas disk where we can neglect the contribution of the
gravitational potential of the stellar mass distribution. Combined with the above considerations, this also suggests that the spectroastrometric curve asymptotic value
at high velocities provides an estimate of the BH position along the slit.

For increasing $M_{BH}$ values the spectroastrometric curve extends to higher
velocities (higher values of $|v-V_{sys}|$) since this has the effect of
increasing the amount of emission in the high velocity bins.
However the limited extension in the velocity axis of the
spectroastrometric curve with real data is due to the presence of noise. At velocities where
the flux of the line is too low with respect to noise we will not be able to
calculate a reliable value of the centroid of the light profile along the slit.

In conclusion the spectroastrometric curve reveals the presence of a pointlike
mass contribution to the gravitational potential when it shows an ``S-shaped''
structure, with a turn-over of the high velocities components that get closer
to the $0\arcsec$. It can then be concluded that the information about the BH
resides predominantly in the ``high velocity'' part of the curve.

\subsubsection{Spatial resolution}\label{s323}

Here we consider the effects of the spatial resolution on the
spectroastrometric curve. The spatial resolution is the width (FWHM) of the
point spread function which, in the model, is approximated by a gaussian
function.

The results of this test are shown in Fig. \ref{fig11} where we compare
spectroastrometric and normal rotation curves. In the left panel we present
the spectroastrometric curves for models differing only in spatial resolution
while in the right panel we display the corresponding rotation curves. The
spectroastrometric curves differ in the ``low velocity'' range (-300 km/s
$\lesssim v \lesssim$ 300 km/s), as they show a steeper gradient and a large
amplitude.

However, in the ``high velocities'' range differences are negligible,
$\sim0.1$ pix at most.  In section \ref{s321} we showed how the information on
the BH mass is encoded in the high velocity range and, in particular, the
presence of the BH is revealed by the fact that the spectroastrometric curve
approaches $0\arcsec$ at high velocities.  The spectroastrometric curve at the
``high velocities'' is almost unchanged by worsening the spectral resolution,
leaving the BH signature unaltered.

For the ``standard" rotation curves the information about the BH is also
encoded in the presence of points at high velocities at low distances from the
center.  However, a lower spatial resolution the BH signature is effectively
canceled in the rotation curves when the sphere of influence of the BH is not
resolved.  In this simulation the apparent radius of the sphere of influence
is $\sim0.6\arcsec$, which is resolved in the case of $0.1\arcsec$ spatial
resolution, partially resolved for $0.5\arcsec$ case, and unresolved for
$1.0\arcsec$.

Clearly, even in the case of the spectroastrometric curves a better spatial
resolution is desirable. In fact, a poorer spatial resolution results in a
broadening of the light profile along the slit and in a decreased accuracy
with which the centroid position (and consequently the BH mass) can be
measured. Note, however, that the spatial resolution is not a free parameter,
since it can be measured directly from the data. Its effects can be 
modeled and taken into proper account.
\begin{figure*}[htbp]
 \centering
  \includegraphics[width=0.465\linewidth]{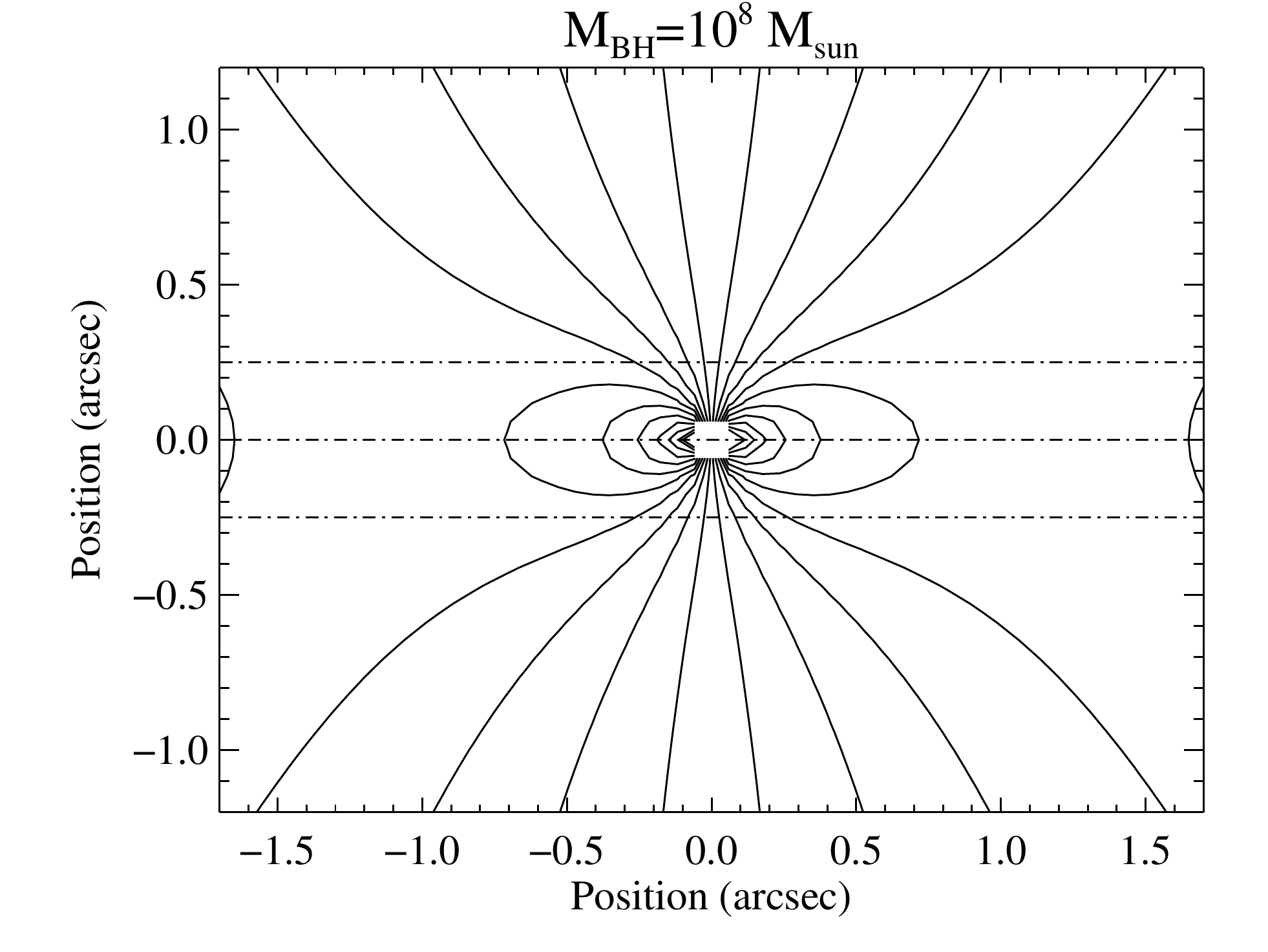} 
  \includegraphics[width=0.465\linewidth]{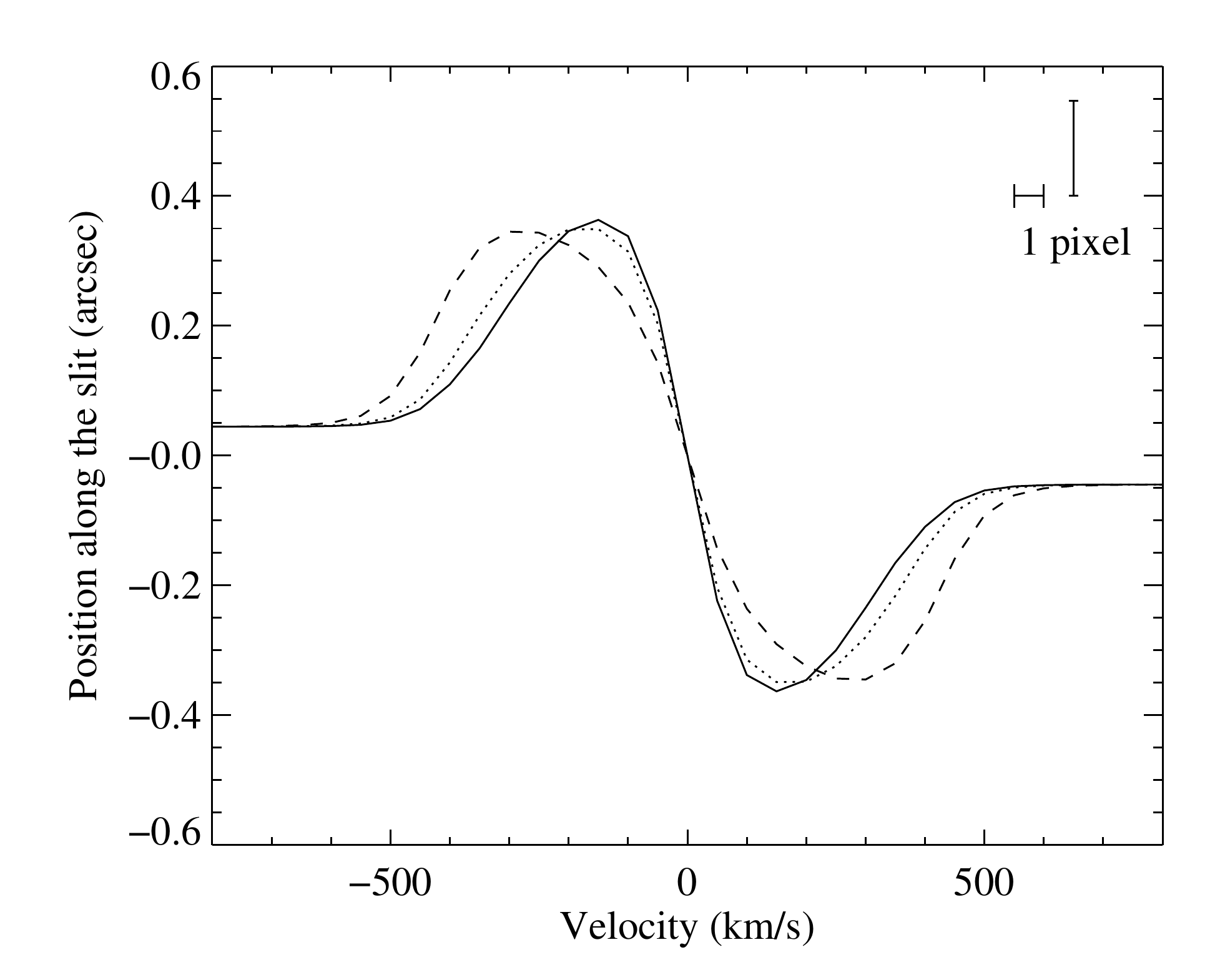}
 \caption{ Left panel: iso-velocity contour map of the line of sight velocity
   field on the plane of sky for a rotating gas disk with the of line of
   nodes along the x axis, with superposed a $0.5\arcsec$ wide slit. Right
   panel: comparison of spectroastrometric curves for simulated spectra
   differing only in slit width (we remind that we assumed a spatial
   resolution of 0$\farcs5$). Solid line: $0.2\arcsec$ wide slit. Dotted
   line: $0.5\arcsec$ wide slit. Dashed line: $1.0\arcsec$ wide slit.}
        \label{fig08}
  \end{figure*}

We have just shown that the the rotation curve changes
drastically when varying the spatial resolution while the spectroastrometric curve
does not. Now we conclude this section with a last set of simulations to
showing in a qualitative but more accurate way how this method can really allow us to
overcome the spatial resolution limit.

The results of these simulations are shown in Fig. \ref{fig12}. In the left
panel we display the spectroastrometric curves for two models with the same
spatial resolution ($0.5\arcsec$) but different BH mass; in the right panel we
display the corresponding rotation curves.

By considering only the rotation curves, the case $M_{BH}=0$ is effectively
indistinguishable from that with $M_{BH}=10^{6.5}M_{\astrosun}$ and indeed the
sphere of influence is unresolved in this case (the apparent
radius of the sphere of influence is $r_{BH}\sim0.02\arcsec$).

Instead, for this particular set of simulations, a
$M_{BH}=10^{6.5}M_{\astrosun}$ mass appears to be still distinguishable from
the $M_{BH}=0$ case; in the high velocity range the differences between the
two models reach a value of $\sim0\farcs2$, i.e. $\sim$ 1.3 pixels. Such
difference can be measured in real data, assuming that we can achieve a
reasonable accuracy of $\sim$ 1 pixel in the measure of the photocenter at
these velocities. In conclusion, according to these simulations we are able to
detect a BH whose apparent size of the sphere of influence is less than
$\sim1/10$ of the spatial resolution.

\subsubsection{Slit position angle}\label{s322}

Here we consider how the spectroastrometric curve is affected by changing the
position angle of the slit. We simulated the spectrum of a particular rotating
gas disk model with different values of the position angle of the slit and
then we get the spectroastrometric curves from this simulated spectra.

In the left panel of Fig. \ref{fig06} we can see the comparison of the
spectroastrometric curves relative to different values of the slit position
angle referred to the disk line of nodes position angle
($\theta_{slit}-\theta_{LON}$ where $\theta_{slit}$ and $\theta_{LON}$ are the
position angles of the slit and disk line of nodes respectively).

We can see that for $\theta_{slit}-\theta_{LON}=0$ (slit aligned with the disk
line of nodes) the spectroastrometric curve has the maximum amplitude. For
increasing values of $\theta_{slit}-\theta_{LON}$ the amplitude decreases
reaching a null value for $\theta_{slit}-\theta_{LON}=90^{\circ}$. For values
larger than $90^{\circ}$ the curve invert itself. This is clearly a
geometrical projection effect just like the one affecting the amplitude of
normal rotation curves as a function of $\theta_{slit}-\theta_{LON}$.  In the
right panel of Fig. \ref{fig06} we show the curve amplitude as a function of
$\theta_{slit}-\theta_{LON}$which is well approximated by a cosine function
(dashed line).

\subsubsection{Slit width}\label{s325}
Another parameter which influences the spectroastrometric curve is the width
of the slit with which the spectra are obtained. Each point of the
spectroastrometric curve represents the centroid of the light profile along
the slit at a given velocity but one can only sample the fraction of
light emitted by the gas at that specific velocity that is intercepted by the
slit.

In Fig. \ref{fig08}, left panel, we show the map of line of sight velocity
field of a rotating gas disk. The gas in a given velocity bin lies in the
locus delimited by two subsequent isovelocity contours.  The impact of
superimposing the slit on the iso-velocity contour map is to artificially
truncate the spatial regions contributing to line emission at a given
velocity that are not confined within the slit extension. This results in
distorted and displaced photo-centers in velocity space.

In Fig. \ref{fig08}, right panel, we show the spectroastrometric curves
obtained from models differing only in slit width. As long as the slit width
is smaller or equal to the spatial resolution of the observations (which is, in this case, $0\farcs5$ FWHM) then differences among spectroastrometric curves in the high velocity range are negligible (e.g., slits with $0.2\arcsec$,
$0.5\arcsec$ width). On the contrary the spectroastrometric curve is
significantly affected for larger slit widths, even in the high velocity range, because of the inclusion of more
extended emission.
This comparison indicates that one should select for the observations
a slit width smaller or at most equal to the spatial resolution of the
observations. As already noted discussing the effects of spatial
resolution, the small residual differences in the spectroscopic curves
corresponding to various slit widths can be effectively modeled out.

\subsubsection{Spectral resolution}\label{s324}
\begin{figure}[!ht]
\centering
\includegraphics[width=0.9\linewidth]{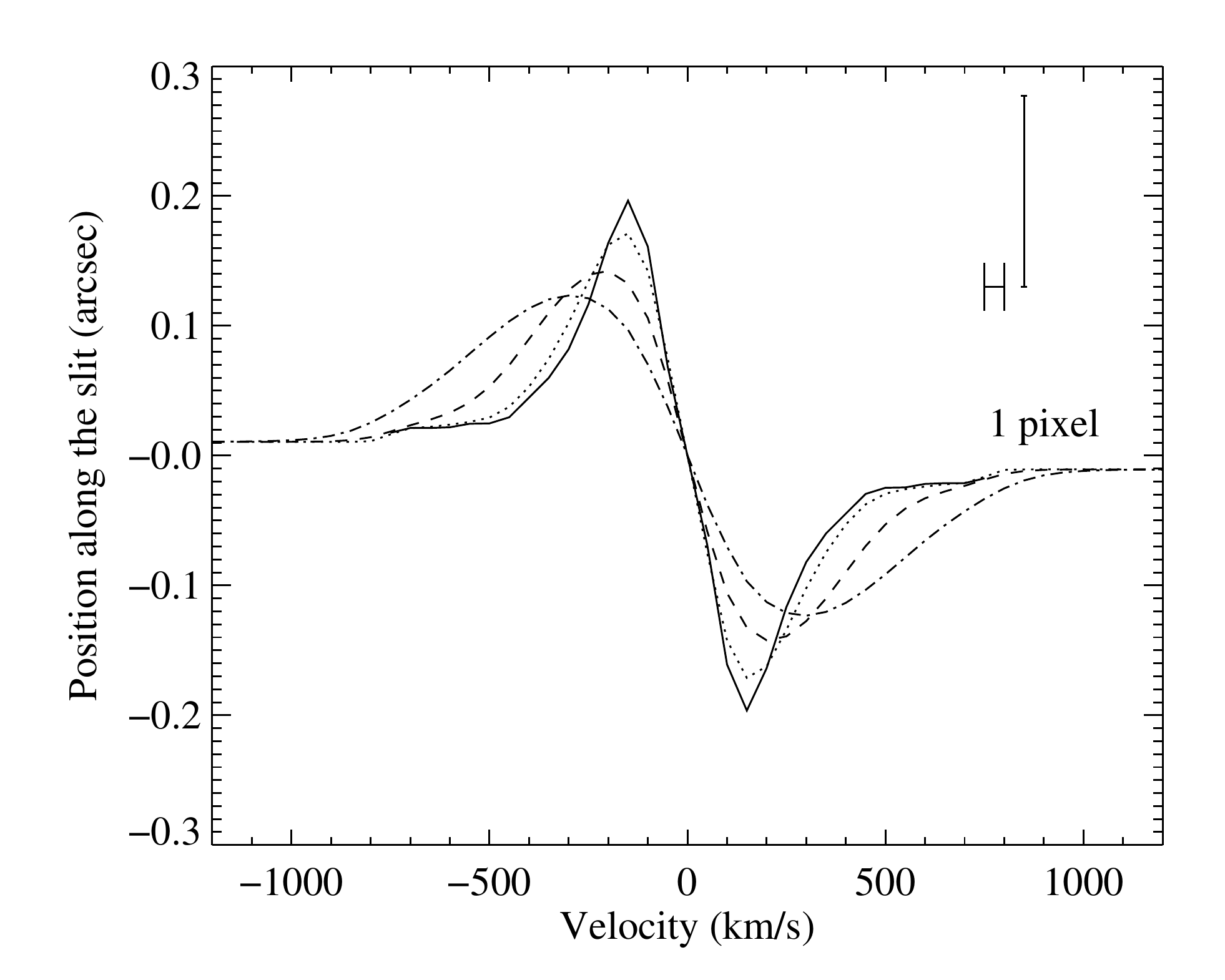} 

\caption{Comparison of spectroastrometric curves for spectra differing only
 in spectral resolution. Solid line: spectral resolution of $10$ km/s.
 Dotted line: spectral resolution of 50 km/s. Dashed line: spectral
 resolution of 100 km/s. Dot-dashed line: spectral resolution of 150
 km/s.}
     \label{fig10}
\end{figure}

We now focus on the effects of the finite spectral resolution of the
observations on the spectroastrometric curves.  In Fig. \ref{fig10} we show
models differing only in spectral resolution. By decreasing the spectral
resolution, the amplitude of the spectroastrometric curves is decreased and
they show, at a given spatial offset, higher velocities. Effectively, due to
the spectral convolution, the curves are stretched along the velocity scale.
To understand this effect we can approximate the true velocity profile and the
instrumental line profile with gaussian functions. The resulting line profile
is the convolution of these two functions and is therefore a gaussian function
with standard deviation
\begin{equation}
\sigma_{wide}=\sqrt{\sigma^2+\sigma_0^2}
\label{3-2}
\end{equation}
where $\sigma_{wide}$ is the resulting standard deviation, $\sigma$ is the
standard deviation of the velocity profile and $\sigma_0$ is the standard
deviation of the instrumental response function.

We here consider as a source of the line broadening the unresolved rotation of
the gas that originates from the fact that one is observing with finite
spatial resolution and is then not able to spatially resolve the high velocity
regions close to the BH.  In our simulations, the line width due to unresolved
rotation reaches $\sigma\sim250$ km/s at the galaxy's center.

As expected, the spectroastrometric curves are almost unchanged at high
spectral resolution ($\sigma_0=10\,,50$ km/s) but they are significantly
altered for the lower spectral resolutions considered ($\sigma_0=100\,,150$
km/s) when $\sigma_0$ approaches the value of $\sigma$. As a consequence, the
BH mass estimates derived from data of insufficient spectral resolution are
systematically overestimated, since at a given position, one measures an
artificially increased velocity.

This result underscores the importance of using data of as high as possible
spectral resolution. The optimal value must be however derived trading-off
with the level of signal-to-noise necessary to build well
defined spectroastreometric curves.

Finally, we note that the value of $\sigma$ can be estimated by modeling the
classical rotation curves. It is then possible to establish, a posteriori,
whether the BH measurement is affected by such an effect and eventually to
validate its value.

\begin{figure}[!ht]
\centering
\includegraphics[width=0.9\linewidth]{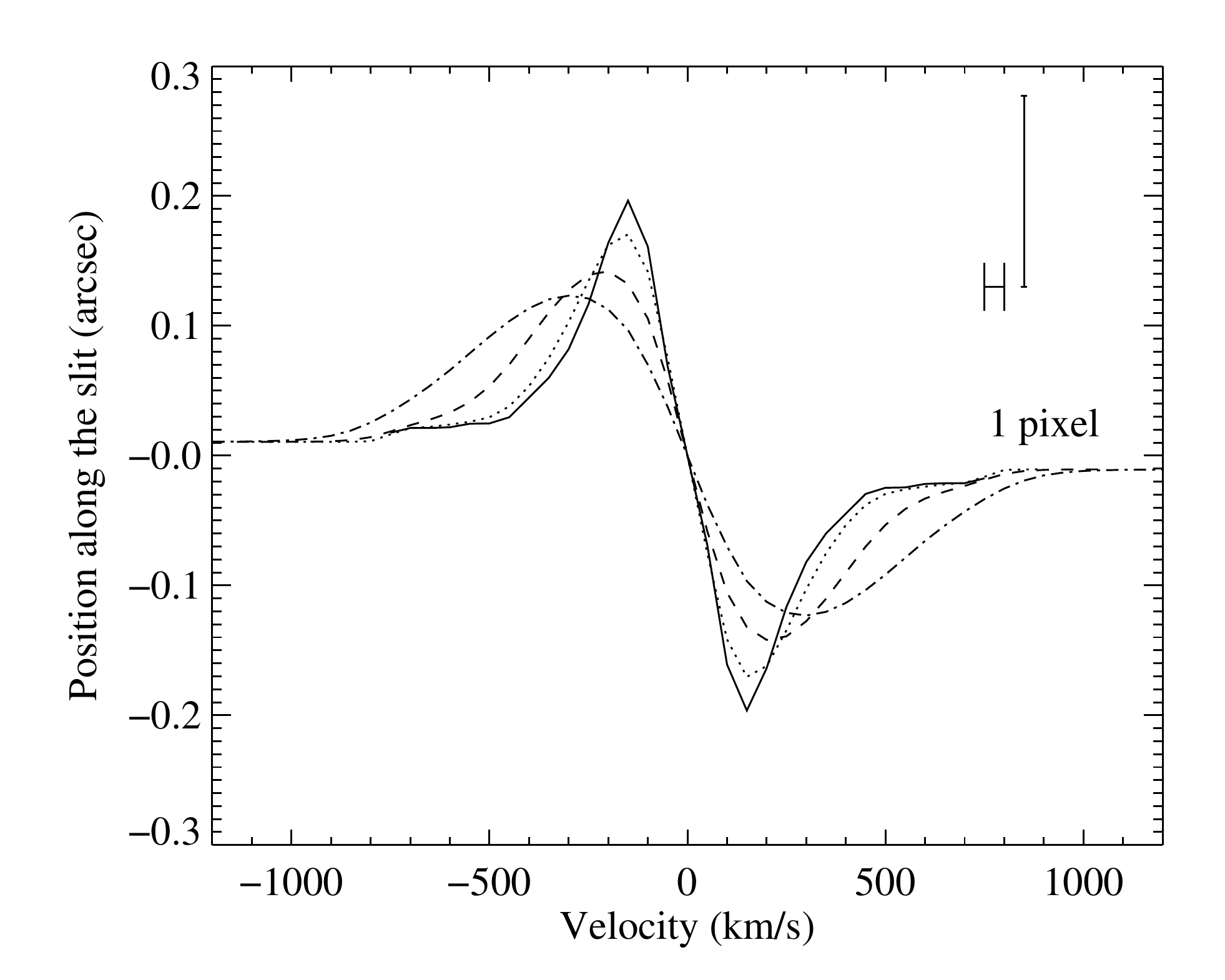} 

\caption{Comparison of spectroastrometric curves for spectra differing only
 in intrinsic velocity dispersion. Solid line: intrinsic velocity dispersion of $10$ km/s.
 Dotted line: intrinsic velocity dispersion  of 50 km/s. Dashed line: s intrinsic velocity dispersion  of 100 km/s. Dot-dashed line:  intrinsic velocity dispersion  of 150 km/s.}
     \label{fig10b}
\end{figure}
\subsubsection{Intrinsic velocity dispersion}\label{s324b}
Another source of line broadening might be due to non circular or chaotic motions in the gas which can be modeled by adding an ``intrinsic'' velocity dispersion to the gas motions (e.g.~see the discussion in \citealt{Marconi:2006}).
In Fig. \ref{fig10b} we show the spectroastrometric curves for models differing
only in intrinsic velocity dispersion. The comparison with Fig. \ref{fig10} clearly shows that, when adopting the same values of $\sigma_0$, no significant difference is found regardless of whether the source of line broadening is poor spectral resolution or an intrinsic velocity dispersion in the source.

In general, when spurious line broadening is present, the high velocity points in the spectroastrometric curve are not
entirely due to the gravitational potential of the BH, and the curve is artificially stretched in the velocity direction
leading to a possible overestimate of the BH mass with the method described in Sec.~\ref{s44}. 

From the simulations presented in Fig. \ref{fig10} and \ref{fig10b} we can verify that the
effect of line broadening on spectroastrometric curves is indeed approximable as 
a x-axis ``stretching''. Following equation \ref{3-2} we verified that can then recover the ``de-stretched'' velocities as:
\begin{equation}F
\left|v-V_{sys}\right|=\left|v-V_{sys}\right|_{obs}\left(1+\frac{\sigma_0^2}{\sigma^2}\right)^{-1/2}
\label{3-3}
\end{equation}
thus correcting the spectroastrometric curve.
A detailed analysis of the effects of spurious line broadening is beyond the scope of this paper and will be presented elsewhere but, briefly, we can in principle estimate the spurious line broadening $\sigma_0$ and that due to unresolved rotation $\sigma$ by modeling the classical rotation curves (e.g.~\citealt{Marconi:2006} and references therein) and then correct the spectroastrometric curves by de-stretching the velocity axis as indicated in eq. \ref{3-3}.
Clearly, the most accurate approach will be that of a combined fit of the classical rotation and spectroastrometric curve, much beyond the simple analysis presented 
here in Sec.~\ref{s44}.

\subsubsection{Flux distribution}\label{s326}

In this section we present simulations based on simple intrinsic flux distribution models, with the aim of showing how sub-resolution variations of the flux distribution influence the spectroastrometric curve.

The basic model adopted for the flux distribution  in these simulations is an exponential function $I(r)=A e^{r/r_0}$ where
$r$ is the radial distance from the symmetry center located at the position $(x_0,y_0)=(0\arcsec,0\arcsec)$ and $r_0$ is a characteristic radius
which we assume equal to $0.05\arcsec$. At the spatial resolution
of our simulations ($FWHM\simeq0.5\arcsec$), this emission is spatially
unresolved.

\begin{figure}[!ht]
\centering
\includegraphics[width=0.9\linewidth]{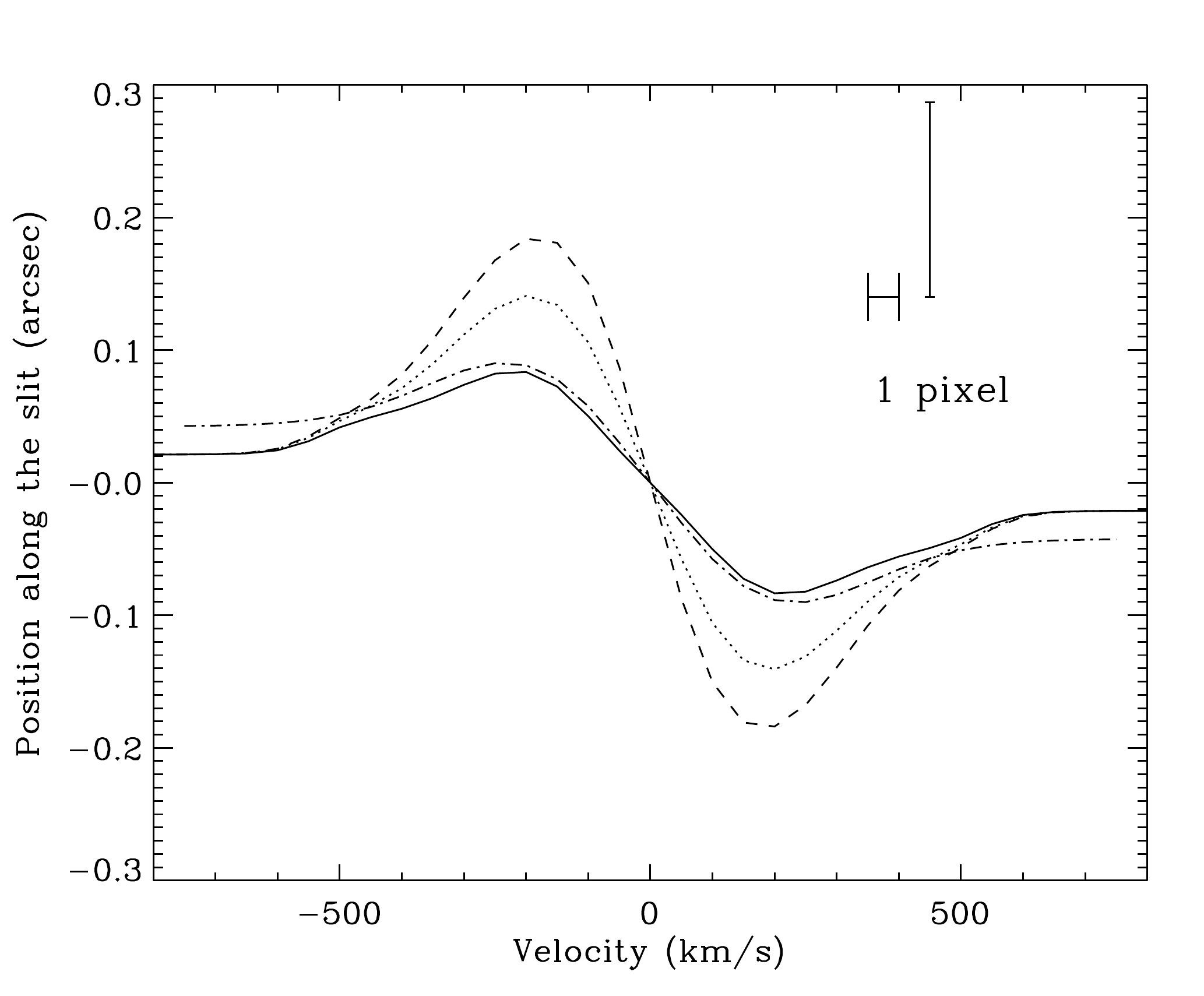}
\caption{Comparison of spectroastrometric curves for spectra differing only in intrinsic flux distribution. Solid line: base model flux distribution (explained in the text). Dotted line: model with $r_0=0.1\arcsec$. Dashed line: model with $r_0=0.15\arcsec$. Dot-dashed line: base model with a central hole of radius $r_h=0.04\arcsec$.}
     \label{fig10bB1}
\end{figure}
\begin{figure}[!ht]
\centering
\includegraphics[width=0.9\linewidth]{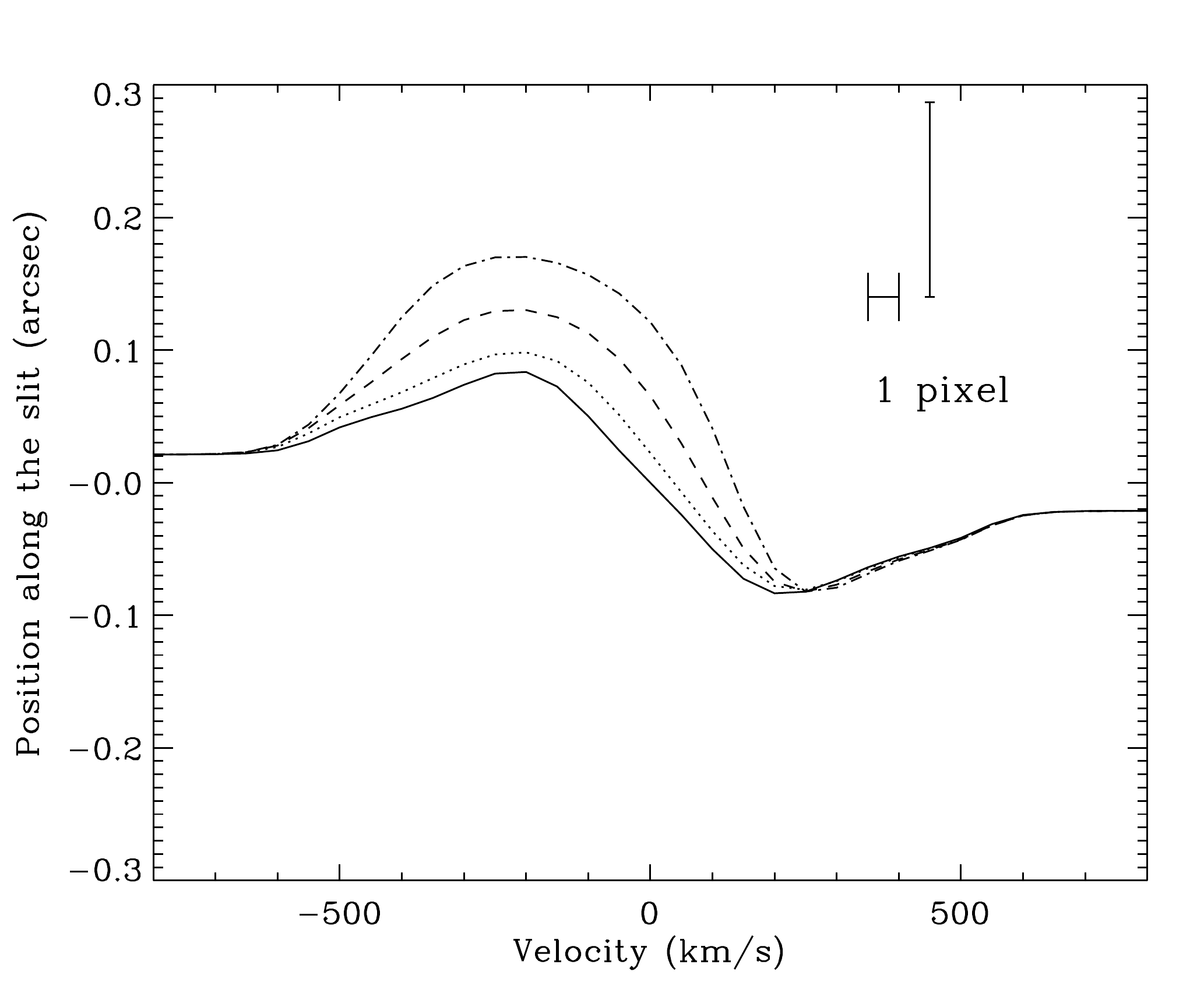}
\caption{Comparison of spectroastrometric curves for spectra differing only in intrinsic flux distribution. Solid line: base model flux distribution (explained in the text). Dotted line: base model centered on $(0.05\arcsec,0.05\arcsec)$ position. Dashed line: base model centered on $(0.1\arcsec,0.1\arcsec)$ position. Dot-dashed line: base model centered on  $(0.15\arcsec,0.15\arcsec)$ position.}
     \label{fig10bB2}
\end{figure}

In the first set of simulations, shown in Fig. \ref{fig10bB1}, we vary the characteristic radius $r_0$ in size and introduce a central hole in the flux distribution with radius $r_h$. In the second set of simulations, shown in Fig. \ref{fig10bB2}, we vary the position of the center of the flux distribution.

In the ``high velocity'' range the variation in the centroid position with respect to the base model is lower than $0.06\arcsec$ and  $0.09\arcsec$ for the first  and second set of simulations, respectively; these values should be compared with the pixel size, $0.125\arcsec$, and the spatial resolution, $FWHM=0.5\arcsec$. As we will show in Sec.~\ref{s44}, these variations will produce only small changes (up to $\pm 0.2$ dex) on the BH mass which can be inferred from spectroastrometric data. On the contrary, much larger variations are present in the ``low velocity'' range.

This behavior, observed also in previous simulations, is explained by the fact that that the gas with lower velocities is located in an extended (spatially resolved) region, and the light profiles along the slit  are influenced by the shape of the flux distribution. Conversely the gas at higher velocities is confined in a small (spatially unresolved) region close to the BH and the light profiles along the slit are not influenced by the shape of the flux distribution.

Only in the case of a central hole in the flux distribution does the largest variation in the spectroastrometric curve occur in the Òhigh velocityÓ range. This behavior is explained by the fact that that the gas with higher velocities (located in a small region close to the BH) is not illuminated because of the central hole of the flux distribution and the centroid position is influenced by the flux distribution at larger distance from the BH. 

\subsubsection{Summary of simulation results}\label{s327}

We can now summarize the results from the above simulations and the
considerations that can be derived.
\begin{itemize}
\item The presence of a black hole is revealed by a turn-over in the
 spectroastrometric curve, with the high velocity components approaching a
 null spatial offset from the center of the galaxy. Conversely, in the case
 of no black hole, the curve shows a monotic behavior.
\item The information about the BH is predominantly encoded in the ``high
 velocity'' range of the spectroastrometric curve which is generated by spatially unresolved emission.
\item The ``high velocity'' range of the curve is not influenced strongly by
 the spatial resolution of the observations, leaving the BH signature
 unaltered. According to our simulations we are able to detect a BH whose
 apparent size of the sphere of influence is as small as $\sim1/10$ of the
 spatial resolution.  Instead, for the ``standard'' rotation curves method
 the information about the BH is effectively canceled when the sphere of
 influence of the BH is not spatially resolved.
\item The amplitude of the spectroastrometric curve decreases by increasing
 the angle between the slit and the line of nodes of the gas disk,
according to a cosine law.
\item The slit width affects the spectroastrometric curve, due to the
 truncation of the iso-velocity contours. This effect can be reduced to a
negligible level selecting for the observations a 
slit width smaller than (or equal to) the spatial resolution.
\item The effect of the finite spectral resolution is to strech (in velocity
 space) the spectroastrometric curve. A robust BH mass estimates requires 
a velocity resolution smaller than the line width due to (spatially) unresolved
rotation.
\item Variations of the intrinsic line flux distribution at sub-resolution scales, do not greatly affect the spectroastrometric curve in the high velocity range.
\end{itemize}

\section{Practical application of the method}\label{s4}
\subsection{The use of multiple spectra of the source: 
the spectroastrometric  map}\label{s42}

In order to improve the constraints on model parameters with the
``standard method'', it is common practice to obtain several long-slit spectra
at different slit orientations.  Also spectroastrometry benefits from this
approach as it is possible to recover for the same source several
spectroastrometric curves, one for each position angle (PA) of the slit. 
With respect to the standard rotation curves method, where spectra taken
along paraller slits have been often used, here the fundamental requirement 
is that the various slits must be centered on the
expected BH position.

\begin{figure}[!ht]
\centering
\includegraphics[width=0.9\linewidth]{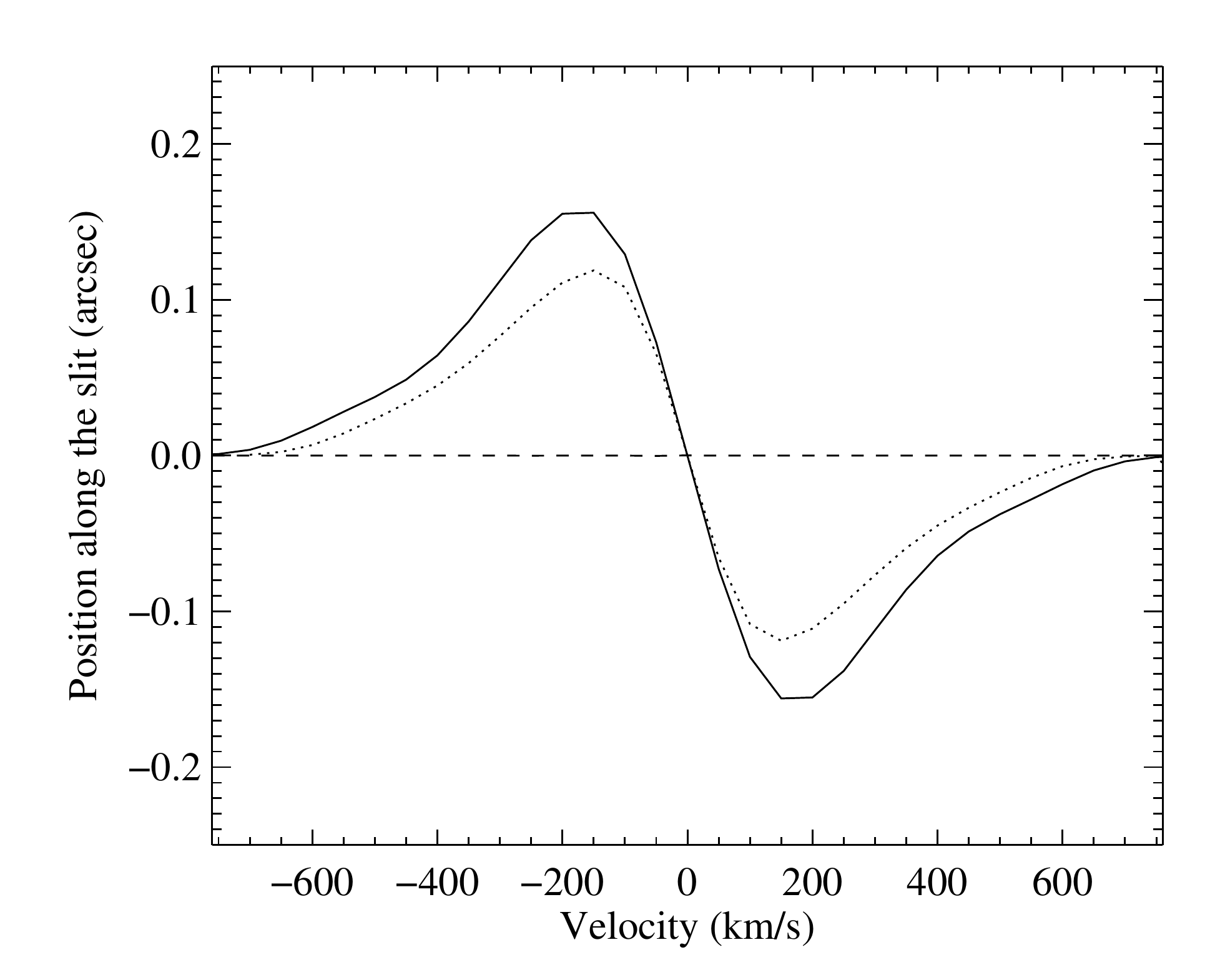} 

\caption{Spectroastrometric curves for three simulated spectra of the same
 model with different PA of the slit. Solid line: slit with PA (referred to
 the disk line of nodes position angle)
 $\theta_{slit}-\theta_{LON}=0^{\circ}$.  Dotted line: slit with
 $\theta_{slit}-\theta_{LON}=45^{\circ}$. Dashed line: slit with
 $\theta_{slit}-\theta_{LON}=90^{\circ}$.}
     \label{fig13}
\end{figure}

We simulated the case of three longslit spectra, oriented parallel (PA1),
perpendicular (PA2) and forming an angle of $45^{\circ}$ with the line of
nodes (PA3). In Fig. \ref{fig13} we can see the spectroastrometric curves
obtained from the three spectra.

\begin{figure*}[!ht]
\centering
\includegraphics[width=0.9\linewidth]{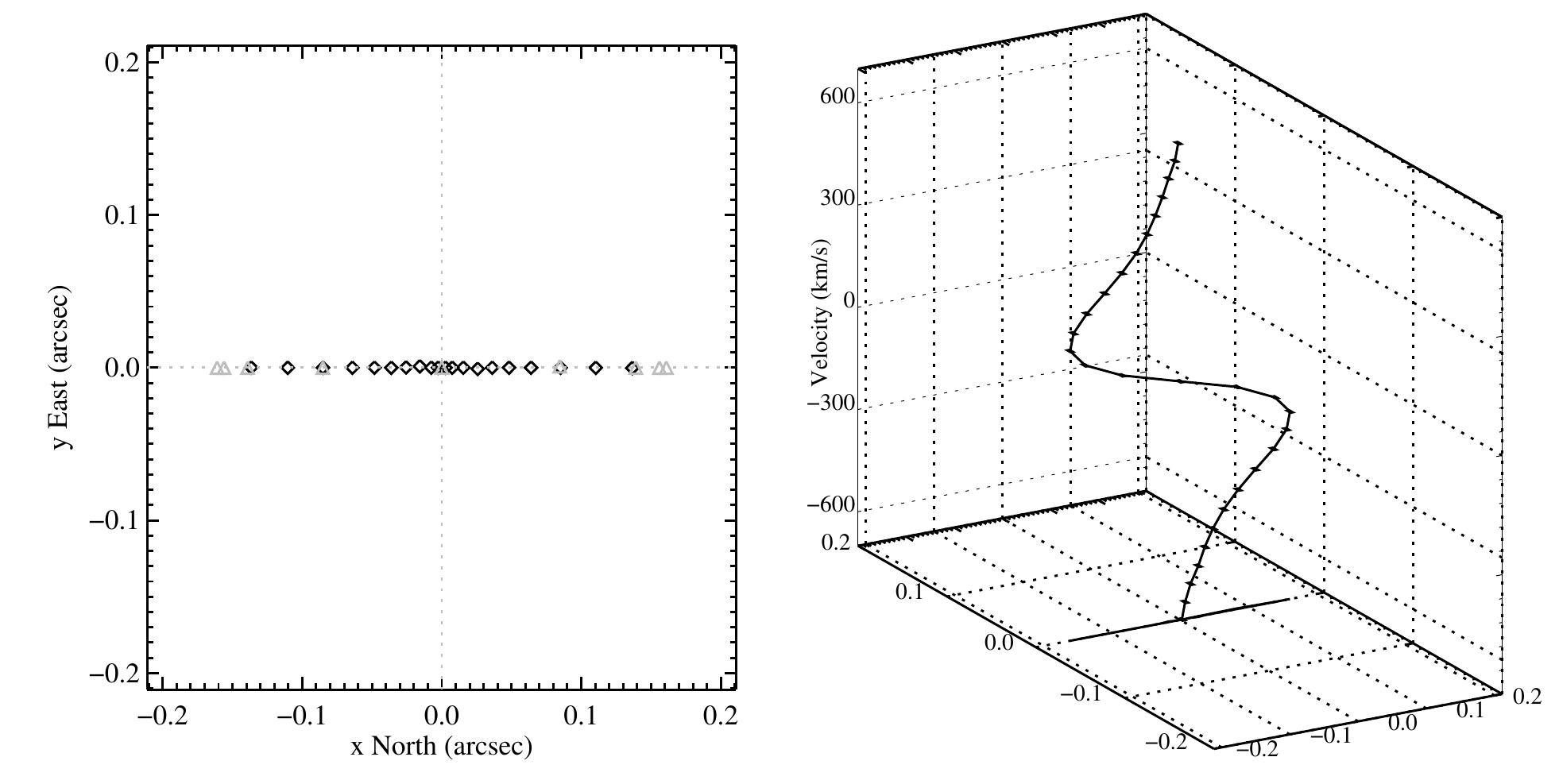}
\caption{Spectroastrometric 2D map derived from the three spectroastrometric
curves of Fig. \ref{fig13}. Left panel: derived photocenter positions on the
sky plane, the black open diamonds are the points actually used for the
$\chi^2$ minimization, the dashed lines indicate the
three slits each with its derived centre location (filled squares). Right
panel: the 3D plot of the map, where the z axis is velocity.}
    \label{fig14}
\end{figure*}

Each spectroastrometric curve provides the photocenter position along one
slit, i.e.~the position of the photocenter projected along the axis identified
by the slit. Combing the spectroastrometric curves we can thus obtain the map
of photocenter positions on the plane of the sky for each velocity bin. In
principle, the spectroastrometric curves from two non-parallel slits should
suffice but we can use the redundant information from the three slits to
recover the 2D sky map as described in more detail in appendix \ref{a3}.

In Fig. \ref{fig14} (left panel) we present the 2D spectroastrometric map on
the plane of the sky. All the points of the map
lies on a straight line which identifies the line of nodes. Indeed, whenever
the intrinsic distribution of line emission is circularly symmetric and
centered on the BH position (as assumed in our simulations), the light
centroids at a given velocity are located on the disk line of nodes. In the
right panel of Fig. \ref{fig14} we show the three-dimensional representation
of the spectroastrometric curve combining the 2D spatial map of the left panel
with the velocity axis. In the 3D representation all the points lie on a plane
parallel to the velocity axis and aligned with the disk line of nodes. On that
plane, the spectroastrometric curve present the usual S shape symmetric with
respect to the zero point of velocities.

We remark that in this paper we only apply the spectroastrometric method to continuum subtracted spectra. When dealing with real data an accurate continuum subtraction will be an important requirement to obtain an accurate spectroastrometric curve, especially for the ``high velocity'' points where the line-continuum contrast is lower. Any residual continuum emission can alter the spatial profile of the emission line, and the "turn over" signature of a point mass in the spectroastrometric curve. In the latter case, the spectroastrometric displacement of the emission line will be shifted towards the continuum photocentre in the line wings no matter the actual spatial distribution of the high velocity gas. The accuracy of continuum subtraction, as well as the signal to noise ratio at high velocities, is therefore a fundamentally limit to the ability of detecting  the central BH. A more detailed analysis of the effects of continuum emission will be presented in a forthcoming paper where we will apply the spectroastrometric method to real data.

\subsection{The effect of noise on spectroastrometry}\label{s43}
 \begin{figure*}[!ht]
 \centering 
  \includegraphics[width=0.48\linewidth]{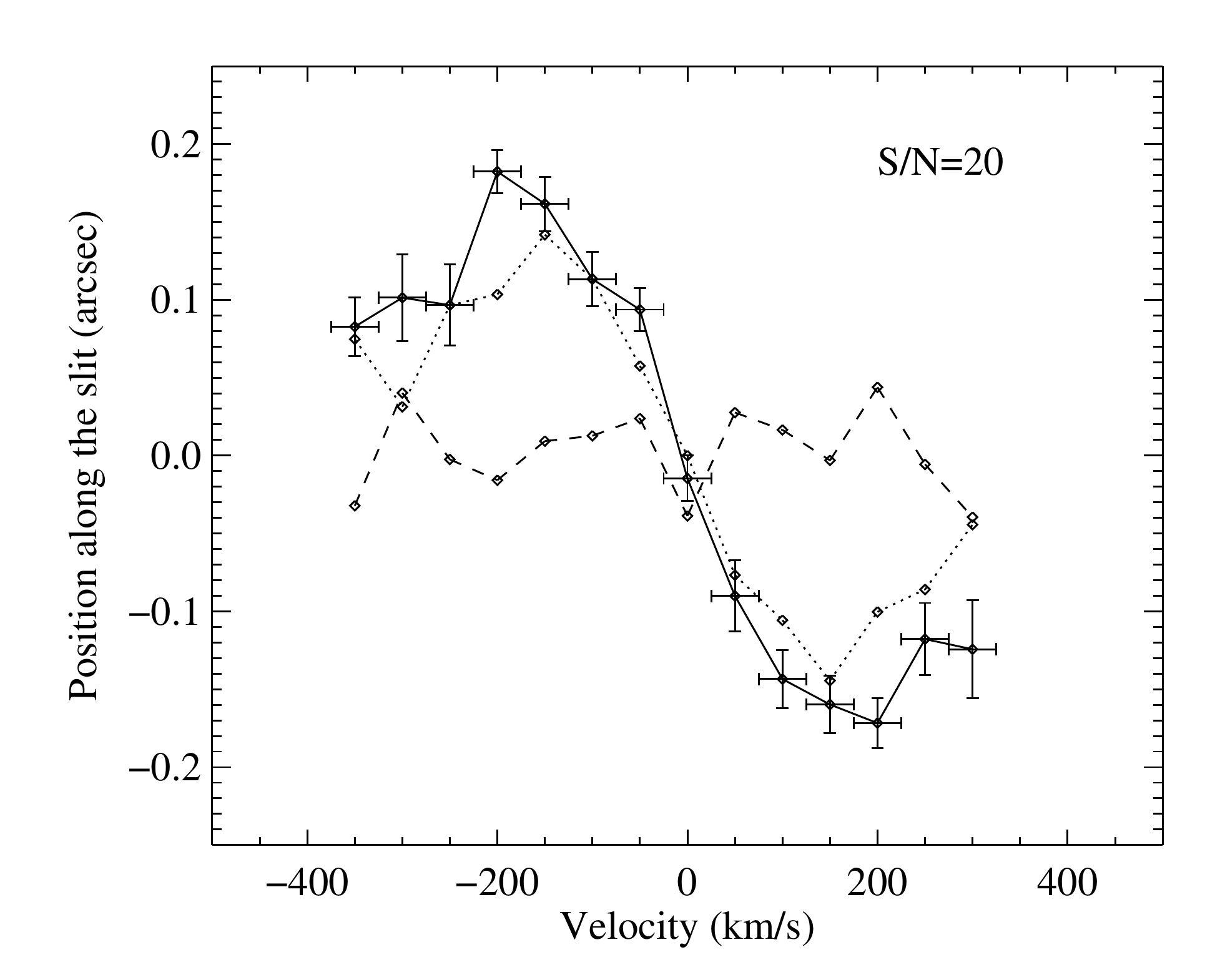}
  \includegraphics[width=0.48\linewidth]{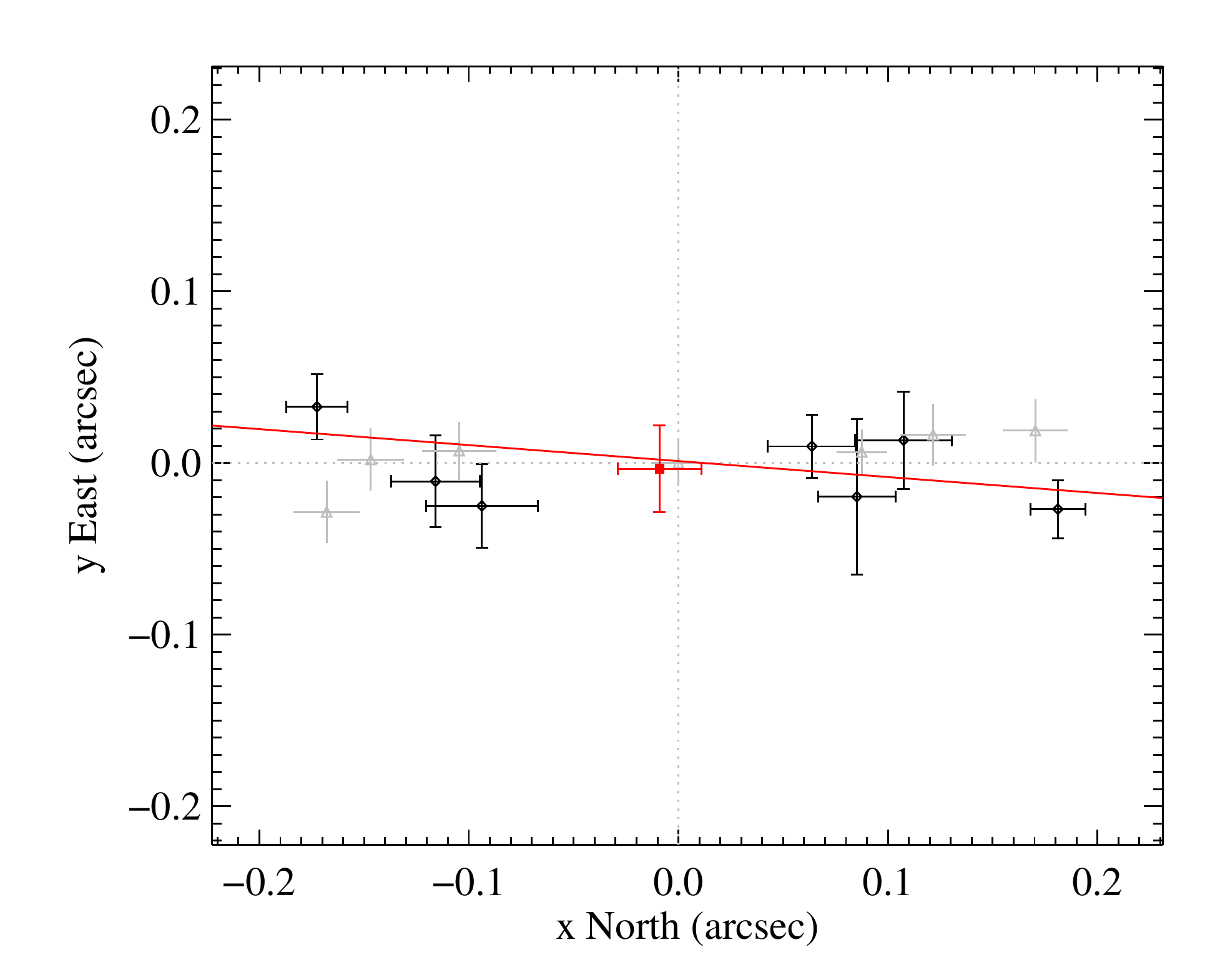}
 \centering 
  \includegraphics[width=0.48\linewidth]{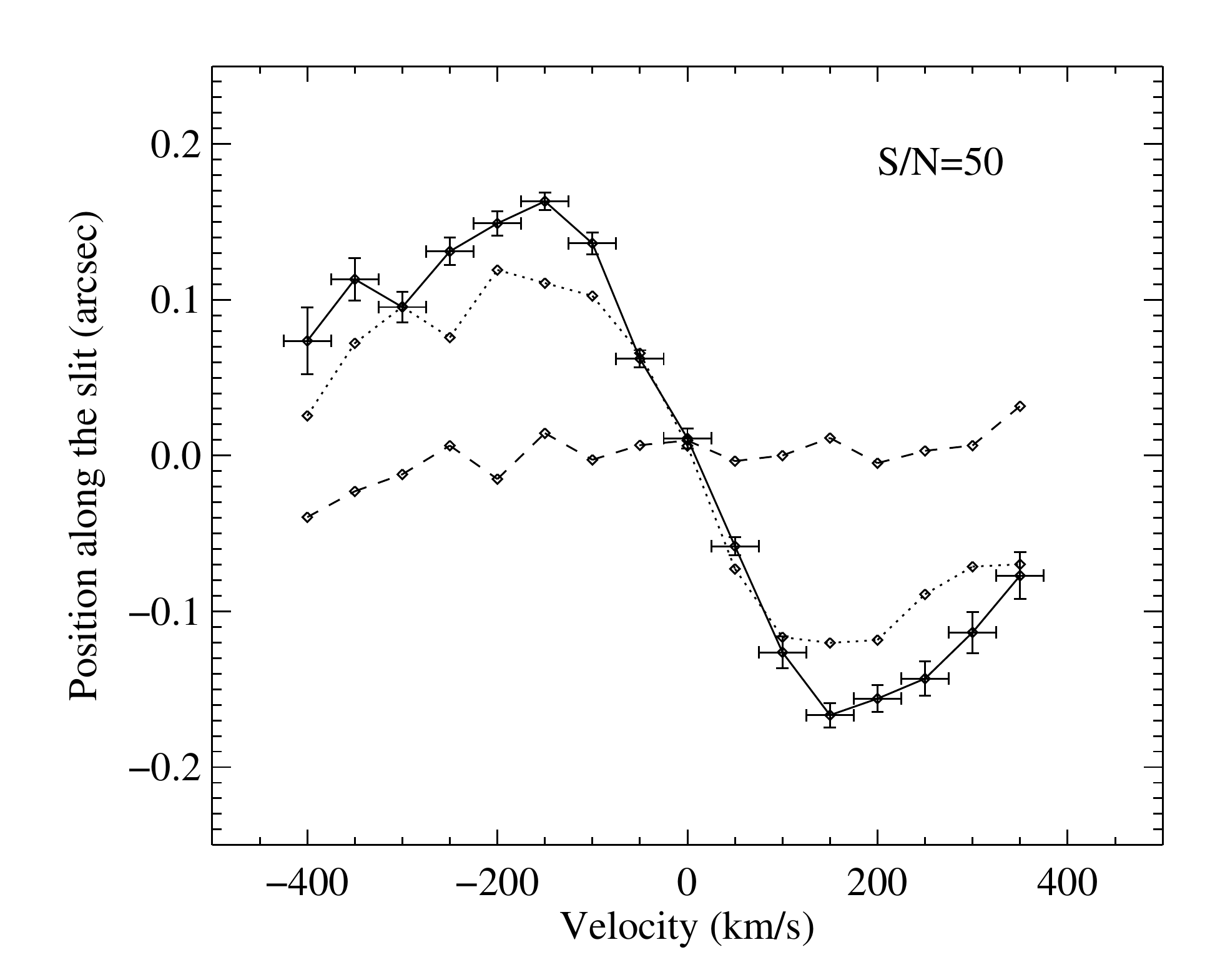}
  \includegraphics[width=0.48\linewidth]{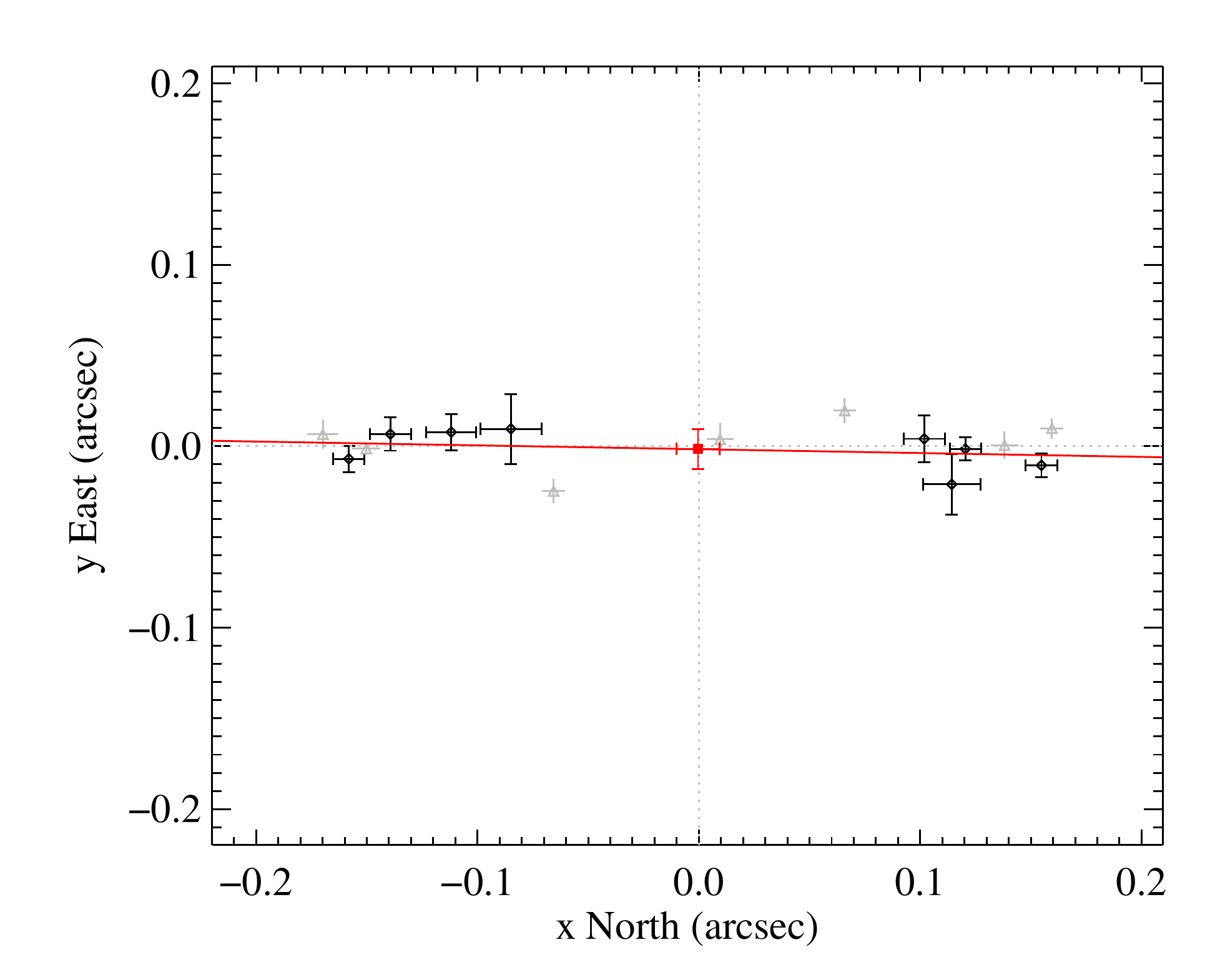}
 \centering 
  \includegraphics[width=0.48\linewidth]{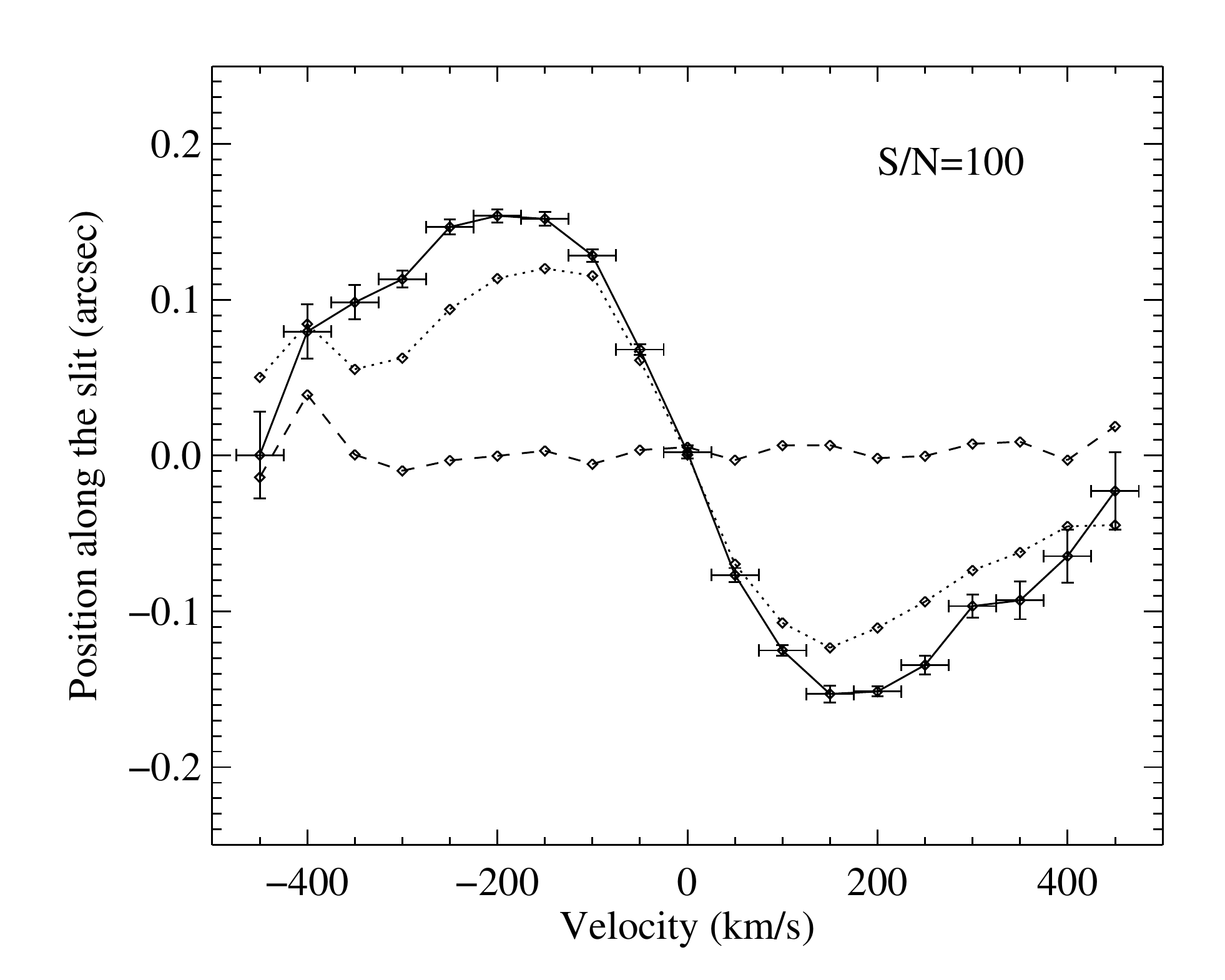}
  \includegraphics[width=0.48\linewidth]{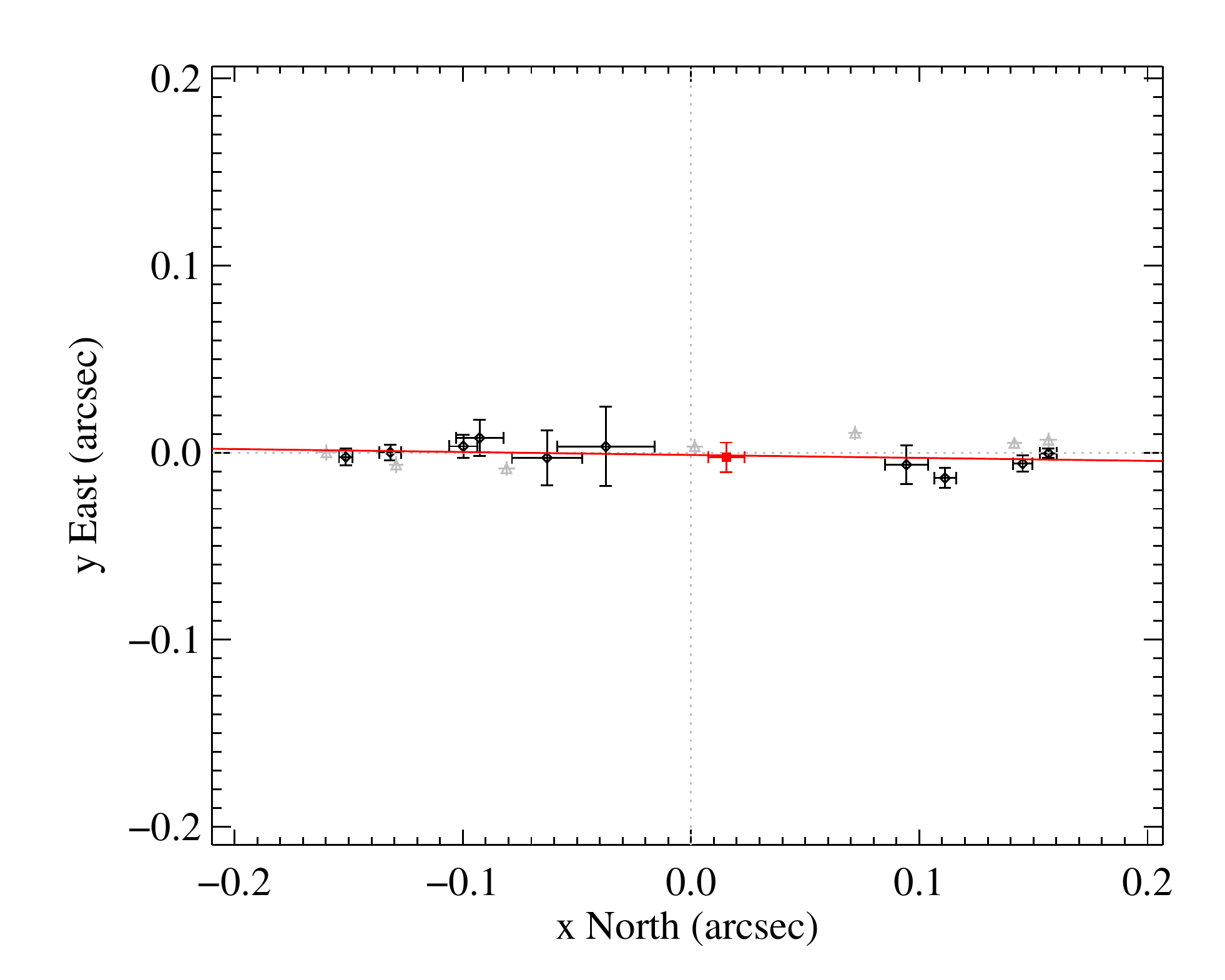}
\caption{Left row: spectroastrometric curves for the model of Fig.
  \ref{fig13} with added noise for a S/N of 20,
  50, and 100 (from top to bottom).  Solid line: slit at PA = $0^{\circ}$.
  Dotted line: slit at PA $45^{\circ}$. Dashed line: slit at PA =
  $90^{\circ}$.  For simplicity we plot the error bars of the points only for
  the PA$=0^{\circ}$ curve. Right row: spectroastrometric 2D map derived from
  the three spectroastrometric curves of the left row.  The red filled circle
  marks the derived position of the BH, while the red solid line represents
  the position angle of the line of nodes ($\theta_{LON}$) obtained by
  fitting a straight line to the X-Y position of the high velocity points.}
        \label{fig12all}
  \end{figure*}

We now consider the effect of noise on spectroastrometric curves and
maps, as in real data. We simulate noise by adding normally distributed
random values to the simulated position-velocity spectral maps. These
random numbers are characterized by zero mean and standard deviation
$\sigma_{noise}$ which is chosen in the following way. We first extract a
"nuclear" spectrum by coadding line emission over an aperture equal to the
PSF FWHM, centered on the nucleus position. The $S/N$ ratio of the
spectrum is then defined on the peak of the line profile, and the actual
$\sigma_{noise}$ value per pixel is selected to provide a given $S/N$.

In Fig. \ref{fig12all} we show the spectroastrometric curves and maps for the
same model presented in Fig. \ref{fig13} to which we added an artificial noise
in order to obtain S/N of $20$, $50$ and $100$, respectively.

A key point to emphasize is that the curves (and maps) with higher S/N tend
to have more measured points.  This is because at the ``high velocities''
we have a low flux with respect to the line core and the presence of noise limits our ability to
obtain a reliable light centroid position. In detail we stop
estimating the light centroids when the mean flux of the light profile for
that particular velocity bin is lower than the noise level. Thus, a
spectrum with higher $S/N$ enables us to obtain more accurate measurements on
individual data-points, but also extends the coverage at higher
velocities, crucial to detect a BH and to measure its mass.

The 2D spectroastrometric maps can be used to estimate
the geometrical parameters of the gas disk. In the following, we will always make use of simulations with noise.

As noted in Sect. \ref{s321}, if the gas kinematics is dominated by
rotation around a pointlike mass (the BH), the position of the light
centroid at the highest velocities approaches the position of the BH. This
consideration, which is also valid for the 2D spectroastrometric map,
allows us to estimate the position on the plane of the sky of the BH as the
average position of the ``high velocity'' points.  In practice, we
calculate this position by first taking the averages of the coordinates of
the points in the blue and red "high velocity'' ranges and then taking the
average coordinates of the ``blue'' and ``red'' positions. The inferred BH
position is marked by a red filled circle in the right row diagrams of Fig.
\ref{fig12all}.

Also, the high velocity points in Fig. \ref{fig12all} (and also Fig.
\ref{fig14}) lie on a straight line which identifies the direction of the
disk line of nodes (see Fig. \ref{fig08} and Sect. \ref{s325}). The red
solid line shown in the right row of Fig. \ref{fig12} represents the
position angle of the line of nodes ($\theta_{LON}$) obtained by fitting a
straight line to the high velocity points.

In the Table \ref{noise} we report the BH position and disk line of nodes
position angle values obtained from the 2D map of the model of Fig.
\ref{fig12all}.  We can observe that the accuracy of these measurements scales
approximately with the square root of the S/N of the spectrum.
Moreover in the maps with lower S/N the scatter of the
points around the line of nodes increases (the points are almost perfectly
aligned on the line of nodes direction in the noise free model of Fig.
\ref{15}). This fact decreases the accuracy of the BH position and line
of nodes PA estimates for lower S/N.

\begin{table}
 \centering
 \caption{Lines of Nodes and BH position estimates from noisy data}
 \label{noise}
 \begin{tabular}{c c c c}
   \hline \hline
Model   & $x_{BH}(\arcsec)$& $y_{BH}(\arcsec)$ & $\theta_{LON} (^{\circ})$  \\
   \hline
   \\
Noiseless   &0.00  &  0.00 &  0.00 \\
S/N=20    & -0.01$\pm$0.02 &0.00$\pm$0.03& -5$\pm$4  \\
S/N=50     & 0.00$\pm$0.01 &0.00$\pm$0.01 & -1$\pm$1  \\
S/N=100    & 0.015$\pm$0.008 &-0.003$\pm$0.008 & -0.9$\pm$0.8  \\
   \hline
 \end{tabular}
\end{table}

\section{Estimate of the BH mass from the spectroastrometric map}\label{s44}
In this section we present a simple and straightforward method to recover the
BH mass from the spectroastrometric measurements. The method is based on the
following assumptions: (i) the gas is circularly rotating in a thin disk
and (ii) each point of the
spectroastrometric map represents a test particle on the disk line of nodes
and is characterized by a line of sight (``channel'') velocity which is given
by the center of the velocity bin where the light centroid was estimated.
Under these assumptions it is trivial to relate the spectroastrometric map to
the BH mass.

The circular velocity of a gas particle with distance $r$ from the BH is given
by:
\begin{equation}
V_{rot}=\sqrt{\frac{G[M_{BH}+M_{star}(r)]}{r}}
\label{9}
\end{equation}
where $M_{star}(r)$ is the stellar mass enclosed in a sphere of radius $r$
(assuming spherical symmetry) and can be written as:
\begin{equation}
M_{star}(r)=M/L\cdot L(r)
\label{10}
\end{equation}
with $L(r)$ representing the radial luminosity density distribution in the galactic
nucleus and $M/L$ is the mass to light ratio of the stars.

The line of sight velocity (hereafter
$V_{ch}$ for ``channel velocity'') of a test particle located on the disk line of nodes at distance $r$ from the BH is then given by
\begin{equation}
V_{ch}=V_{rot}sin(i)+V_{sys}
\label{11}
\end{equation}
where $i$ is the inclination of the disk plane and $V_{sys}$ is the systemic
velocity of the galaxy.

The centroid positions on the spectroastrometric map are denoted by $(x_{ch},
y_{ch})$ and, in general, they are not perfectly aligned along the line of
nodes as described in Sect. \ref{s42} and \ref{a3}. We select the "high
velocity" points and we then identify the line of nodes by a linear fit of the
$(x_{ch}, y_{ch})$ spectroastrometric map. To reduce the correlation between
the values of intercept and slope of the fitted line we consider
\begin{equation}
y=(x-x_{mean})\,tan(\theta_{LON})+b
\label{12}
\end{equation}
where $x_{mean}$ is the mean of the $x_{ch}$ positions. $\theta_{LON}$ and $b$
denote respectively the position angle and the intercept of the line of nodes
to be determined by the fit.  Then $\theta_{LON}$ and $b$ are recovered by
minimizing the following $\chi^2$ where we take into account both errors in
$x$ and $y$:
\begin{equation}
\chi^2=\sum_{ch}{\frac{\left[y_{ch}-(x_{ch}-x_{mean})tan(\theta_{LON})-b\right]^2}{\Delta y_{ch}^2+tan(\theta_{LON})^2\Delta x_{ch}^2}}
\label{13}
\end{equation}

where $\Delta x_{chan}$ and $\Delta y_{chan}$ are the uncertainties on
the position of the points of the spectroastrometric map and, as observed in
Sect. \ref{s42}, the sum is extended only over the ``high velocities''
range.
\begin{figure}[!ht]
\centering
\includegraphics[width=0.9\linewidth]{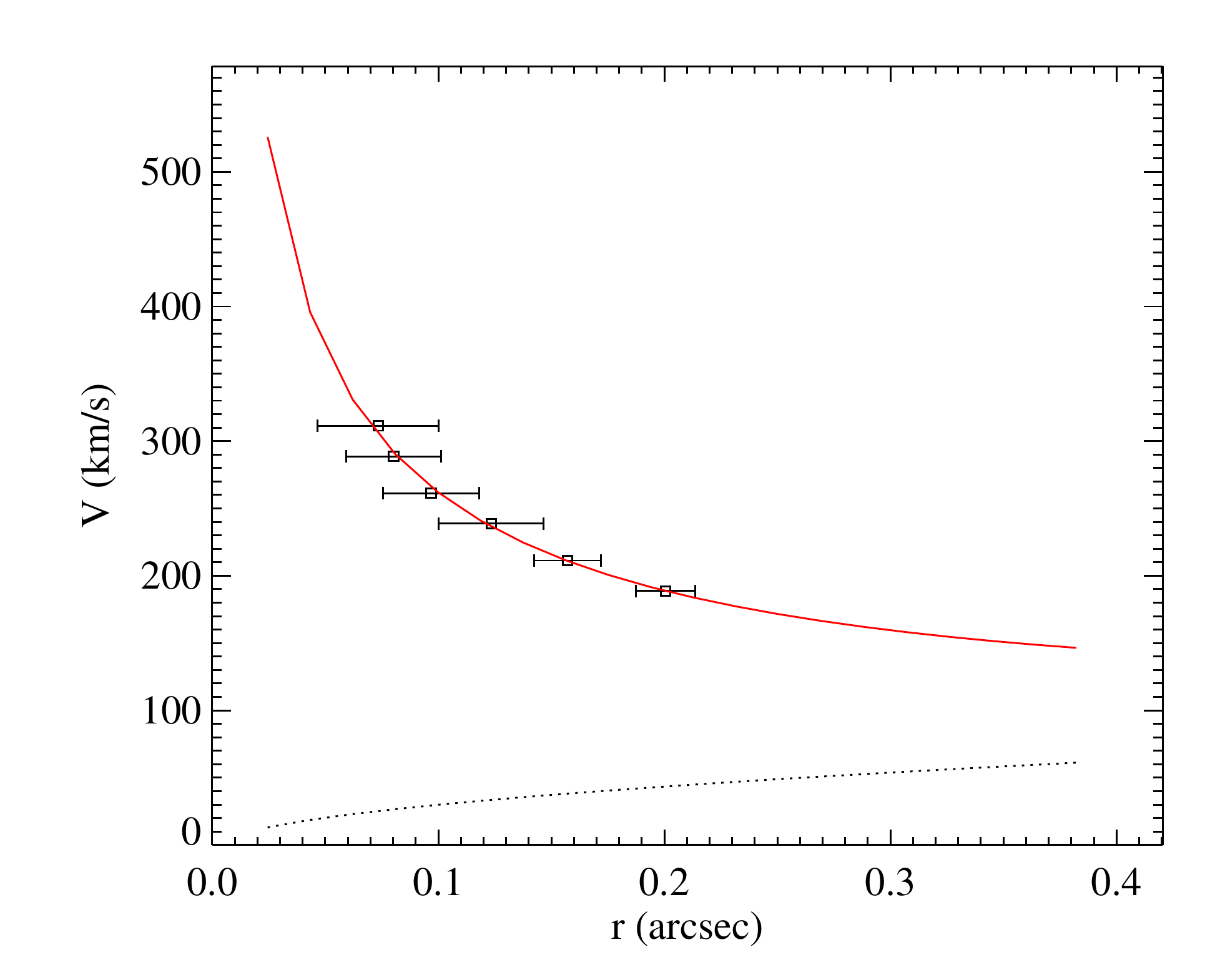}
\includegraphics[width=0.9\linewidth]{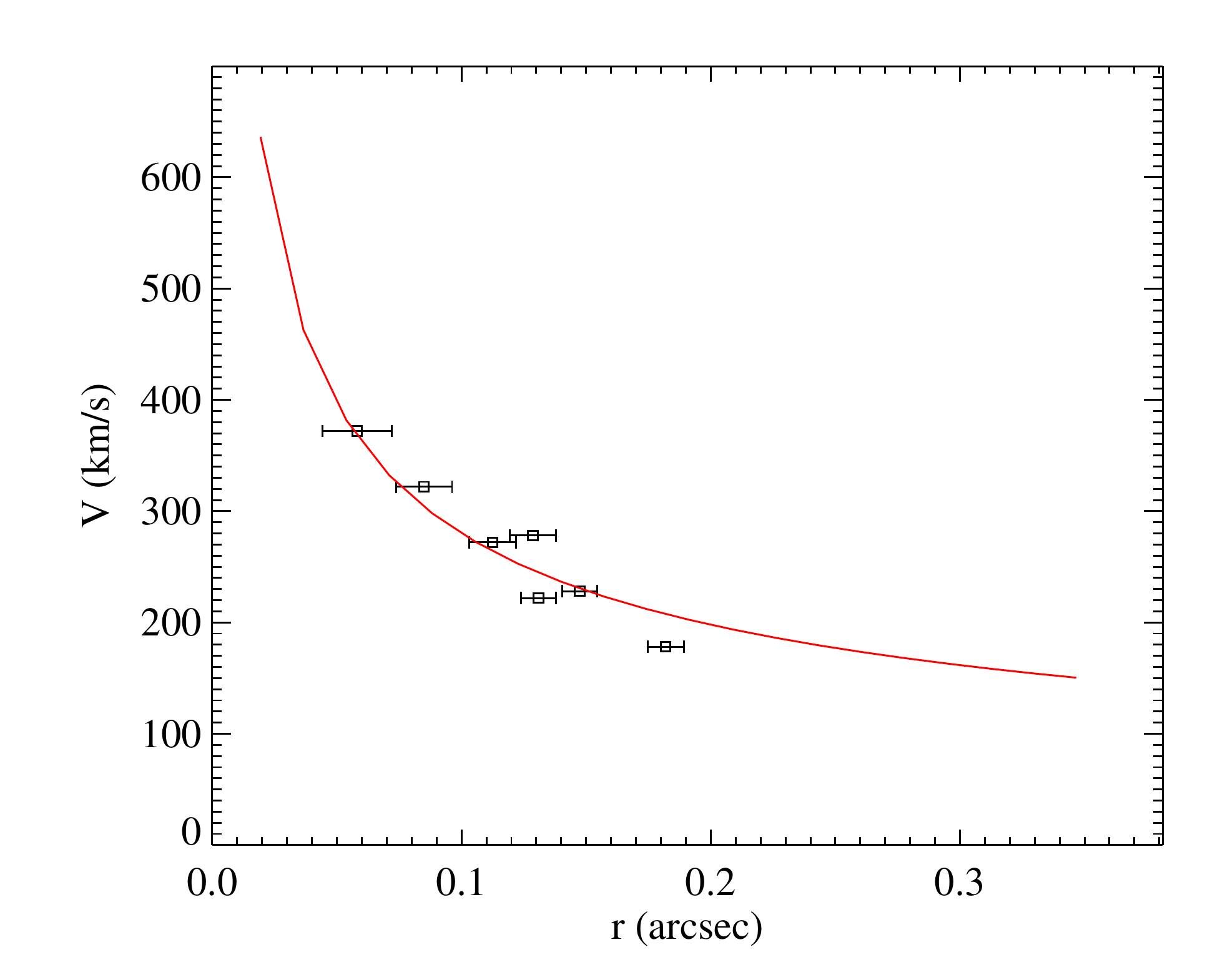} 
\includegraphics[width=0.9\linewidth]{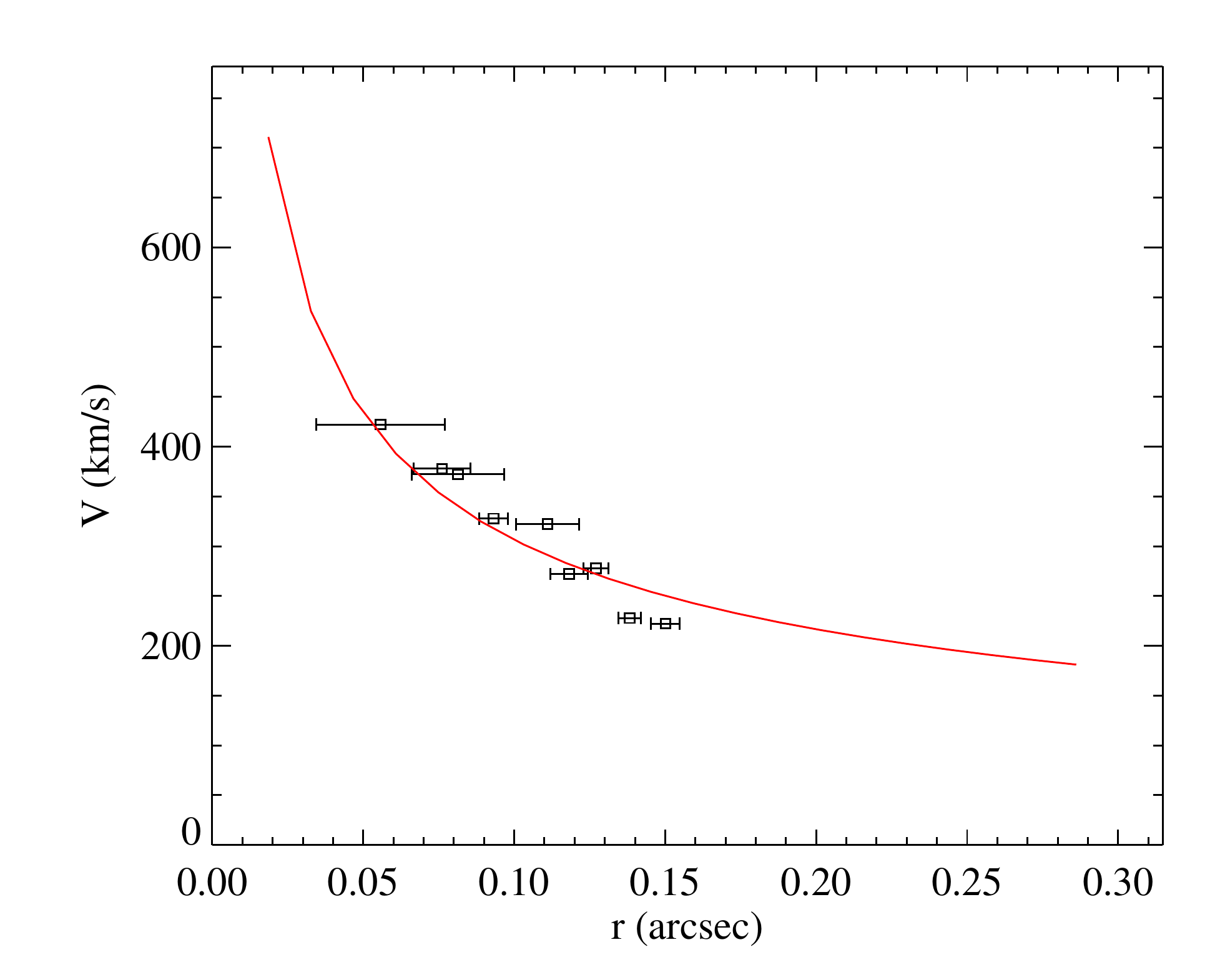}
\caption{Fit results for models of Fig.~\ref{fig12all} 
with $S/N=20$ (top panel), $S/N=50$ (middle) and $S/N=100$ (bottom). Points with error bars are the result of "observations". The red line is the best fit model and the dotted line is the expected contribution from the stellar mass.}
     \label{fig22}
\end{figure}
\begin{figure}[!ht]
\centering
\includegraphics[width=0.9\linewidth]{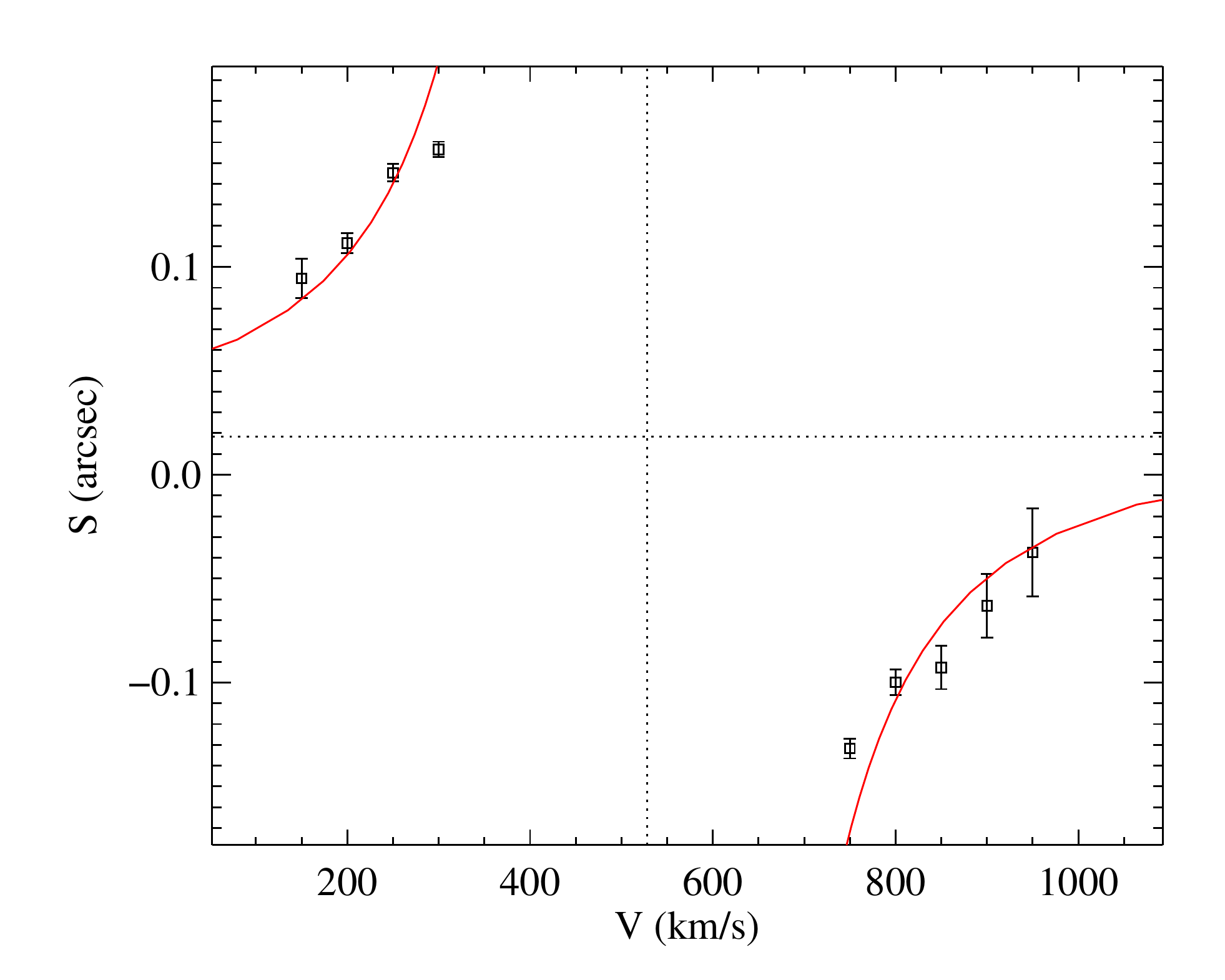}
\caption{Observed and model spectroastrometric curves at $S/N=100$ (e.g.  $S_{chan}$ vs  $V_{chan}$). The horizontal dotted line is the BH ``S''
 coordinate $S_0$. The vertical dotted line is the systemic velocity $V_{sys}$.}
     \label{fig24}
\end{figure}

The second step requires to associate each velocity bin to its position
within the rotating disk. As described in Sect.
\ref{s42}, the 2D map points lie with good
approximation on the line of nodes, particularly 
in the ``high velocities'' range.
We then project the $(x_{ch},y_{ch})$ position of the 2D-map points on the
line of nodes and calculate their coordinate ``S'' with respect to this
axis of reference, with ``S'' defined as:
\begin{equation}
S_{ch}=x_{ch}\,cos(\theta_{LON})+[y_{ch}-b+x_{mean}tan(\theta_{LON})]sin(\theta_{LON})
\label{14}
\end{equation}
Each $S_{ch}$ has an uncertainty $\Delta S_{ch}$ estimated by taking into
account the uncertainties on the 2D map points position $(\Delta x_{ch}$ and
$\Delta y_{ch}$ as well as the uncertainties on the line of nodes parameters
$(\Delta \theta_{LON}, \Delta b)$ resulting from the fit.  The distance $r$ of
a test particle from the BH used in Eq. \ref{9} differ from $S_{ch}$ only by a
constant $S_0$, the unknown coordinate of the BH along the line of nodes, i.e.
\begin{equation}
r_{ch}=|S_{ch}-S_0|
\label{15}
\end{equation}
Combining Eqs. \ref{9}, \ref{10}, \ref{11}, and \ref{15} we finally obtain the expected velocity in channel $ch$:
\begin{equation}
\bar{V}_{ch}=\sqrt{\frac{G[M_{BH}+M/L\cdot L(|S_{ch}-S_0|)]}{|S_{ch}-S_0|}}sin(i)+V_{sys}
\label{16}
\end{equation}
The unknown model parameters are:
\[
  \begin{array}{lp{0.8\linewidth}}
     M_{BH}       & mass of the BH     \\
     M/L          & mass to light ratio of the nuclear stars\\
     S_0      & line of nodes coordinate of the BH\\
     V_{sys}      & systemic velocity of the galaxy\\
     i            & inclination of the gas disk \\
  \end{array}
\]
and can be determined by minimizing the following $\chi^2$
\begin{equation}
\chi^2=\sum_{ch}{\left[\frac{|V_{ch}-V_{sys}|-|\bar{V}_{ch}(par)-V_{sys}|}{\Delta(S_{ch}; par)}\right]^2}
\label{17}
\end{equation}

 \begin{figure*}[!ht]
 \centering
  \includegraphics[width=0.4\linewidth]{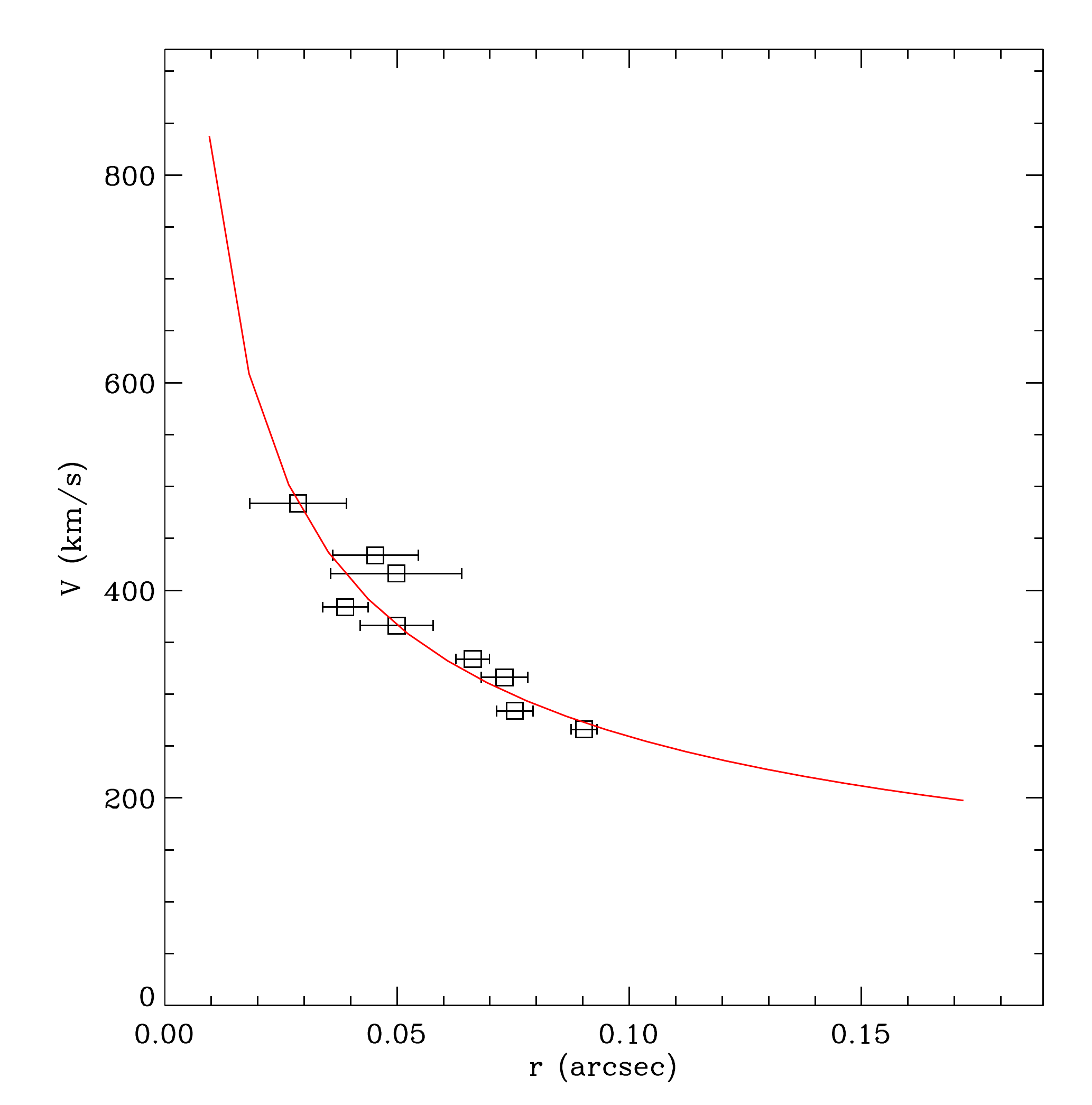}
  \includegraphics[width=0.4\linewidth]{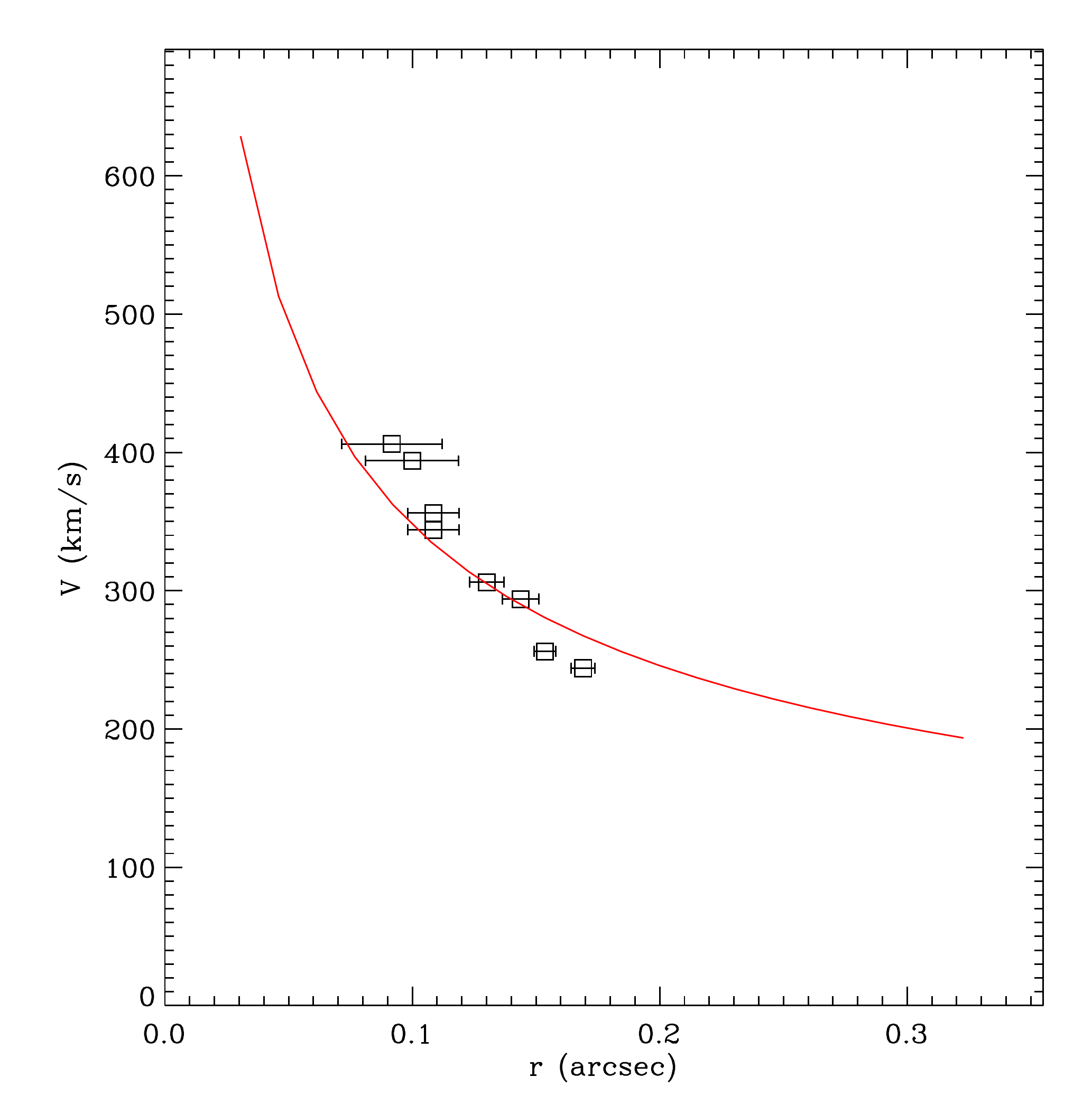}
 \includegraphics[width=0.4\linewidth]{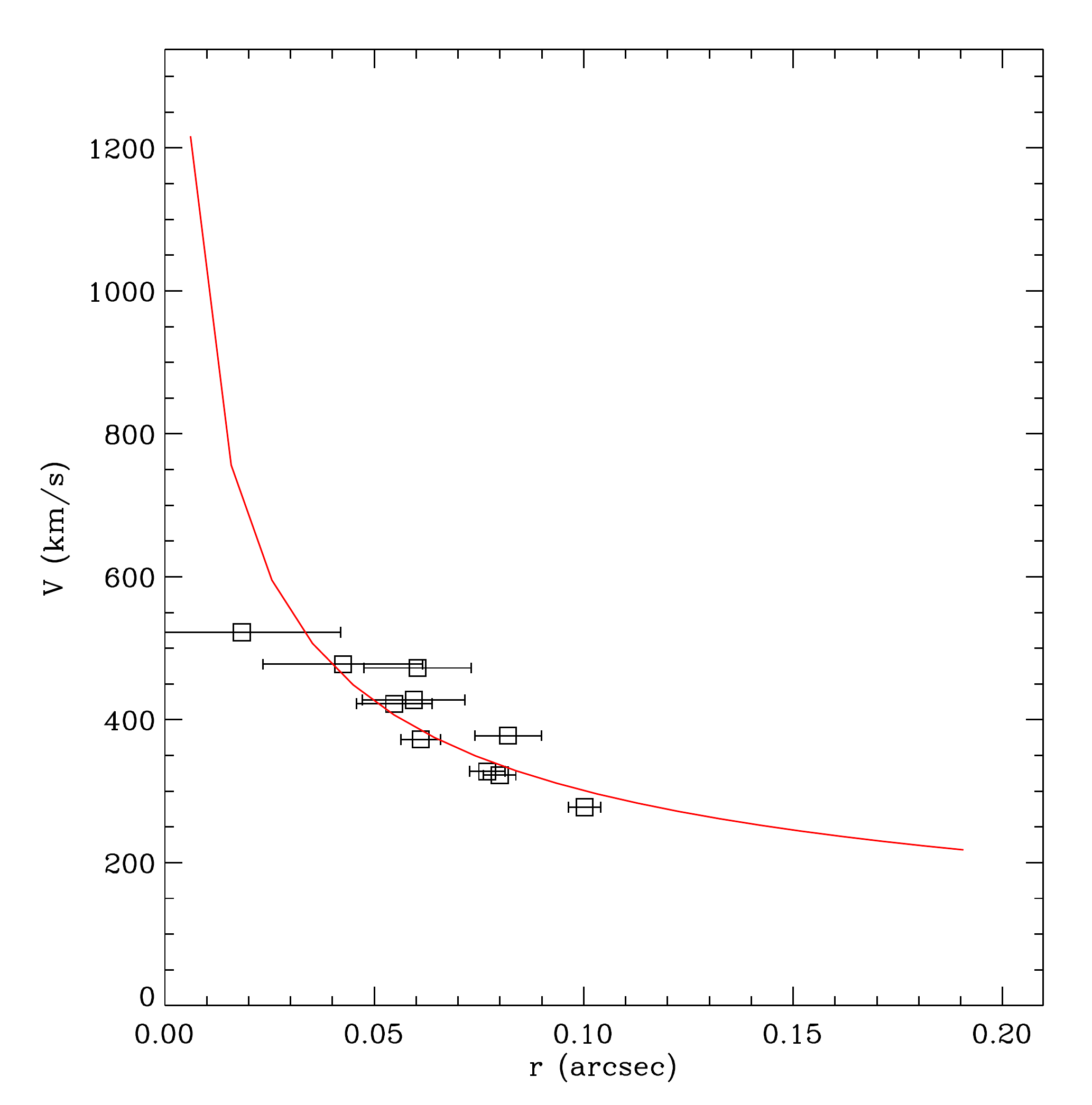}
 \includegraphics[width=0.4\linewidth]{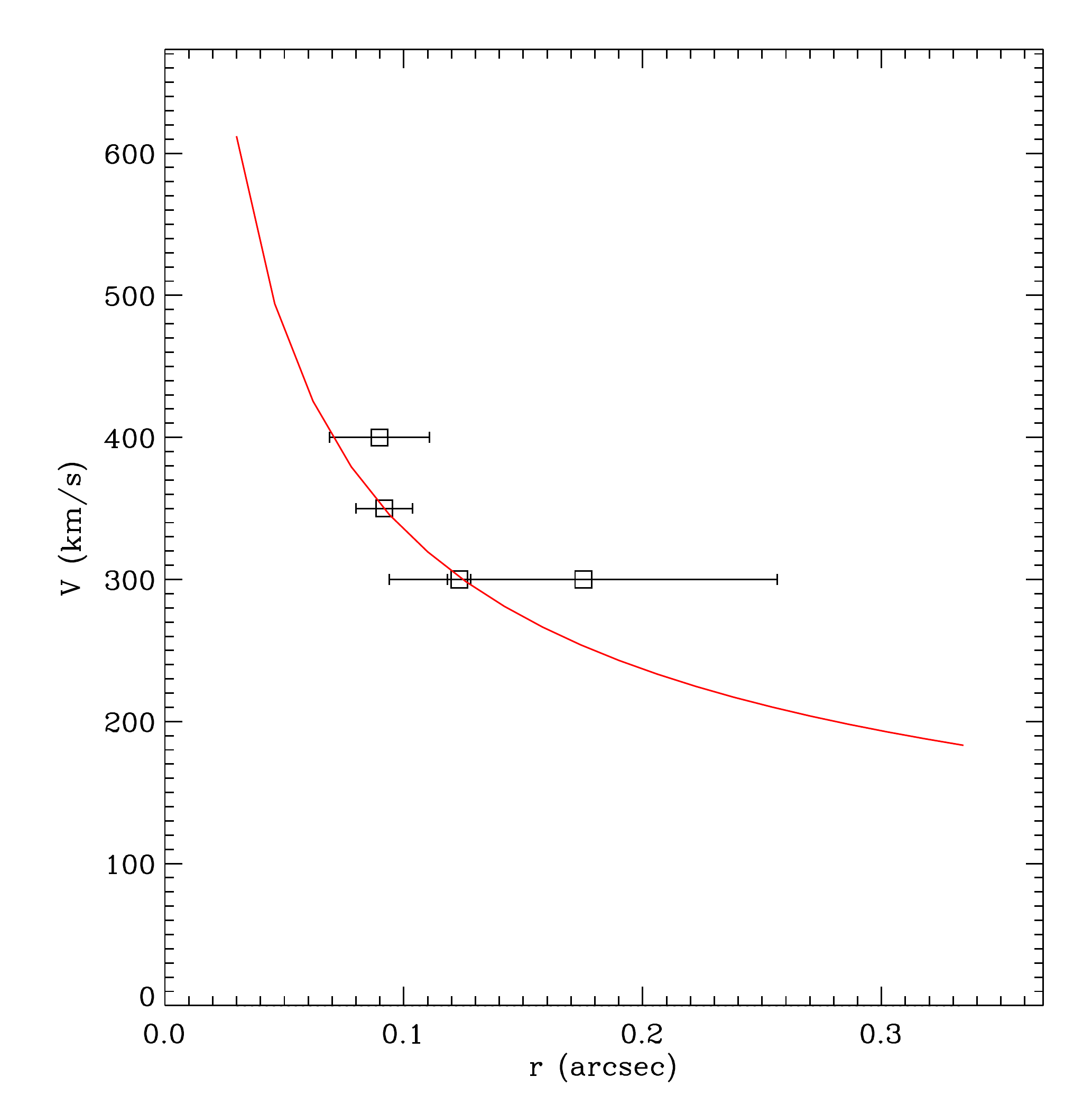}
    \caption{Effect of varying the intrinsic flux distribution at sub-resolution scales: results for various intrinsic flux distributions. Top left panel: exponential flux distribution centered on $(0\arcsec,0\arcsec)$ with $r_0=0.05\arcsec$. Top rigth panel: exponential flux distribution centered on $(0\arcsec,0\arcsec)$ with $r_0=0.15\arcsec$. Bottom left panel: exponential flux distribution centered on $(0\arcsec,0\arcsec)$ with $r_0=0.05\arcsec$ and a central hole of radius $r_h=0.04\arcsec$. Bottom rigth panel: exponential flux distribution centered on $(0.1\arcsec,0.1\arcsec)$ with $r_0=0.2\arcsec$.}
       \label{fig25B}
 \end{figure*}
 \begin{figure*}[!ht]
 \centering
  \includegraphics[width=0.45\linewidth]{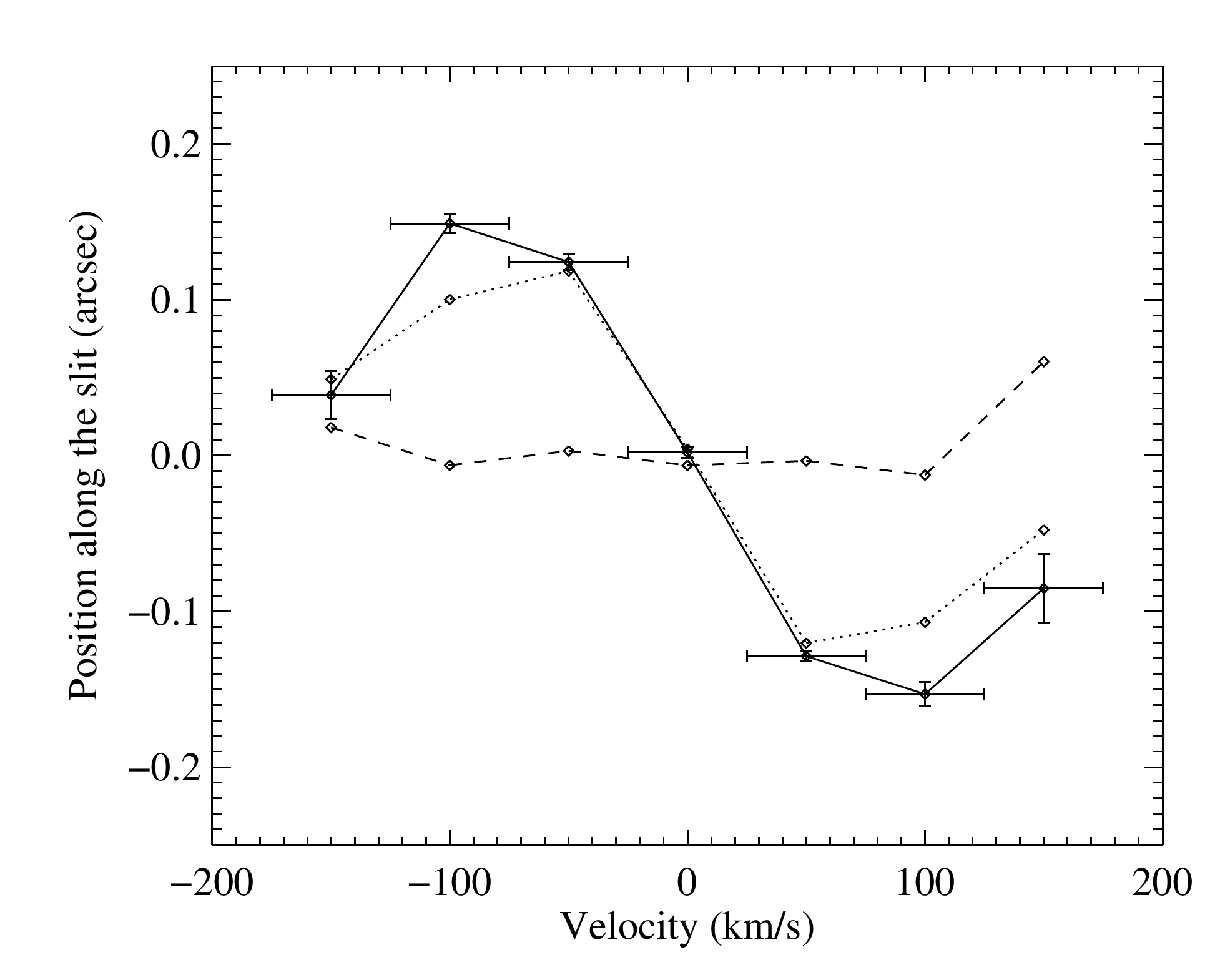}
  \includegraphics[width=0.45\linewidth]{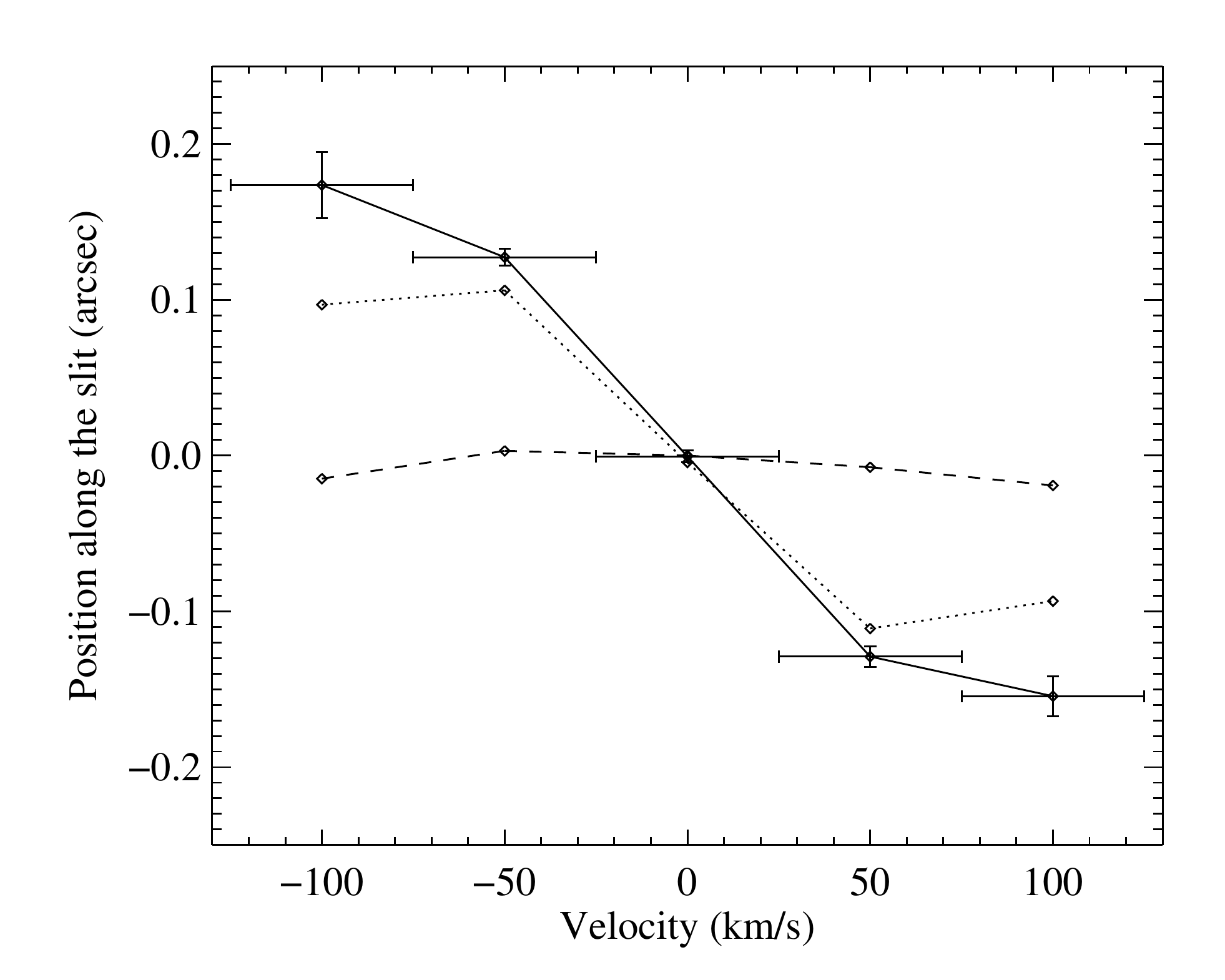}
 \includegraphics[width=0.45\linewidth]{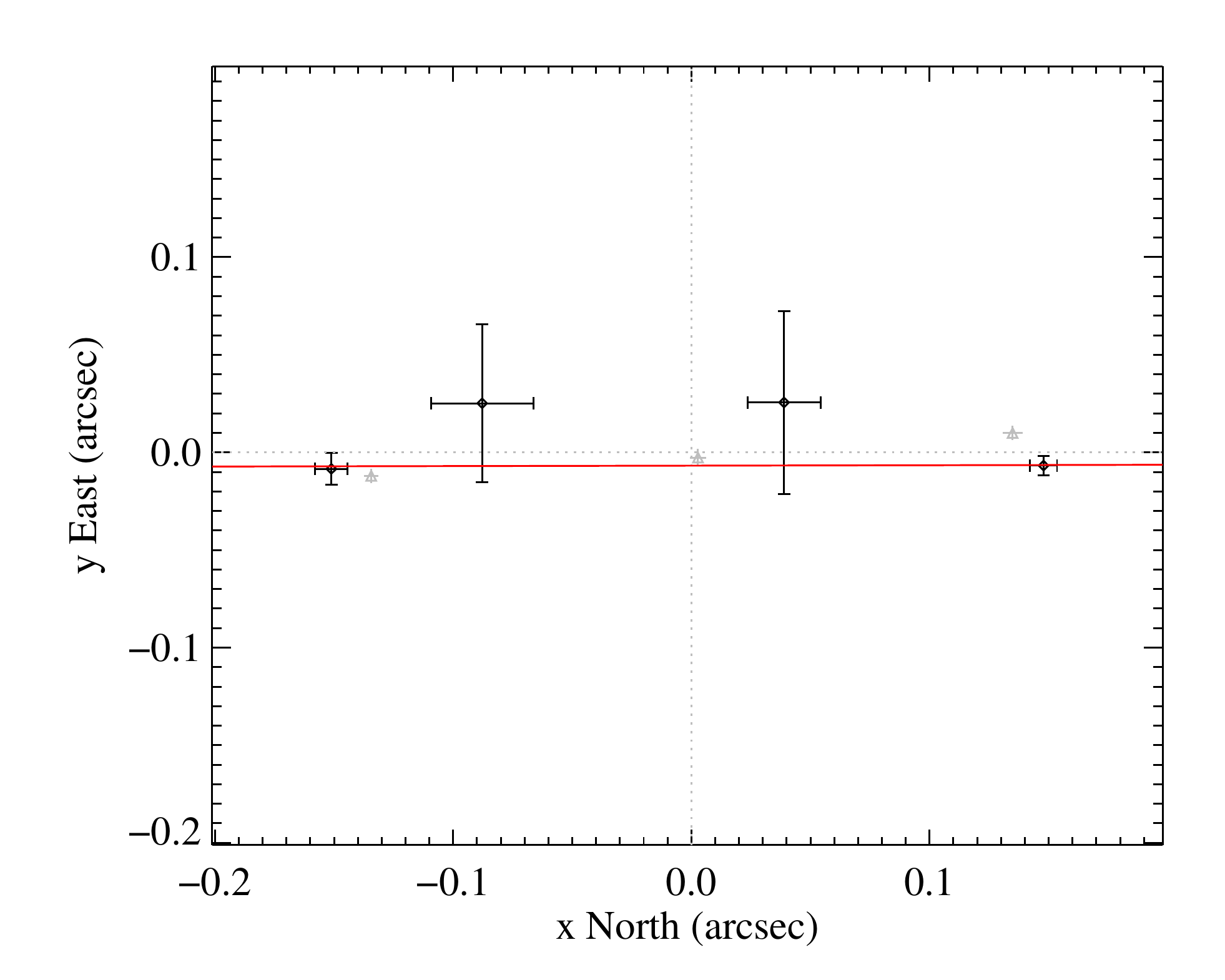}
 \includegraphics[width=0.45\linewidth]{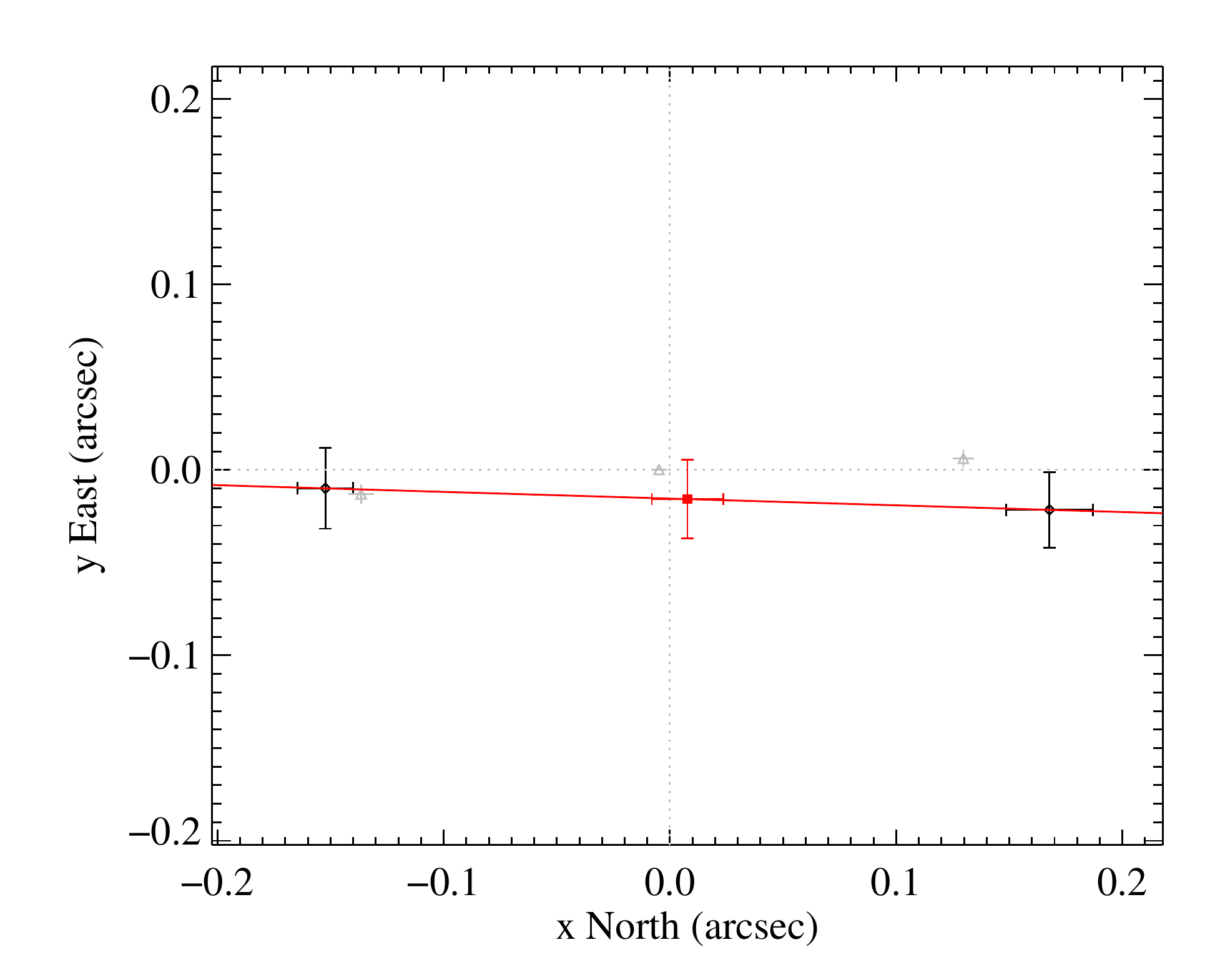}
 \includegraphics[width=0.45\linewidth]{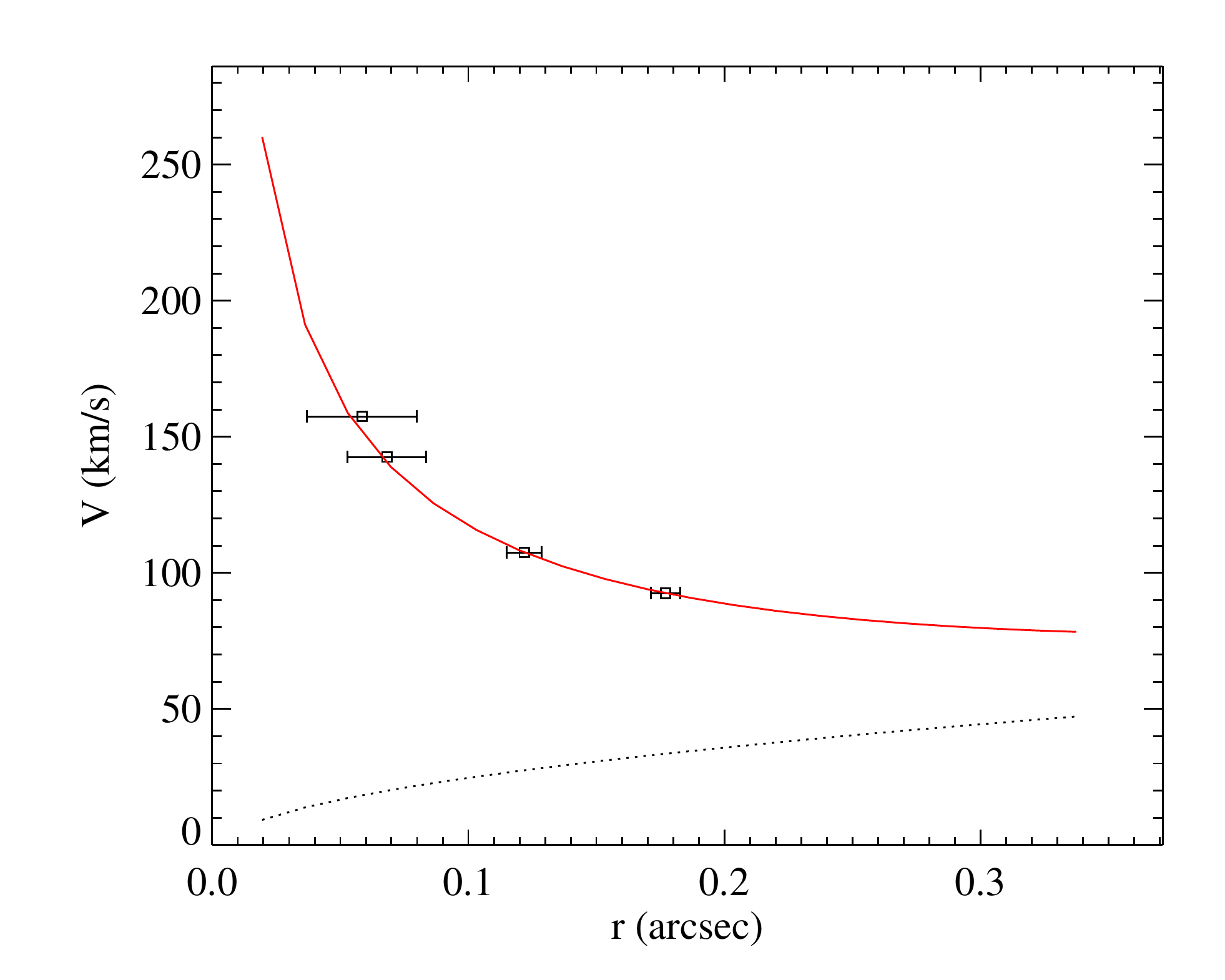}
 \includegraphics[width=0.45\linewidth]{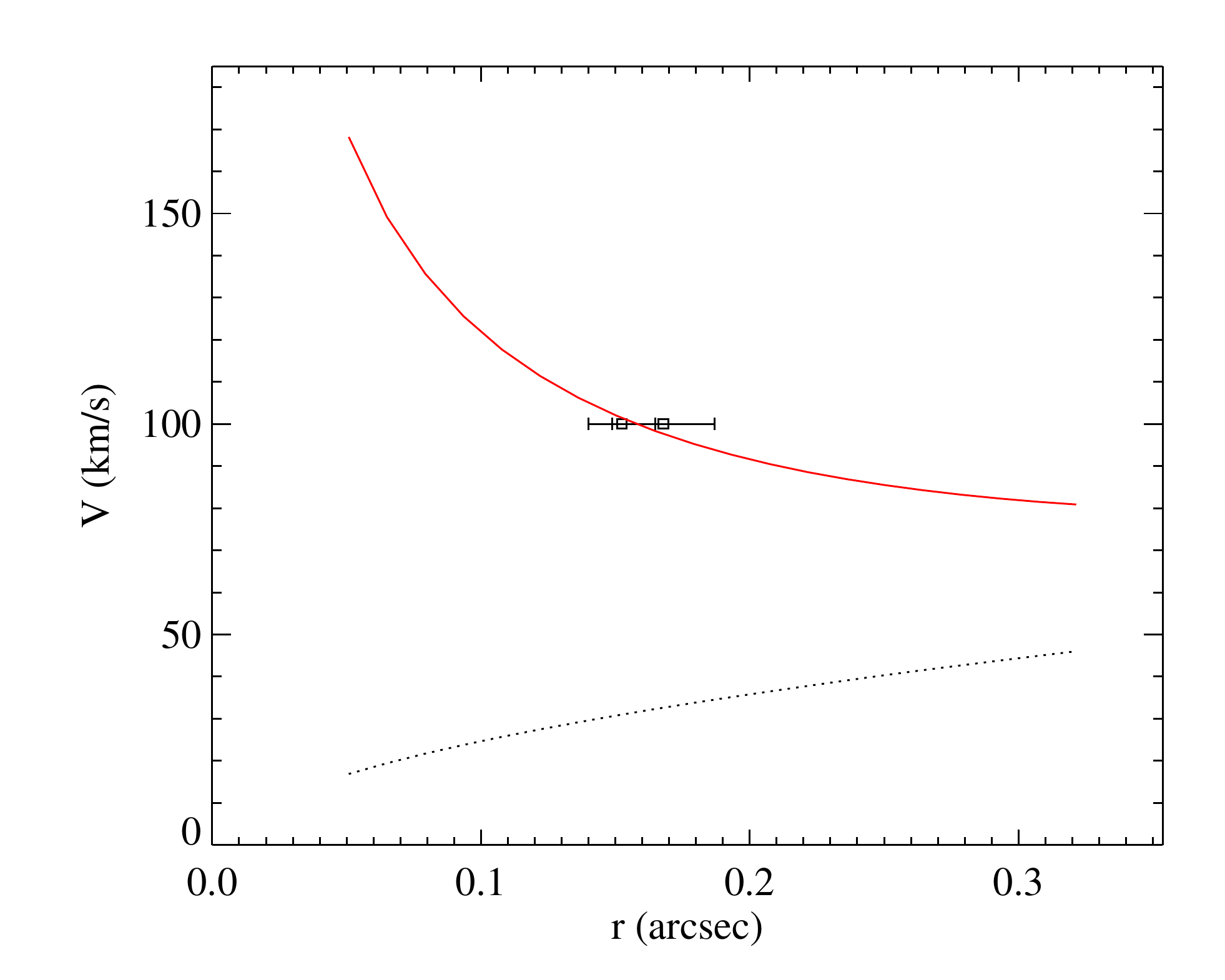}
    \caption{Application of the method to low BH masses: results from simulations with $S/N=100$, $M_{BH}=10^7M_{\astrosun}$ (left) and $10^{6.5}M_{\astrosun}$ (right) . Top: spectroastrometric curves. Middle: 2D spectroastrometric maps. Bottom: fit results.}
       \label{fig25}
 \end{figure*}

where $\Delta(S_{ch}; par)$ is the uncertainty of the numerator computed as a
function of the unknown parameters values ($par$) and the uncertainties
$\Delta S_{ch}$ on the positions along the line of nodes. We remark that the
channel velocity $V_{ch}$ has no associated unceratinty since it is not a
measured quantity but is the central value of a velocity bin where the
spectroastrometric curve is measured.  As discussed in the previous sections,
we restrict the fit (i.e. the sum over the velocity channels) to the ``high
velocities'' range.

The $\Delta(S_{ch}; par)$ factor in equation
\ref{17} is much smaller for the points at lower velocities (i.e. smaller $|V_{ch}-V_{sys}|$) which are closer to the peak of the line profile and have therefore much larger $S/N$ than the points at ``high velocity".  However, from the discussion in the previous sections, we know that the spectroastrometry is less reliable for the determination of BH properties like position and mass.
Therefore, to avoid being biased by potentially faulty points we add in quadrature a constant
error $\Delta_{sys}$ to $\Delta(S_{ch}; par)$. The value of $\Delta_{sys}$ is found by imposing that the reduced $\chi^2$ is equal to 1.
Therefore, if the points at lower velocity with higher $S/N$ are problematic they will not bias the final fit results, because their high weight will be greatly reduced by the addition od of a much larger $\Delta_{sys}$.

After determining the best fit values of the free parameters, the position of the BH on the plane of the sky is simply given by

\begin{eqnarray}
\lefteqn{x_{BH}=S_0\,cos(\theta_{LON})}\\ 
\lefteqn{y_{BH}=S_0\,sin(\theta_{LON})+b-x_{mean}\,tan(\theta_{LON})}
\label{18}
\end{eqnarray}
We have performed the fit of the spectroastrometric data for the models of
Figs. \ref{fig12all}, which are computed with the same parameters but different
spectrum $S/N$. The results of the fitting procedure are presented in Fig.~\ref{fig22} where we plot the high velocity points of the
spectroastrometric map in the $|V_{ch}-V_{sys}|$ vs $r$ ($|S_{ch}-S_0|$)  
diagram. The solid red lines represent the curves expected from the rotating disk model.  
In Fig. \ref{fig24} we present the direct comparison between the observed spectroastrometric 
curve along the line of nodes (e.g.  $S_{chan}$ vs  $V_{chan}$) with the best fit model (for $S/N=100$).

In all cases, for each ``test particle", we are able to measure photocenter
positions on the plane of the sky with an accuracy of better than 0\farcs025,
corresponding to $\simeq 1/20$ of the spatial resolution ($FWHM = 0\farcs5$).

The comparison between best fit and 
simulation parameters is shown in Table \ref{tab1}.
In particular, we can recover the BH mass value with an accuracy of $\sim0.1$ dex.
The $M/L$ value is not well constrained by the models because with the adopted parameters the contribution to the
gravitational potential of the stellar mass is negligible.  

In the test cases considered here ($M_{BH}=10^{8.0}M_{\astrosun}$, stellar
velocity dispersion of $\sigma_{star}\simeq200$ km/s and $D\simeq3.5Mpc$), the
radius of the BH sphere of influence is $r_{BH}\simeq 10$ pc corresponding to
$\simeq0\farcs6$, barely resolved with the adopted $0\farcs5$ resolution. With
the spectroastrometric technique, we have instead probed down to radii of the
order of $0.05\arcsec$ ($\sim$ 1 pc), a factor $\sim10$ smaller than $r_{BH}$.
This clearly demonstrates that with spectroastrometry we can recover
information on spatial scales which are much smaller than the limit imposed by
the spatial resolution, thus opening the possibility of probing the
gravitational potential inside smaller BH spheres of influence.

\begin{table*}
  \caption[!ht]{Fit results from the baseline model with different $S/N$.}
  \label{tab1}
  \centering
  \begin{tabular}{l c c c c}
    \hline
    \noalign{\smallskip}
    Parameter & \multicolumn{3}{c}{Fit result} & Model value\\
    \noalign{\smallskip}
    & S/N=100 & S/N=50 & S/N=20 & \\
   \noalign{\smallskip}
   \hline
   \noalign{\smallskip}
   $\theta_{LON}\ \ \ [^{\circ}]\ \ \ \ (LON\  fit)$             & $-0.9\pm0.8$  & $-1 \pm 1$ & $-5 \pm4$ & $0.00$\\
   $b\ \ \ [\arcsec]\ \ \ \ \ (LON\  fit)$                               & $-0.001\pm0.002$  & $-0.002\pm 0.003$ & $0.000\pm0.009$ & $0.00$\\
   \noalign{\smallskip}
   \hline
   \noalign{\smallskip}
   $log_{10}(M_{BH}/M_{\astrosun}) $ & $8.05\pm 0.03$ & $8.00\pm 0.06$ & $7.9\pm 0.1$ & $8.00$\\
   $log_{10}(M/L)\ \ \ [ M_{\astrosun}/L_{\astrosun} ]\ \ \ \ \ $ & $-9.02^a$ & $-7.90^a$ & $0.2\pm2.6$ & $0.00$\\
   $S_0\ \ \ [\arcsec]$                    & $0.02\pm0.02$  & $-0.03 \pm0.03$ & $-0.02 \pm0.03$ & $0.00$\\
   $V_{sys}\ \ \ [km/s]$                              & $528\pm 26$ & $478 \pm 33$ & $489 \pm 23$ & $500$\\
   $i\ \ \ [^{\circ}]$                                & $35.0\ ^b$ & $35.0\ ^b$ & $35.0\ ^b$ & $35.0\ ^b$\\
   $\Delta_{sys}\ \ \ [km/s]$                             & $25$ & $28$ & $0$ & $-$ \\
   \noalign{\smallskip}
   \noalign{\smallskip}
   reduced $\chi^2 \ \ \ (\chi^2/D.O.F.)$ & $1.003\ \ (5.12/5)$ & $0.992\ \ (2.976/3)$ & $0.015\ \ (0.031/3)$ & $-$\\
   \noalign{\smallskip}
   \hline
   \noalign{\smallskip}
   $x_{BH}\ \ \ [\arcsec]$                           & $0.02 \pm 0.02$ & $-0.03 \pm 0.03$ & $-0.02 \pm 0.03$ & $0.00$  \\
   $y_{BH}\ \ \ [\arcsec]$                             & $-0.001\pm 0.002$ & $-0.001 \pm 0.004$ & $-0.002 \pm 0.009$ & $0.0088$ \\
   \noalign{\smallskip}
   \hline
 \end{tabular} 
\begin{list}{}{}
\item[$^a$] Parameter not constrained from the fit
\item[$^b$] Parameter hold fixed
\end{list}
\end{table*}
\begin{table*}
  \caption[!ht]{Effect of varying the intrinsic flux distribution at sub-resolution scales: fit results.}
  \label{tab2B}
  \centering
  \begin{tabular}{l c c c c c}
    \hline
    \noalign{\smallskip}
    Parameter & \multicolumn{4}{c}{Fit result} & Model value\\
    \noalign{\smallskip}
    & flux distribution A$^f$ & flux distribution B$^f$ & flux distribution C$^f$ & flux distribution D$^f$ &\\
   \noalign{\smallskip}
   \hline
   \noalign{\smallskip}
   $\theta_{LON}\ \ \ [^{\circ}]\ \ \ \ (LON\  fit)$             & $-1.0\pm1.3$  & $0.9\pm1.1$ & $-4.1\pm1.3$ & $5.7\pm8.3$ & $0.00$\\
   $b\ \ \ [\arcsec]\ \ \ \ \ (LON\  fit)$                               & $0.001\pm0.002$  & $0.000\pm0.002$ & $0.002\pm0.003$ & $0.02\pm0.01$ & $0.00$\\
   \noalign{\smallskip}
   \hline
   \noalign{\smallskip}
   $log_{10}(M_{BH}/M_{\astrosun}) $ & $7.91\pm 0.02$ & $8.16\pm 0.03$ & $8.04\pm0.03$ & $8.13\pm 0.10$ & $8.00$\\
   $log_{10}(M/L)\ \ \ [ M_{\astrosun}/L_{\astrosun} ]\ \ $ & $-8.23\ ^a$ & $-7.41\ ^a$ & $-8.15\ ^a$  & $-6.68\ ^a$ & $0.00$\\
   $S_0\ \ \ [\arcsec]$                    & $0.012\pm0.008$  & $-0.01\pm0.03$ & $-0.01\pm0.01$ & $0.02\pm0.03$ & $0.00$\\
   $V_{sys}\ \ \ [km/s]$                              & $534\pm 18$ & $494\pm32$ & $478\pm35$ & $500\ ^b$ & $500$\\
   $i\ \ \ [^{\circ}]$                                & $35.0\ ^b$ & $35.0\ ^b$ & $35.0\ ^b$ & $35.0\ ^b$ & $35.0$\\
   $\Delta_{sys}\ \ \ [km/s]$                             & $8$ & $28$ & $22$ & $0$ & $-$ \\
   \noalign{\smallskip}
   \noalign{\smallskip}
   reduced $\chi^2 \ \ \ (\chi^2/D.O.F.)$ & $1.05\ \ (5.25/5)$ & $1.05\ \ (4.19/4)$ & $1.05\ \ (6.28/6)$ & $1.006\ \ (1.006/1)$ & $-$\\
   \noalign{\smallskip}
   \hline
   \noalign{\smallskip}
   $x_{BH}\ \ \ [\arcsec]$                           & $0.012\pm 0.008$ & $-0.01 \pm 0.03$ & $-0.01\pm0.01$ & $0.02\pm0.03$ & $0.00$  \\
   $y_{BH}\ \ \ [\arcsec]$                             & $0.001\pm 0.002$ & $0.001 \pm 0.003$ & $0.001\pm 0.002$ & $0.02\pm 0.02$ & $0.00$ \\
   \noalign{\smallskip}
   \hline
 \end{tabular} 
\begin{list}{}{}
\item[$^f$] Flux distribution. A: exponential centered on $(0\arcsec,0\arcsec)$ with $r_0=0.05\arcsec$. B: exponential centered on $(0\arcsec,0\arcsec)$ with $r_0=0.15\arcsec$. C: exponential centered on $(0\arcsec,0\arcsec)$ with $r_0=0.05\arcsec$ and a central hole of radius $r_h=0.04\arcsec$. D: exponential centered on $(0.1\arcsec,0.1\arcsec)$ with $r_0=0.05\arcsec$.
\item[$^a$] Parameter not constrained from the fit
\item[$^b$] Parameter hold fixed
\end{list}
\end{table*}

\begin{table*}
  \caption[!ht]{Application of the method to low BH masses: fit results.}
  \label{tab2}
  \centering
  \begin{tabular}{l c c c c c}
    \hline
    \noalign{\smallskip}
     Parameter & \multicolumn{2}{c}{$M_{BH}=10^7M_{\astrosun}$ model} & \multicolumn{3}{c}{$M_{BH}=10^{6.5}M_{\astrosun}$ model}\\
    \noalign{\smallskip}
    \hline
    \noalign{\smallskip}
     & Fit result & Model parameters & Fit result $^A$ & Fit result $^B$ & Model parameters \\
    \noalign{\smallskip}
    \hline
    \noalign{\smallskip}
    $\theta_{LON}\ \ \ [^{\circ}]\ \ \ \ (line\ of\ nodes\ fit)$ & $0\pm2$  & $0.00$ & $-2 \pm5$ & $-0.5\pm0.7$&$0.00$\\
    $b\ \ \ [\arcsec]\ \ \ \ \ (line\ of\ nodes\ fit)$                   & $-0.007\pm0.005$&$0.00$ & $-0.02 \pm0.02$ & $-0.001\pm0.002$&$0.00$\\
    \noalign{\smallskip}
    \hline
    \noalign{\smallskip}
    $log_{10}(M_{BH})\ \ \ [log_{10}(M_{\astrosun})]$ & $7.20\pm 0.03$ & $7.00$ & $7.23\pm 0.04$ & $6.82\pm0.08$ &$6.50$\\
    $log_{10}(M/L)\ \ \ [log_{10}(M_{\astrosun}/L_{\astrosun})]\ \ \ \ $ & $0.00\ ^b$ & $0.00$ & $0.00\ ^b$ & $0.00\ ^b$ &$0.00$\\
    $S_0\ \ \ [\arcsec]$                              & $-0.03\pm 0.02$ &  $0.00$ & $0.00\ ^b$ & $0.025\pm0.00\ ^c$& $0.00$\\
    $V_{sys}\ \ \ [km/s]$                              & $493\pm 6$ &  $500$ & $500\ ^b$ & $503\pm4$& $500.0$\\
    $i\ \ \ [^{\circ}]$                                & $35.0\ ^b$ & $35.0$ & $35.0\ ^b$ & $35.0\ ^b$ &$35.0$\\
    $\Delta_{sys}\ \ \ [km/s]$                             & $0$ & $-$& $0$ & $8$& $-$\\
    \noalign{\smallskip}
    \noalign{\smallskip}
    reduced $chi^2 \ \ \ (\chi^2/D.O.F.)$ & $0.04\ \ (0.04/1)$ & $-$ & $0.26\ \ (0.26/1)$ & $1.13\ \ (1.13/1)$ &$-$\\
    \noalign{\smallskip}
    \hline
    \noalign{\smallskip}
    $x_{BH}\ \ \ [\arcsec]$                           & $-0.03 \pm 0.02$ & $0.00$ & $0.00 \pm 0.00\ ^d$ & $0.025 \pm 0.00\ ^d$& $0.00$ \\
    $y_{BH}\ \ \ [\arcsec]$                             & $-0.007 \pm 0.005$ & $0.00$& $-0.02 \pm 0.02$ & $-0.001 \pm 0.003$& $0.00$\\
    \noalign{\smallskip}
    \hline
  \end{tabular} 
\begin{list}{}{}
\item[$^A$] Simulation of the model with $50km/s$ velocity bins
\item[$^B$] Simulation of the model with $10km/s$ velocity bins
\item[$^b$] Parameter hold fixed
\item[$^c$] Parameter at the edge of allowed range
\item[$^d$] Parameter calculated from fixed parameters
\end{list}
\end{table*}
We now present a set of simulations where we vary the intrinsic flux distribution of the gas at sub-resolution scales and observe how this affects the recovery of the input model parameter values from the fit. As in section \ref{s326} we adopt  a basic model for the flux distribution that is an exponential function ($I(r)=A e^{r/r_0}$) where
the symmetry is center located at the position $(x_0,y_0)=(0\arcsec,0\arcsec)$ and the characteristic radius is set to $r_0=0.05\arcsec$. In Figs. \ref{fig25B} we show the cases of: the basic flux distribution, flux distribution with characteristic radius $r_0=0.15\arcsec$, flux distribution with central hole of radius $r_h=0.04\arcsec$, flux distribution with characteristic radius $r_0=0.05\arcsec$ but centered at the position $(x_0,y_0)=(0.1\arcsec,0.1\arcsec)$, all with a simulated noise for a S/N of $100$.  In Table \ref{tab2B} we report the best fit values of the free parameters. These simulations are chosen to represent extreme cases of sub resolution variation of the intrinsic flux distribution.

We can observe that with the fit we can recover the correct BH mass value with an accuracy of better than $\sim0.16dex$, showing that even "extreme" variations of the flux distribution at sub-resolution scales, do not greatly affect the BH mass estimate.

Finally we consider a set of simulations where we decrease the
$M_{BH}$ value. In Figs. \ref{fig25} we show the cases of BH
masses $M_{BH}=10^7M_{\astrosun}$ and $M_{BH}=10^{6.5}M_{\astrosun}$ respectively, both
with a simulated noise for a S/N of $100$.  In Table \ref{tab2} we report the
best fit values of the free parameters.

The case of lower black hole mass is particularly significative. 
We choose a $M_{BH}=10^{6.5}M_{\astrosun}$ value because, as
observed in section \ref{s327}, the standard
rotation curve method cannot distinguish between a case of 
$10^{6.5}M_{\astrosun}$ BH and no BH. By looking at Fig. \ref{fig25}
we note that the BH signature is still present in a realistic
spectroastrometric curve.

  \begin{figure}[!ht]
  \centering
  \includegraphics[width=0.9\linewidth]{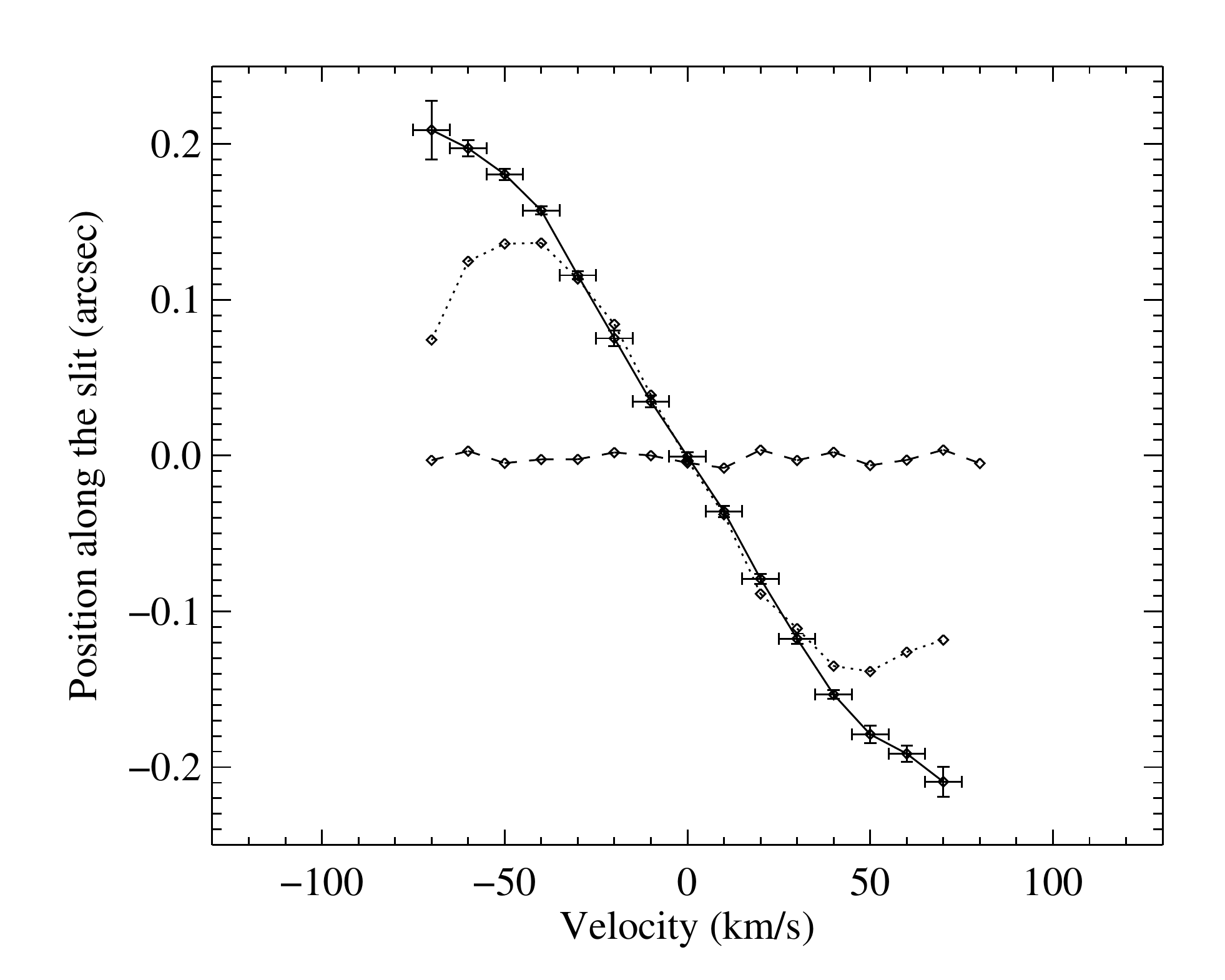}
\includegraphics[width=0.9\linewidth]{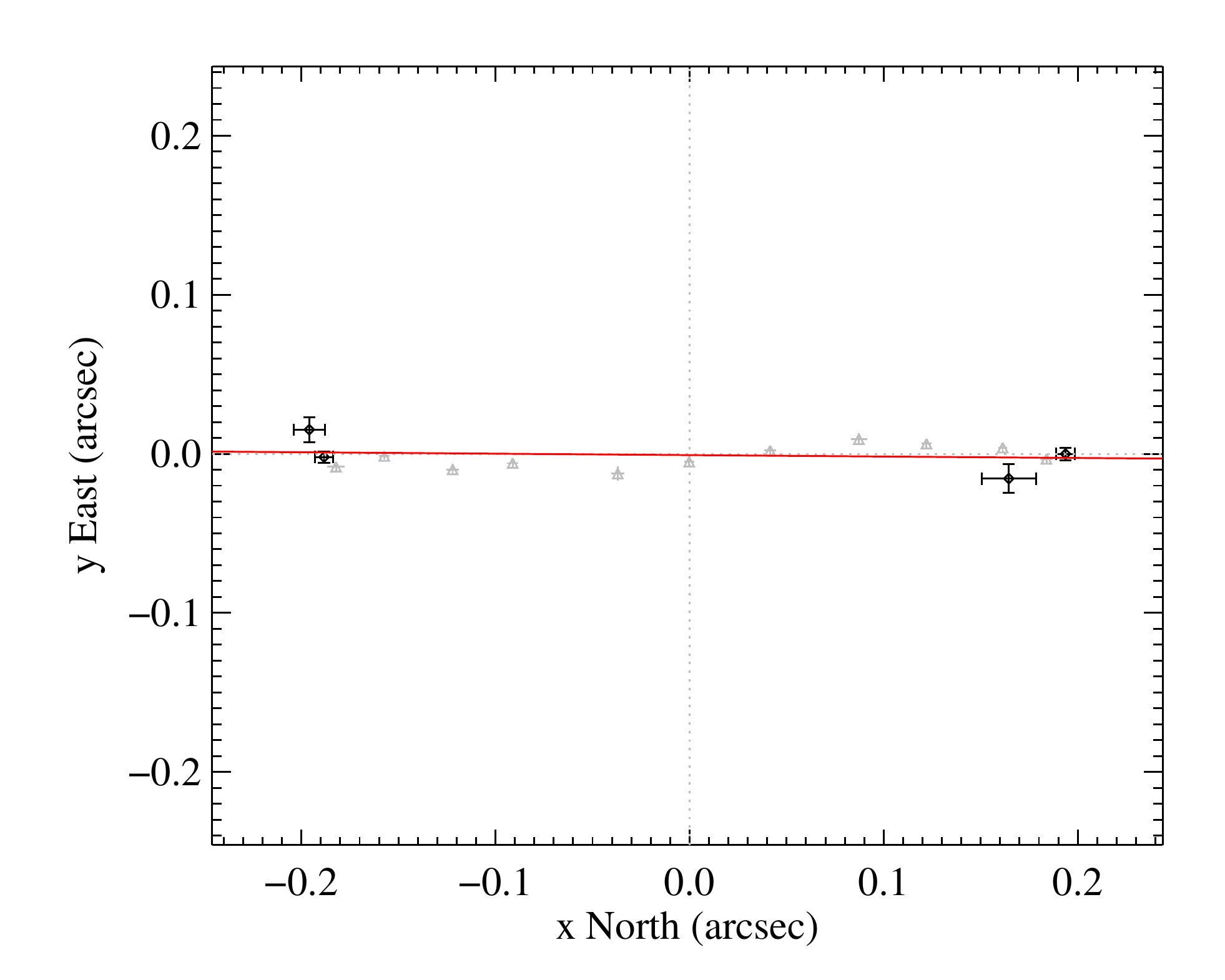}
\includegraphics[width=0.9\linewidth]{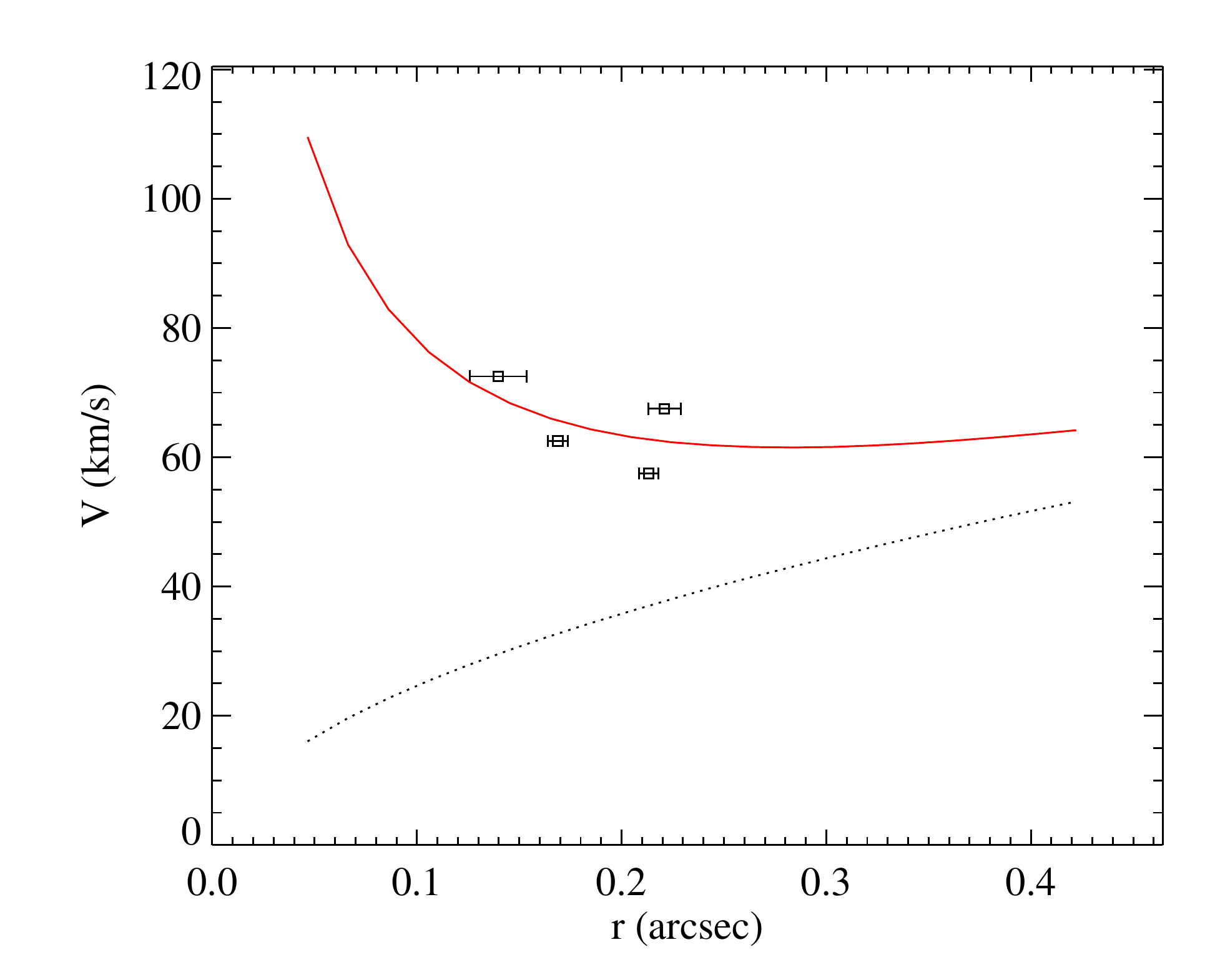}
     \caption{Application of the method to low BH masses: results from the simulations with BH mass of
 $M_{BH}=10^{6.5}M_{\astrosun}$ and velocity bins of $10km/s$. Top: Spectroastrometric curves (for simplicity we plot the error bars of the points only for the PA$=0^{\circ}$ curve). Middle:  spectroastrometric map. Lower panel: The result of the fit.}
        \label{fig27}
  \end{figure}

Nonetheless, the fit is not well constrained by the few datapoints.
Furthermore, we obtain a value of $M_{BH}$ of $10^{7.2}M_{\astrosun}$ (Table
\ref{tab2}), substantially larger than the model mass of
$10^{6.5}M_{\astrosun}$.

This is due to the fact that in our simulation the size of the velocity bins
in the PVD diagram is 50 km/s, matched to the assumed spectral resolution (R
$\sim$ 6000). This value is insufficient to adequately resolve the emission
line profile. As observed in sec. \ref{s324}, the spectroastrometric curves
are stretched along the velocity axis with artificially high velocity values
which result in an overestimated BH mass.

This problem can be overcome by increasing the spectral resolution of the
simulated data to a higher value, R $\sim$ 30000, still well within reach of
existing spectrographs.  We then repeated the
simulation setting the size of the velocity bin of the longslit spectra to 10
km/s. In Fig. \ref{fig27} we show the spectroastrometric
curves and the result of the fit for this model.  The sampling of the
spectroastrometric curves is largely improved and the fit is consequentely
better constrained. The derived BH mass value, 
$\log (M_{BH}/M_{\astrosun}) = 6.8\pm 0.1$
is now in better agreement with the model value.
The results of these simulations confirm that to assess the accuracy of BH masses from the spectroastrometric method one should first check that the line widths in the central emission region are well resolved spectrally.

In the last fits reported in table \ref{tab2} we hold fixed the mass-to-ligth ratio $M/L$  to the model value. The number of ''high velocity'' points in the 
spectroastrometric curves is small (only $2$ for the model $M_{BH}=10^{6.5}M_{\astrosun}$ with $50 km/s$ velocity bin) and it is not possible to constraint the radial dependence of the mass distribution. In general, when dealing with real data with very few ''high velocity" points, one can only measure the total mass enclosed in the smaller radius which can be estimated from spectroastrometry. If it is possible to analyze the ``classical'' rotation curve, one could determine the 
mass-to-ligth ratio $M/L$ from that and use it in the spectroastrometric analysis. The fixed mass-to-ligth ratios used in the above analysis should be interpreted as resulting from a classical analysis of the extended rotation curves.

\section{Integral Field Spectroscopy}   
\label{ifu}
The extension of the spectroastrometric technique from multi-slit observations to Integral
Field Units (IFU) is straightforward, and carries with it many advantages. With IFUs
the complex issue of the limited spatial coverage of the slits (leading to the
problem of velocity truncation) is partially removed and a full 2D map can be directly derived.
Another substantial advantage is the reduced observing time requested to
obtain a given S/N level, since there is no need to obtain observations of the
same galaxy at different silt position angle.

The analysis of IFU data now reduces to fitting a 2D gaussian to each channel
map in turn yielding the X,Y photocenters as a function of velocity, i.e. the
spectroastrometric map. Armed with these derived photocenters one proceeds
directly to the application of Sect. \ref{s44}.

In practice it is important to determine the accuracy of each photocenter.
This can easily be accomplished by using monte carlo realizations drawn
from the original data in each channel giving a distribution of i values for X
and Y from which one can determine a median and inter-quartile spread. 

In order to provide a quick analysis of advantages and disadvantages in the use of IFUs ve longslit spectra, we first recall the mandatory requirements needed to obtain an accurate and useful spectroastrometric curve.
The spectra must be characterized by good spectral resolution, signal-to-noise ratio and spatial sampling.
The emission line profile must be well spectrally resolved otherwise, as discussed in Sec.~\ref{s324}, one would overestimate the  BH mass. 
High signal-to-noise and spatial sampling are required for a robust and accurate robust estimate of the spatial centroid of  the emission line, as discussed in detail in appendix \ref{a1}.

Compared to longslit spectrographs, IFUs have the advantage of a two-dimensional spatial covering of the source which provides a more direct and accurate determination of the spectroastrometric map on the plane of the sky.
However, for a given number of detector pixels, this is usually done at the expense of spectral resolution thus limiting the application of the spectroastrometric analysis to larger BHs.
As a compromise between field-of-view and spectral resolution, an IFU can have a lower spatial sampling compared to longslit spectra which limits the accuracy of the photocenter determination.
However, with IFUs it is possible to integrate longer, since one does not need to obtain spectra at different position angles.
 
As an example, let us compare the seeing-limited performances of ISAAC and SINFONI (both at the ESO VLT) for spectroastrometric use. 
ISAAC has a spatial sampling along the slit of $0.147\arcsec$, where for SINFONI  in the seeing limited mode spatial pixels (spaxels) have angular dimensions of $0.125\arcsec\times0.25\arcsec$. There exist modes with finer spatial sampling ($0.05\arcsec\times0.1\arcsec$) but at the expense of a smaller field of view which might be insufficient in seeing limited observations.
Overall, the spatial sampling in ISAAC and SINFONI are not extremely different, at least for moderately good seeing ($\sim 0\farcs5$).
Regarding the spectral resolution, let consider the case of the K (J) band, at $2.2\mu m$ ($\sim1.2\mu m$). ISAAC, in the Short Wavelength Medium Resolution configuration offer a spectral resolution of $8900$ ($10500$) while SINFONI with the K (J) grating offers a spectral resolution of $4000$ ($2000$).
Clearly, the advantage of the 2D coverage provided by SINFONI is obtained at the expense of spectral resolution, and therefore SINFONI is not a good choice to detect small BHs (or large BHs in more distat galaxies).

In conclusions, IFUs have the advantage of the two-dimensional spatial covering which allows longer integrations on source and more accurate spectroastrometric maps. On the other hand, longslit spectrographs can provide a better combination of high spectral resolution and spatial sampling for the detection of smaller BHs.

Regardless of the use of IFUs or longslit spectrographs, the super-resolution provided by spectroastrometry 
allows BH mass determinations in galaxies at distances substantially higher than those that can be studied with the ``standard'' rotation curves method.

\section{Combining Spectroastrometric and Rotation Curves}

In this paper we have shown that by means of spectroastrometric curves we can recover information at scales smaller than the spatial resolution of the observations and we have provided a simple method to estimate BH masses. However, spectroastrometry is not a replacement of the standard rotation curve method. As shown in Sec.~3 (e.g.~Fig.~1), the spectroastrometric curve is complementary to the rotation curve and it is clear that any kinematical model must account for both curves at the same time. For example, any contribution from extended mass distributions (e.g.~stars) can be well constrained with rotation curves but not with spectroastrometric curves which sample very small scales. A detailed analysis of the constraints posed simultaneously by spectroastrometric and rotation curves will be discussed in forthcoming papers.

The  reader might wonder why a combined modeling of spectroastrometric and rotation curves should be better than than modeling the full Position-Velocity diagram.
The reason is very simple and is related to the fact that the full PVD depends on the unknown intrinsic flux distribution of the emission line. \cite{Marconi:2006} showed that line profiles do depend on the assumed flux distribution while, on the contrary, such dependence is much weaker on the first moment of  the line profile  (i.e.~the mean velocity). In this paper we have shown that the spectroastrometric curve depends on the assumed flux distribution in the low velocity range, while such dependence almost disappears in the high velocity range. Therefore, by selecting the rotation curve and the spectroastrometric curve we can greatly diminish the effects of the unknown intrinsic flux distribution which plague the full PVD.

\section{Summary and Conclusions}\label{s5}

In this paper we have discussed the application of the spectroastrometric method in the context of kinematical studies aimed at measuring the masses of supermassive black holes in galactic nuclei and we have presented its advantages compared to classical "rotation curves" method. 

We have conducted an extensive set of simulations which have shown that the presence of a supermassive black hole is revealed by a turn-over in the spectroastrometric curve, with the high velocity component approaching a null spatial offset from the location of the galaxy nucleus.
All the relevant information about the BH is encoded in the "high velocity" range of the spectroastrometric curve, which is almost independent of the spatial resolution of the observations.  According to our simulations, the use of spectroastrometry can 
 allow the detection of BHs whose apparent size of the sphere of influence is as small as $\sim 1/10$ of the spatial resolution.
 
We have then provided a simple method to estimate BH masses from spectroastrometric curves.
This method consist in the determination of the spectroastrometric map, that is the positions of emission line photocenters in the available velocity bins. This is trivially obtained from IFUs but can also be obtained by combining longslit spectra centered on a galaxy nucleus and obtained at different position angles.
From the "high velocity" points in the spectroastrometric map one can then  obtain a rotation curve to trivially estimate the BH mass.
We have provided practical applications of this method to simulated data but considering noise at different levels.
From this analysis, we confirm that with seeing limited observations ($\sim 0\farcs5$) we are able to detect a BH with mass $10^{6.5}\,M_\odot\, (D/3.5\,Mpc)$, where $D$ is the galaxy distance. 
This is a factor $\sim 10$ better than can be done with the classical method based on rotation curves.

Finally, we have discussed advantages and disadvantages of IFU vs longslit spectrographs concluding that  IFUs have the advantage of the two-dimensional spatial covering which allows longer integrations on source and more accurate spectroastrometric maps. On the other hand, longslit spectrographs can provide a better combination of high spectral resolution and spatial sampling for the detection of smaller BHs.
Regardless of the adopted type of spectrograph,  the super-resolution provided by spectroastrometry 
allows BH mass determinations in galaxies at distances substantially higher than those that can be studied with the ``standard'' rotation curves method.

\bibliographystyle{aa} 

\appendix
\section{The measure of the light profile centroid}\label{a1}

\begin{figure}[!th]
  \centering
  \includegraphics[width=0.9\linewidth]{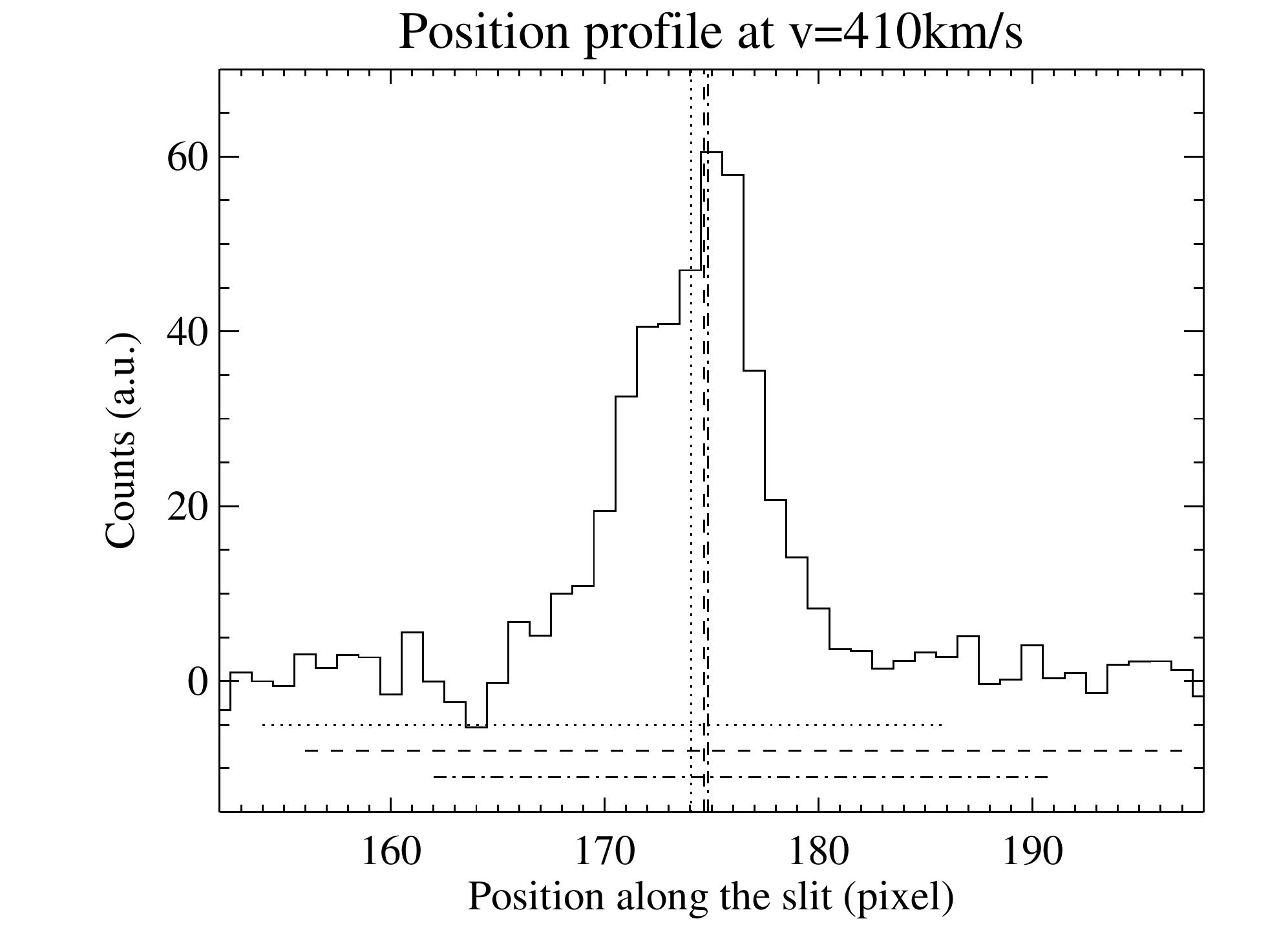} 

     \caption{Example of a light profile along the slit for a given velocity bin for real data. Three different position ranges are indicated as horizontal lines and the corresponding centroid positions are indicated with vertical lines with the same style. }
        \label{fig03}
  \end{figure}

The first fundamental step in the application of the spectroastrometric method is to find the best possible way to estimate the centroid position of the light profile for a given wavelength (or velocity).
In principle one could simply estimate the centroid position as the weighted mean of the position with the corresponding emission line flux. If $y_i$ denotes the position and $I_i$ the flux of the i-th pixel along the slit (for a given velocity), the centroid position $y_{cent}$ is simply:
\begin{equation}
y_{cent}=\frac{\sum_i{y_iI_i}}{\sum_i{I_i}}
\label{ea2}
\end{equation}
The problem in using equation \ref{ea2} resides in the presence of noise and in the fact that the value of $y_{cent}$ depends on the choice of the position range used to compute the weighted mean.
In Fig. \ref{fig03} we exemplify the effect of choosing different ranges to calculate the centroid of real data. This effect will not be negligible especially because one  is looking for precisions which are much lower than the spatial resolution (e.g., $\sim2.5$ pixels for the data shown in the figure).

Another possibility to estimate the centroid position is provided by a parametric fitting of the light profile with an appropriate function. For instance, we have tried a Gauss-Hermite expansion:
\begin{equation}
F(x)=Ae^{\frac{y^2}{2}}[1+h_3H_3(y)+h_4H_4(y)]
\label{ea2b}
\end{equation}
where $y=(x-x_0)/\sigma$ and $H_3$ and $H_4$ are the Gauss-Hermite polynomials of the third and fourth order.

The advantage of the fitting method is that with the parametric fit the results are much less sensitive to noise and choice of the position range for the analysis
but it is difficult to find an appropriate parametric function which can reproduce the profile shape without a large increase of the free parameters which, in turn, decreases the accuracy of the centroid position. The light profile along the slit has often a complex shape, sometimes with multiple peaks and there is not any physical reason to suggest a particular parametric function.

Moreover, the shape of the light profile for a given velocity bin is determined not only by velocity field of the rotating gas disk but also 
by many other factors, like the intrinsic light distribution of the emission line, the slit width and position and many others. Therefore one has to investigate what is the signature which is most related to the velocity field.

The piece of information we seek is the average position of the gas rotating at a given observed velocity which can be converted into the gravitational potential in the nuclear region, with the usual assumption of a thin, circularly rotating disk.

As explained above, the use of the entire light profile along the slit for a given velocity bin to calculate the photocenter position is extremely complex, moreover such position might be influenced by many features which do are not connected with the rotation of the disk.
After testing the method with real and simulated data, we have reached the conclusion that the quantity we need to measure is not the centroid of the whole light profile but only of the central spatially unresolved component which is related to gas located very close to the galaxy nucleus (e.g.~gas rotating around the BH).
Thus, the problem of measuring the light centroid along the slit can be transformed into measuring the position of the peak of the central unresolved emission.

  \begin{figure}[!th]
  \centering
  \includegraphics[width=0.9\linewidth]{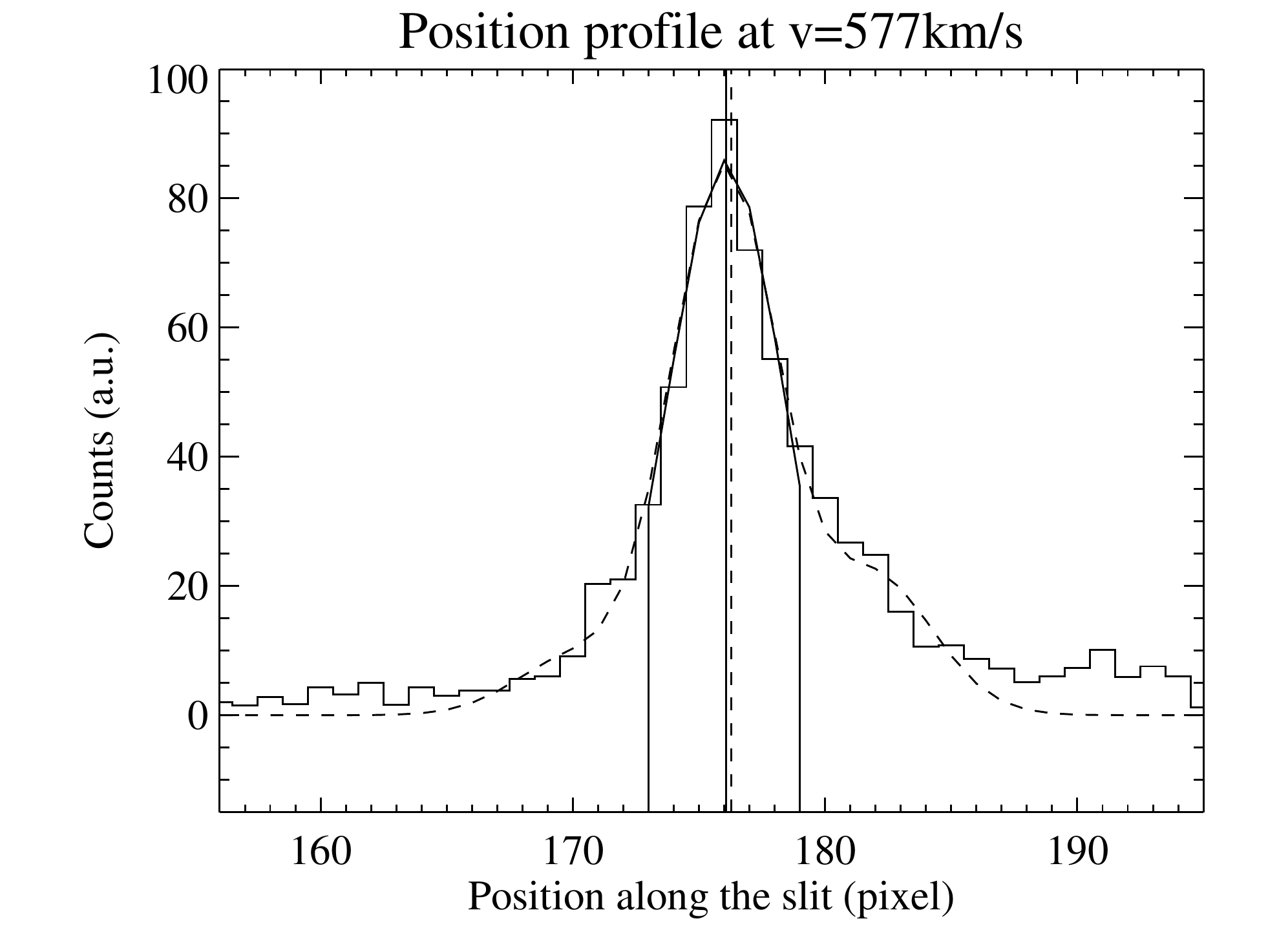}
  \includegraphics[width=0.9\linewidth]{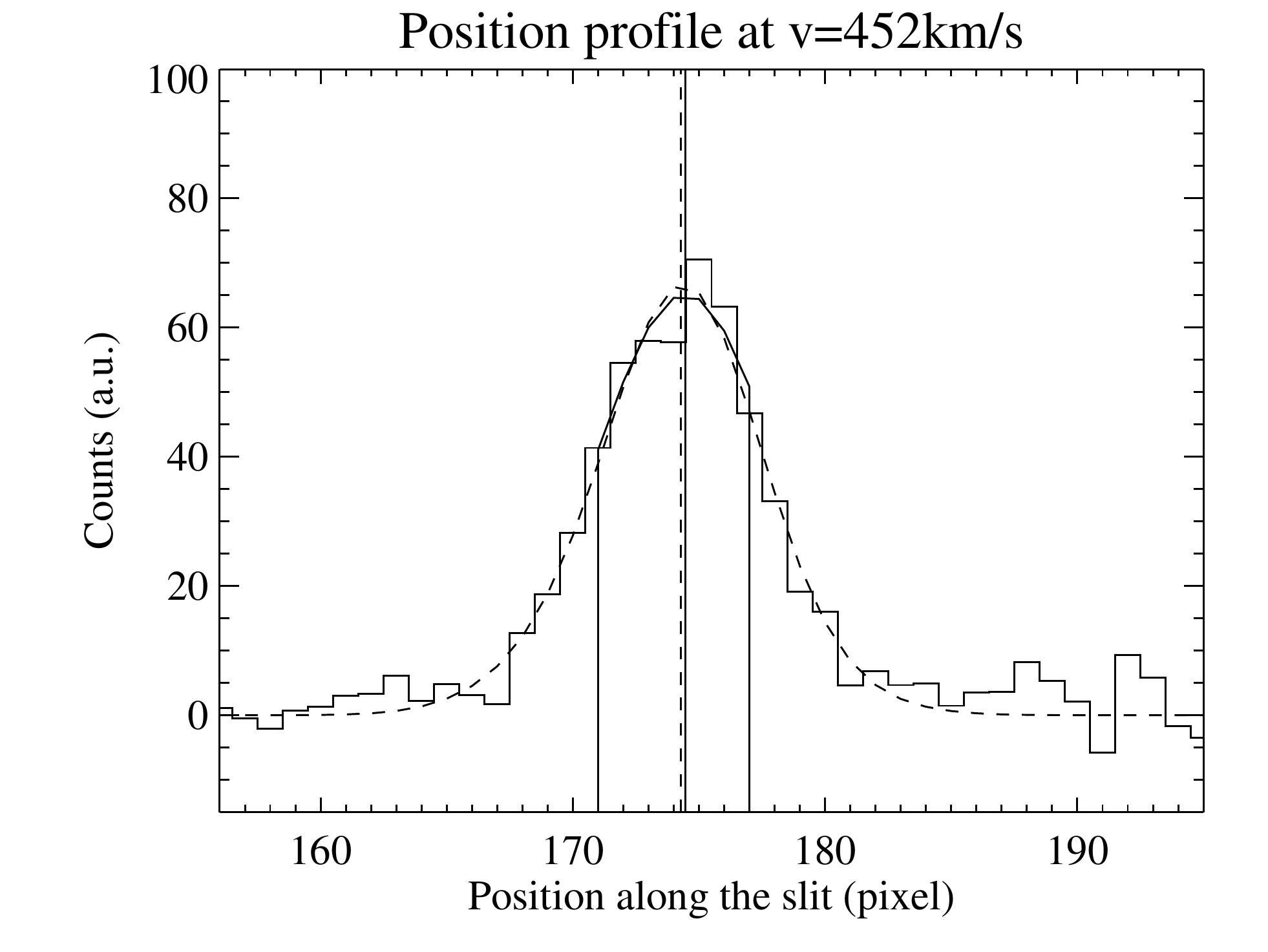}
 
     \caption{Example of centroid determination on a real spectrum. The dashed line denotes the fitted gauss-hermite functions and the relative centroid position (vertical dashed line). The solid lines denotes the gaussian function which has been fitted in a window around the principal peak as described in the text. The vertical solid lines represent the edges of the window and the gaussian center respectively.}
        \label{fig04}
  \end{figure}

The profile peak at the location of the nucleus is due to the spatially unresolved emission from the innermost parts of the rotating gas disk. Because it is spatially unresolved, the width of central peak is expected to be of the order of the spatial resolution, and its shape can be well approximated with a gaussian function. This component is easy to identify and fit. 
Moreover this unresolved component is emitted from the innermost parts of the gas disk, where the gravitational potential of the BH is strongest with a larger influence on the gas kinematics.
Conversely, the spatially resolved features are very complex to fit and are influenced by many factors not directly connected with the gravitational potential.

In practice, we first obtain an initial guess of the principal peak position and we 
select the light profile within a box centered on this initial guess and whose width is of the order of the spatial resolution.
We then accurately locate the peak position by fitting a simple gaussian function to this selected portion of the light profile.
In details, we usually set the width of the box as $k\sigma_0$ where $\sigma_0$ is the spatial resolution (standard deviation of the gaussian approximating the point spread function) and $k$ a constant between $\simeq 1$ and $\simeq 3$. Using values of $k$ too close to $1$ results in too few fit points but 
$k$ should also not be too large if we want to consider only the unresolved component. From our tests, the best choice has come out to be $k\simeq2$.

This method used to obtain the light centroid provides us with an important control parameter, that is the width of the fitted gaussian: if this is much larger than the spatial resolution, this is an indication that line emission is dominated by a spatially extended component and the obtained value of the centroid is less reliable.

In Fig. \ref{fig04} we show an example of centroid determination obtained by fitting a gauss-hermite function to the whole profile and a simple gaussian in a $2\sigma_0$ wide window centered around the principal peak. The centroid values obtained with the two methods might or might not be consistent but it is clear that the centroid value with the first method is shifted to the right by the contribution of the profile wings, which, moreover, are not even well reproduced by the gauss-hermite function. In the bottom panel of Fig. \ref{fig04} none of the two methods can well reproduce the profile peak, but the width of the fitted gaussian turns out to be $\sim2.5$ times larger than the spatial resolution indicating that the principal peak is partially resolved, and the value of the centroid is less reliable.

  \begin{figure}[!th]
  \centering
  \includegraphics[width=0.9\linewidth]{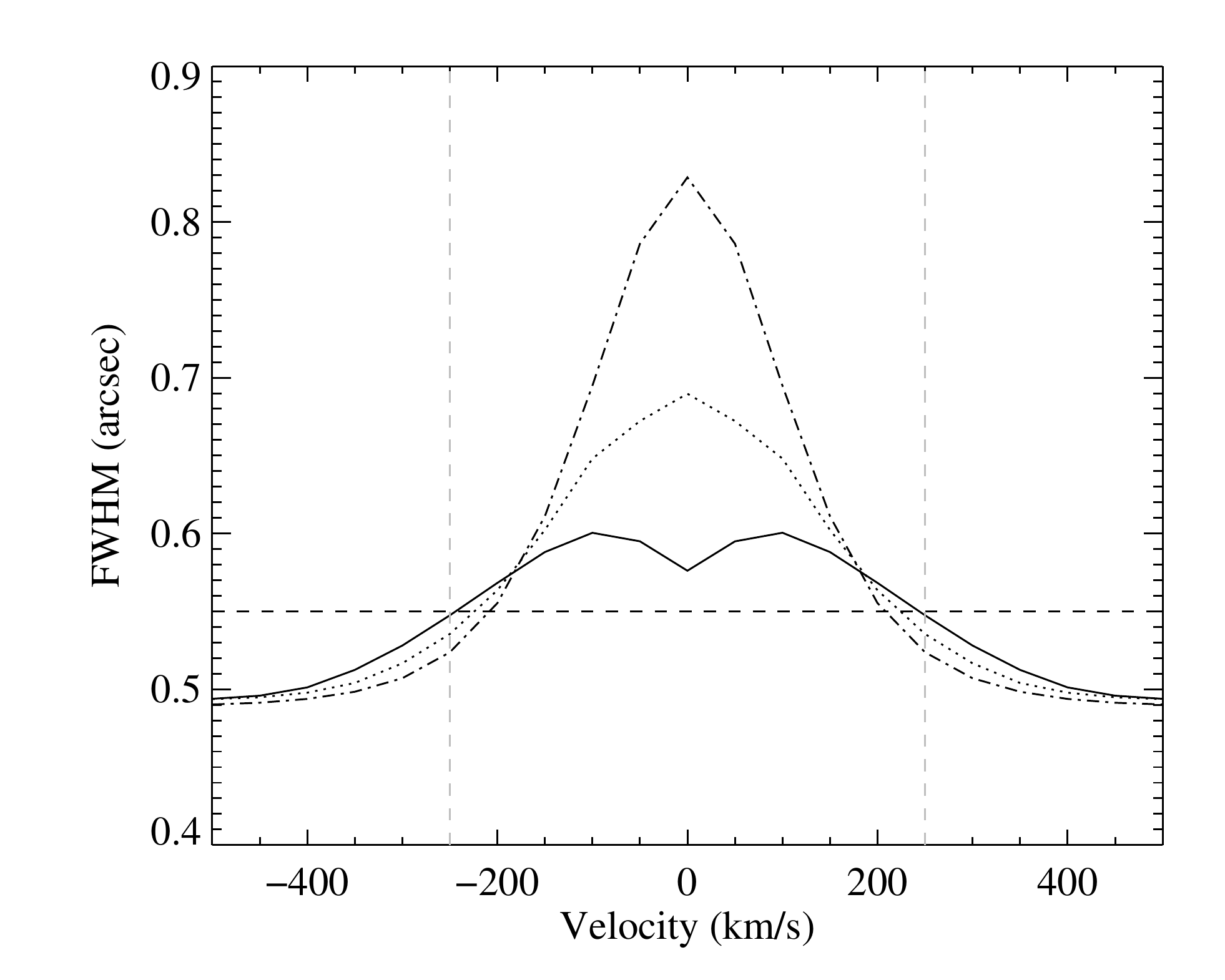}

     \caption{FWHM of the gaussian fitted to the principal peak of the light profile principal peak for spectroastrometric curves of fig. \ref{fig13}. Solid line: PA1 curve. Dotted line: PA2 curve. Dot-dashed lines: PA3 curve. The horizontal dashed line denotes $1.1$ times the PSF FWHM. The gray dashed lines are the estimated limits of the ``high velocities''.}
        \label{fig15}
  \end{figure}
Following these considerations, we found that the most robust and objective way of defining the ``high velocity'' range is through the width of the gaussian fitted to the principal peak of the light profile: in Fig. \ref{fig15} we display the
FWHMs (Full Width Half Maxima) of the gaussian fitted to the principal peak of
the light profile for spectroastrometric curves of Fig. \ref{fig13}. We can
observe that at low velocities the FWHM increases because the peak is no
longer unresolved. We compare the FWHM to the spatial resolution ($\sim0.5
\arcsec$) because the FWHM of the light profile of an unresolved source should
be of the order of the spatial resolution.  We selected the ``high velocity''
range by imposing that the FWHM is lower than $1.1$ times the spatial
resolution (FWHM of the PSF) resulting in the range $v\lesssim-250$ km/s and
$v\gtrsim250$ km/s for the spectroastrometric curves.

\section{Building the spectroastromertic map from multiple slits observations.}\label{a3}

In this section we describe how to build a 2D spectroastrometric map, i.e. as
to estimate the photocenter position on the plane of the sky for each velocity
bin, from multiple slits observations.

We first consider a reference frame in the plane of the sky which is centered
on the center of PA1 slit and which has the X axis along the North direction.
For a given velocity bin $v_i$ the position of the light centroid on the sky
plane is $[x(v_i)$, $y(v_i)]$; the expected photocenter position along a given
slit (i.e.~a point of the spectroastrometric curve) is simply the position
$[x(v_i)$, $y(v_i)]$ projected along the slit direction that is:

\begin{equation}
P^{slit}_{theor}(v_i)=[x(v_i)-x^{slit}_0]cos(\theta_{slit})+[y(v_i)-y^{slit}_0]sen(\theta_{slit})
\label{4}
\end{equation}

where $x^{slit}_0$ and $y^{slit}_0$ are the coordinates of the slit centre and
$\theta_{slit}$ is the position angle of the slit referred to the X axis
(i.e.~the North direction as in the usual definition of PA).

The spectroastrometric curve provides the measured centroid position along a
given slit $P^{slit}_{obs}(v_i)$. Thus, in order to determine the free
parameters $x(v_i), y(v_i), x^{slit}_0, y^{slit}_0$ ($\theta_{slit}$ is
known), one can minimize the following $\chi^2$:

\begin{equation}
\chi^2_i=\sum_{slit}{\left[\frac{P^{slit}_{theor}(v_i)-P^{slit}_{obs}(v_i)}{\Delta P^{slit}_{obs}(v_i)}\right]^2}
\label{5}
\end{equation}

It is worth noticing that the relevant quantities are not the absolute
positions of the slit centers, but the relative ones. Indeed we have chosen
the center of the reference frame coincident with the center of the PA1 slit.
The number of free parameters is then 6 (2 slit centers and the photocenter
position on the sky, i.e.~6 coordinates) compared to 3 data points (the
photocenter positions from the 3 slits).  However, the problem is not
undefined since many velocity bins are available and the slit center positions
must be the same for all velocity bins.  Therefore, for each set of slit
center positions $x^{slit}_0, y^{slit}_0$ ($slit=1,2,3$) we minimize
separately all $\chi^2_i$. The best $x^{slit}_0, y^{slit}_0$ values are then
those which the minimize

\begin{equation}
\chi^2=\sum_i \chi^2_i(x^{slit}_0, y^{slit}_0)
\label{7}
\end{equation}

if $N$ points are available from each spectroastrometric curve, the number of
degrees of freedom is then $d.o.f.=3N-4-2N=N-4$, where 4 is the number of
unknown slit center coordinates and 2N is the number of unknown photocenter
positions. Since N is larger than $4$ the problem is well posed. The final
spectroastrometric map on the plane of the sky is that given by the best
fitting set of slit centers.

The sum over $v_i$ in (\ref{7}) is extended only over
the ``high velocity'' range. Indeed, in Sect. \ref{s32} we concluded that
the ``high velocity'' range of the spectroastrometric curve is more robust,
and less affected by slit losses which artificially change the photocenter
position in the low velocity range. As explained in appendix \ref{a1}, where we discuss in detail, centroid determination, we find the spectroastrometric curve by fitting a gaussian to the principal peak of line emission along the slit (that centered on the nucleus position). Then the "high velocity range" is selected by imposing that the FWHM of the fitted gaussian is lower than $1.1$ times the spatial resolution (FWHM of the PSF). This ensures that we consider oinly velocities where the line emission along the slit is spatially unresolved.

\end{document}